\documentclass[letterpaper, 12pt]{article}

\usepackage{amsmath,amssymb,mathrsfs}
\usepackage{graphicx}
\usepackage{color}
\usepackage{ushort}
\usepackage{url}
\usepackage{lscape}
\usepackage{subcaption}
\usepackage{pdflscape}
\usepackage{bbm}
\usepackage{epstopdf}
\usepackage{amsthm}
\usepackage[onehalfspacing]{setspace}
\usepackage{natbib}
\usepackage{enumerate}
\usepackage{libertine}
\usepackage[T2A, T1]{fontenc}
\usepackage[utf8]{inputenc}
\usepackage[english]{babel}
\usepackage{soul}
\usepackage{tcolorbox}
\usepackage{xcolor,colortbl}
\usepackage{threeparttable,multirow}
\usepackage{xfrac}
\usepackage{rotating}
\usepackage{tabularx,ragged2e,booktabs}
\usepackage{subfiles}
\usepackage[margin=1in]{geometry}
\usepackage{bm}
\usepackage{diagbox}

\definecolor{red}{rgb}{0.9,0.1,0.1}

\definecolor{Burgundy}{RGB}{144,0,32}
\usepackage[colorlinks=true,allcolors=Burgundy]{hyperref}
\graphicspath{{figures/}}
\newtheorem{alemma}{Lemma}[section]
\title{Multiple-Player War of Attrition\\with Asymmetric Private Values\thanks{I acknowledge financial support from Peking University. I am grateful for the helpful comments from many. Particular thanks to Hao Wang and Xi Weng. Others who made valuable suggestions are Tan Gan, Johannes H{\"o}rner, Shinsuke Kambe, Xiaosheng Mu, and Daniel Vincent.}}
\author{Hongcheng Li\thanks{Yale University, \href{mailto:hongcheng.li@yale.edu}{hongcheng.li@yale.edu}}}
\date{\today}
\newtheorem{Lemma}{Lemma}
\newtheorem{Proof of Lemma}{Proof of Lemma}
\newtheorem{Proposition}{Proposition}
\newtheorem{Proof of Proposition}{Proof of Proposition}
\newtheorem{Theorem}{Theorem}
\newtheorem{Proof of Theorem}{Proof of Theorem}
\newtheorem{Corollary}{Corollary}

\begin{document}

\maketitle
\begin{abstract}

\noindent This paper studies a war of attrition over public good provision among multiple ex-ante asymmetric privately informed players. In equilibrium, asymmetry leads to a stratified behavior pattern in which at most one player provides the public good at time zero with a positive probability, while the other players can become active one after another, each at a player-specific activation time, before which even his highest type does not provide the good. The equilibrium is essentially unique: the types who never provide make the players' costs of providing a saddle point of the equilibrium dynamics. Comparing two players in the same war, I show that the one with lower cost, more impatience, or a value distribution tilted toward high values in the sense of a higher revelation rate provides earlier at every value, other things equal. In large societies, the cost of delay is determined by the highest value across players. \\


\end{abstract}

\newpage

\section{Introduction}\label{introduction}

When cooperation and coordination are prohibitively costly, public goods can only be provided privately, through wars of attrition. Strategic delay in such situations happens on a daily basis and is a major source of social inefficiency.

One important question is whether \emph{asymmetry} matters in such conflicts. Individuals in different economic, political, or social positions usually differ in their motives and abilities, and in how others perceive these attributes. The question comes up in many applications. While socioeconomic groups attempt to offload social responsibilities such as fiscal stabilization, incumbent entities often feel more pressured to act. Similarly, when countries face riots at their common border, different levels of domestic development affect the costs and benefits of restoring order. In seeking climate change agreements, the United Nations faces countries whose diverse incentives influence their willingness to commit voluntarily.\footnote{For the first example, \citet{alesina1991stabilizations} model delayed fiscal stabilization as a war of attrition between interest groups, each privately informed about its cost of living with the unstabilized economy; stabilization occurs when one group concedes and bears a larger share of the burden. The second example corresponds to the Golden Triangle area, which is the common border of Thailand, Laos, and Myanmar and is also close to China. The area's rampant drug dealing and violent activities are the consequence of delayed and loose regulation by the neighboring countries. Behavior there is likely asymmetric, as Thailand appears to enforce stricter regulation than the others. Many other famous drug-trading areas, such as the Golden Crescent and the Silver Triangle, are also the common borders of several countries. The Paris Agreement illustrates the last example: while China has shown its willingness, the United States announced in 2017 that it would withdraw.} Asymmetry also arises from how people perceive one another: impressions based on race, gender, age, and other categories of social identity shape what each person is believed to value.

This paper asks how asymmetry changes behavior and how an asymmetric behavior pattern differentiates players' roles in causing delay and thus social inefficiency. To answer these questions, I develop a war of attrition model that combines (i) multiple players, (ii) incomplete information, and (iii) ex-ante asymmetry. I also consider continuous type spaces to enrich the equilibrium structure. Continuous private values that differ across players are the defining feature of the model, and they set it apart from the closest work, which I discuss in Section~\ref{discussion}.

In my model, multiple players are involved in a continuous-time war of attrition where each player is privately informed about his value (type) and chooses when to provide a costly public good to maximize his expected payoff. Once someone provides the good, the game ends and everyone earns a lump-sum value. Importantly, I allow players to differ in all their primitives: provision cost, discount rate, and prior distribution of values.

I characterize the essentially unique equilibrium (Lemma~\ref{Characterization}) and find that heterogeneous players behave in a \emph{stratified} pattern with two features: (i) \emph{instant exit} and (ii) \emph{delayed activation}. Instant exit, which can already arise with two players, means that a player exits the game (in my case, by providing the good) with positive probability at the very beginning. At most one player exits instantly. Delayed activation means that a player, regardless of his type, does not provide the good until a positive time, his \emph{activation time}. It can arise only with more than two players, because a player active alone would face no provision by the others and gain nothing by waiting, so at least two players are active from time 0.\footnote{\citet{kambe2019n}, who instead considers public values and binary private cost types, also shows that, with more than two asymmetric players, players can become active at different times.}

In the generic case, as illustrated in Sections~\ref{characterization} and~\ref{ltd war}, the equilibrium behavior begins with some probability of one player's instant exit. Next comes a \emph{stage} during which only two players are active. Then a third player becomes active, and in this manner stages with increasing numbers of active players follow one another until, after a sufficiently long time, all players are active. In this sense, instant exit could be construed as a ``one-player stage'' whose length is zero.

The stratified pattern grows out of the trade-off that every player faces at each moment. By waiting a little longer, a player saves the provision cost if someone else provides the good in the meantime, but delays a good he wants if no one does, so he keeps waiting only as long as the others are likely enough to provide soon. A player is less willing to wait when he values the good more or is more impatient, since he then loses more from the delay, or when his provision cost is lower, since he then has less to save by waiting.

Asymmetry turns this trade-off into the stratified pattern because players who differ are not equally willing to wait. When one player's value is high enough, he provides before another player would, even at that player's highest possible value. The other player then stays inactive at first, because early in the war the others are likely enough to provide soon to keep him waiting, even at his highest possible value. When a player's value is so high that he would provide before every other player, even at their highest possible values, the others are never likely enough to provide soon to keep him waiting, and he provides at time zero. At most one player can exit instantly in this way, because if two did, each would rather wait a moment and let the other provide. With identical players, by contrast, two players with the same value provide at the same time, so no player ever provides before another player would, even at that player's highest possible value. Every player is then active from the start, and no one exits instantly.

Comparing two players in the same war, I show that, other things equal, the one with a lower provision cost or greater impatience provides earlier at every value (Proposition~\ref{Asy Comp Sta}), just as the trade-off above suggests. The more interesting comparison concerns what the players believe about each other. When two players' values have the same range but one player's value is more likely to be high, the other player expects him to provide soon and free rides on this expectation by waiting longer. Facing this slower provision, the first player finds waiting less attractive and provides earlier, even when the two players have the same value. Being believed to care more about the good thus makes a player bear its cost sooner: whoever is perceived as valuing a public good more is pushed to provide it first. More generally, the comparison runs through the \emph{revelation rate} (the density at each value conditional on the value not exceeding it), and a player with a higher revelation rate provides earlier at every value.

As a technical contribution, I show that an equilibrium exists and is essentially unique (Theorem~\ref{exi and uni}). Uniqueness relies on \emph{stubborn} types, who value the good less than the cost of providing it and so never provide it, and I assume that every player is stubborn with some positive probability. As his other types provide over time, a player who is still waiting becomes more and more likely to be stubborn, so the share of his types that can still provide keeps falling. I show that every equilibrium must make all these shares approach zero without ever reaching it. However, doing so is delicate, because a small imbalance between players feeds on itself: if one player provides a little too slowly, the others must provide faster to keep him willing to wait, and keeping them willing to provide in turn requires him not to speed up, so he falls further behind. This feedback forces the shares to approach zero in one particular proportion, which pins down how the war is played in the long run. Formally, Lemma~\ref{asymp property} shows that the players' costs are a saddle point of the equilibrium dynamics, so that exactly one path of play converges to them, and I construct the whole equilibrium by working backward from this long-run play.\footnote{Stubborn types play the same role in \citet{fudenberg1986theory} and \citet{ponsati1995war}. With two players, these papers select the equilibrium path by ranking candidate solutions and by linearizing the planar orbit equation, respectively, but neither two-player argument extends to three or more players, so I rely on the saddle point at the costs instead.}

Beyond behavior, I study how ex-ante asymmetry affects welfare, and the clearest result is about large societies. In such a society, all players have the same cost and discount rate, but they form groups whose value distributions can take any shape. As the population grows with the groups' proportions fixed, the \emph{cost of delay}, the share of the good's value lost to waiting, converges to $c/\max_i\overline{v}_i$, the common cost divided by the highest value that any player can have, and every type loses this share of his value (Lemma~\ref{Aligned Welfare}). Efficiency and the welfare of every type therefore depend on the value distributions only through this highest value. The reason is that in a large society each player matters little on his own, so the good is provided at the pace that keeps a player with this highest value just willing to wait, and, since everyone provides slowly, this pace hardly changes over time. Theorem~\ref{U-LTD equivalency} shows that the same holds from every moment of the war onward, with the highest value that any player may still have at that moment in place of the highest value overall.

With finitely many players, the highest value also plays a leading role in \emph{upper-truncated-distribution (UTD)} wars, in which players have the same cost and discount rate and draw their values from one distribution cut off at different heights, so that they differ only in the highest value each can have, his \emph{upper bound}. Raising the highest upper bound only adds provision at time 0 and makes no one wait longer. Raising a lower upper bound, by contrast, gives that player new high types who make early provision more likely, but the other players then provide more slowly, so every type below the new bound waits longer, which works against the gain. When values are uniformly distributed, the two effects cancel exactly, and efficiency depends on the upper bounds only through the highest one (Proposition~\ref{U-LTD expect rate}).\footnote{In uniform UTD wars, I measure efficiency by $E[e^{-\rho rt_m}]$, the expected discount factor at rate $\rho r$ of the provision time $t_m$, where $r$ and $c$ are the common discount rate and cost, $\underline{v}<c$ is the lowest value, and $\rho:=1-\underline{v}/c$. When $\underline{v}=0$, this measure is the expected discount factor $E[e^{-rt_m}]$ itself, and the cost of delay is then $c/\max_i\overline{v}_i$ for every number of players.}

\emph{Outline.} Section~\ref{model} describes the model setup and the equilibrium concept. Section~\ref{equilibrium analysis} first characterizes the equilibrium and proves its existence and essential uniqueness. It then compares two players in the same war, and finally introduces a special case, the UTD war, to illustrate the behavior features and the welfare implications formally discussed later. Section~\ref{large societies and asymmetry} shows the relationship between ex-ante asymmetry and social welfare in large societies. Section~\ref{discussion} relates the paper to the literature and applies its results to an example. Appendix~\ref{appendix preliminaries} collects preliminaries, and Appendices~\ref{appendix equilibrium analysis} and~\ref{appendix large societies} contain the proofs for Sections~\ref{equilibrium analysis} and~\ref{large societies and asymmetry}.

\section{Model}\label{model}

\paragraph{Players and information.}
There is an indivisible public good potentially valuable to $N{}\geq2$ players, indexed by $i\in I_N$, where $I_N:=\{1,2,\dots,N\}$. Player $i$ knows exactly his provision cost $c_i>0$, the rate $r_i>0$ at which he exponentially discounts his payoffs, and his value of the public good $v_i$, also called his \emph{type}. The costs, discount rates{} and value distributions of all players are common knowledge, whereas each value $v_i$ is private information independently drawn from a cumulative distribution function $F_i:[\underline{v}_i,\overline{v}_i]\to[0,1]$ with $\underline{v}_i<c_i<\overline{v}_i<+\infty$ for all $i$. Assume that each $F_i$ yields a density function $f_i:[\underline{v}_i,\overline{v}_i]\to\mathbb{R}^+$ that is continuous and bounded away from $0$. I call $\overline{v}_i$ player $i$'s \emph{upper bound}, or his highest type. Player $i$'s types $v\leq c_i$, for whom providing the good never pays, are \emph{stubborn}, and I call $\underline{v}_i<c_i$ the \emph{stubborn-type assumption}. Under it, player $i$ is stubborn with probability $F_i(c_i)>0$.

\paragraph{Game and payoffs.}
A continuous-time war of attrition begins at $t=0$. Each player chooses a stopping time, at which he provides the public good if no one has provided it yet. Player $i$'s (pure) strategy is a Borel measurable function $T_i:[\underline{v}_i,\overline{v}_i]\to[0,+\infty]$ that gives the stopping time $T_i(v)$ chosen by each type $v$, where $T_i(v)=+\infty$ means that he never provides. The good is provided at the \emph{provision time} $t_m:=\min_{i\in I_N}T_i(v_i)$. At $t_m$, all players earn their values, while each provider additionally pays his own cost divided by the number of providers. Payoffs are discounted to time $0$, with $e^{-\infty}:=0$, so if all players choose to wait forever, each earns zero.

Let $F_{-i}^{\min}(t):=\operatorname{Prob}(\min_{j\neq i}T_j(v_j)\leq t)$ be the probability that at least one player other than $i$ provides the good by time $t$. If type $v$ of player $i$ chooses the stopping time $t\in[0,+\infty]$, his \emph{expected payoff} is therefore:
\begin{equation}\label{expected gain}
\begin{split}
R_i(t|v):={}&v\int_{[0,t)}e^{-r_is}\,dF_{-i}^{\min}(s)+(v-c_i)e^{-r_it}\big(1-F_{-i}^{\min}(t)\big)\\
&{}+e^{-r_it}E\Big[\Big(v-\frac{c_i}{1+\#\{j\neq i:T_j(v_j)=t\}}\Big)\mathbbm{1}\big\{\min_{j\neq i}T_j(v_j)=t\big\}\Big],
\end{split}
\end{equation}
where the first term covers the case in which some other player provides the good before $t$, the second term the case in which no other player provides by $t$ and thus player $i$ pays the cost, and the third term a tie at $t$, in which he pays only his share of the cost.

\paragraph{Equilibrium.}
A player learns nothing before the good is provided except that it has not been provided yet, so this game is strategically static. I seek to characterize its pure-strategy Bayesian equilibria. Since players' beliefs about one another's values are pinned down by the strategy profile wherever Bayes' rule applies, I describe these equilibria by their strategy profiles alone. A strategy profile $(T_i)_{i\in I_N}$ is an \emph{equilibrium} if, for every player $i$ and $F_i$-almost every type $v$, $T_i(v)$ maximizes $R_i(\cdot|v)$, as is standard with a continuum of types; see \citet{milgrom1985distributional} and \citet[p.~878]{athey2001single}. Two profiles that agree $F_i$-almost everywhere for every $i$ induce the same distributions of stopping times and the same expected payoffs, and either both are equilibria or neither is. I identify such profiles. Accordingly, an equilibrium is \emph{essentially unique} if every other equilibrium differs from it only on sets of types of probability zero.

\paragraph{Curves, activation and hazards.}
The following objects are defined for every strategy profile. For $t\geq0$, player $i$'s \emph{inverse strategy curve}, or \emph{curve} for short, is
\begin{equation*}
\Phi_i(t):=F_i^{-1}\big(\operatorname{Prob}(T_i(v_i)>t)\big),
\end{equation*}
so $F_i(\Phi_i(t))$ is the probability that player $i$ has not provided by $t$, and $\Phi_i$ is nonincreasing. When $T_i$ is nonincreasing, as in every equilibrium up to null sets of types (Lemma~\ref{Characterization}), $\Phi_i(t)$ is the highest type of player $i$ who has not provided by $t$. I call it his \emph{current type} at $t$. By independence,
\begin{equation*}
\operatorname{Prob}(t_m>t)=\prod_{j\in I_N}F_j(\Phi_j(t)),\qquad 1-F_{-i}^{\min}(t)=\prod_{j\neq i}F_j(\Phi_j(t)).
\end{equation*}

Player $i$'s \emph{activation time} is $d_i:=\inf\{t\geq0:\Phi_i(t)<\overline{v}_i\}$: he provides with probability zero before $d_i$, and with positive probability by every later time. He \emph{becomes active} at $d_i$, is \emph{active} at $t$ if $t\geq d_i$, and is \emph{inactive} before. Let $A(t):=\{i\in I_N:d_i\leq t\}$ be the set of active players at $t$, and $M(t):=\#A(t)$ the number of active players. The distinct values of $d_1,\dots,d_N$ are the \emph{activation times}, and a \emph{stage} is an interval $[d,d')$ from one activation time $d$ to the next $d'$, or $[d,+\infty)$ from the last one. On each stage, the set of active players is constant.

Let $u_i:=\Phi_i(d_i)$, and say that player $i$ has an \emph{atom} at $t$ if he provides at $t$ with positive probability. There are three kinds of player. Following \citet{myatt2005instant}, player $i$ \emph{exits instantly} if he has an atom at $0$, that is, if $\Phi_i(0)<\overline{v}_i$. He is then an \emph{instant-exit player}, and his types above $u_i=\Phi_i(0)$ are his \emph{instant-exit types}. Player $i$ is \emph{initially inactive} if $d_i>0$, a case I call \emph{delayed activation}. Every other player is active from time $0$ but provides at time $0$ with probability zero.

Player $j$ provides at hazard rate $-\frac{d}{dt}\ln F_j(\Phi_j(t))$ at $t$. Here and below, $\frac{d}{dt}$ denotes a right derivative, where it exists. Summed over all players, these rates give the \emph{provision hazard} $h(t)$, and summed over the opponents of player $i$, they give his \emph{opponents' provision hazard} $h_{-i}(t)$:
\begin{equation*}
h(t):=-\sum_{j\in I_N}\frac{d}{dt}\ln F_j(\Phi_j(t)),\qquad h_{-i}(t):=-\sum_{j\neq i}\frac{d}{dt}\ln F_j(\Phi_j(t)).
\end{equation*}
By the products above, $h(t)$ is the hazard rate of $t_m$ at $t$, and $h_{-i}(t)$ is that of $\min_{j\neq i}T_j(v_j)$.

The \emph{indifference level} of type $v$ of player $i$ is $W_i(v):=\frac{r_i}{c_i}(v-c_i)$. At every $t>0$ at which the derivatives in $h_{-i}(t)$ are two-sided, $R_i(\cdot|v)$ is differentiable at $t$ (Lemma~\ref{payoff derivative} in Appendix~\ref{appendix preliminaries}), and (\ref{expected gain}) and the products above give
\begin{equation}\label{payoff slope}
\frac{dR_i(t|v)}{dt}=e^{-r_it}\big(1-F_{-i}^{\min}(t)\big)\big[c_ih_{-i}(t)-r_i(v-c_i)\big]=c_ie^{-r_it}\big(1-F_{-i}^{\min}(t)\big)\big[h_{-i}(t)-W_i(v)\big].
\end{equation}
By waiting a little longer, type $v$ saves the cost $c_i$ if an opponent provides meanwhile, which happens at rate $h_{-i}(t)$, but delays the net gain $v-c_i$, which costs him $r_i(v-c_i)$ per unit of time. So $W_i(v)$ is the opponents' provision hazard at which type $v$ is indifferent between providing at $t$ and waiting a little longer: his expected payoff rises in $t$ while $h_{-i}(t)>W_i(v)$ and falls while $h_{-i}(t)<W_i(v)$. I call $h_{-i}(t)=W_i(v)$ his \emph{indifference condition}. Since $W_i(v)$ rises with $v$, a type of player $i$ who values the good more is willing to wait only if his opponents provide faster. Player $i$'s higher types therefore provide no later than his lower types in every equilibrium (Lemma~\ref{Characterization}(v)).\footnote{In a war of attrition over a prize, by contrast, high values hold out longest; see, for example, \citet{myatt2005instant}.}
\section{Equilibrium Analysis}\label{equilibrium analysis}

In this section, I first show in Section~\ref{characterization} a set of sufficient and necessary conditions that reveal the stratified behavior pattern of an equilibrium. In Section~\ref{existence and uniqueness}, I show existence and essential uniqueness. Section~\ref{comparative statics} then compares two players in the same war. Finally, Section~\ref{ltd war} introduces a special case, the UTD war, to illustrate some significant behavior and welfare insights which are formally discussed later.

\subsection{Characterization}\label{characterization}

I present the set of conditions for an equilibrium:

\begin{Lemma}\label{Characterization}
	A profile $(T_i)_{i\in I_N}$ is an equilibrium if and only if, after changes on null sets of types, it satisfies:
	\begin{enumerate}[(i)]
		\item For all $i\in I_N$, $T_i(v_i)=+\infty$ on $[\underline{v}_i,c_i]$, and $T_i(v_i)<+\infty$ on $(c_i,\overline{v}_i]$;
		\item For all $i\in I_N$, $\lim_{v_i\downarrow c_i}T_i(v_i)=+\infty$;
		\item At least two players have $d_i=0$;
		\item There is at most one instant-exit player, and no player has an atom at any $t>0$;
		\item For all $i\in I_N$, $T_i(v_i)$ is nonincreasing, equal to $d_i$ on $[u_i,\overline{v}_i]$, and continuous and strictly decreasing on $(c_i,u_i]$. Thus, by (ii), on $[d_i,+\infty)$ the curve $\Phi_i$ is the inverse function of $T_i$ on $(c_i,u_i]$, and it is also continuous and strictly decreasing;
		\item For every $t>0$ that is not an activation time, every curve $\Phi_i$ with $i\in A(t)$ is differentiable at $t$ and satisfies:
		\begin{equation}\label{main equations}
		\Phi_i'(t)=\frac{F_i(\Phi_i(t))}{f_i(\Phi_i(t))}\Big[W_i(\Phi_i(t))-\frac{1}{M(t)-1}\sum_{k\in A(t)}W_k(\Phi_k(t))\Big].
		\end{equation}
	\end{enumerate}
\end{Lemma}

\emph{Proof}. See Appendix~\ref{proof Characterization}.
\medskip

Every equilibrium agrees, up to null sets of types, with exactly one profile that satisfies (i)--(vi) (Appendix~\ref{proof Characterization}, Steps~11 and~12). I call this profile the \emph{canonical representative} of the equilibrium and describe the equilibrium by it.

Lemma~\ref{Characterization}(v) shows the monotonicity property explained in Section~\ref{model}; that is, the more player $i$ values the public good, the weakly earlier he chooses to provide it. Moreover, Lemma~\ref{Characterization}(i)--(v) together demonstrate an intriguing feature of the asymmetric equilibrium involving multiple players: the war of attrition starts with at least two active players, of whom at most one exits instantly. If some player's activation is delayed, the first stage ends with the start of a stage with more active players. Likewise, players leave the inactive state sequentially as the game proceeds, one or more of them at each later activation time, so that after a sufficiently long time, everyone becomes active.

\begin{figure}[htbp]
	\centering
	\includegraphics[width=.6\textwidth]{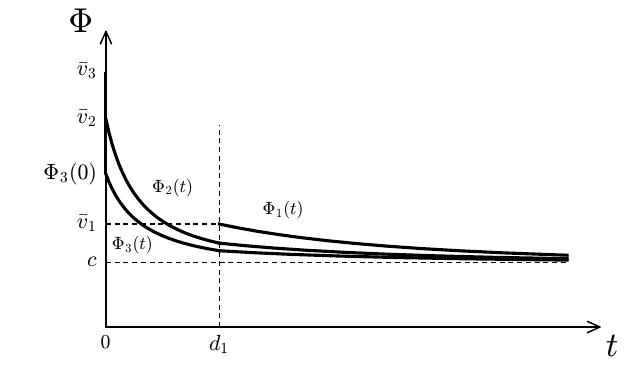}
	\caption{\footnotesize Player 3's curve $\Phi_3(t)$ starts at $\Phi_3(0)<\overline{v}_3$, corresponding to an instant exit, while player 1's curve $\Phi_1(t)$ stays at $\overline{v}_1$ until his activation time $d_1$. Before and after $d_1$, the active curves obey the two-player and three-player versions of (\ref{main equations}), respectively, and they are continuous at $d_1$, but the curves of players 2 and 3 kink there and are flatter after $d_1$, because these players provide at lower hazard rates once player 1 becomes active. All curves converge to $c$.}
	\label{Three-player Example}
\end{figure}

For a visual impression, consider three players with identical provision costs, $c_i=c$, and an information structure asymmetric enough that both instant exit and delayed activation occur. I denote the initially inactive player by 1 and his activation time by $d_1$. I depict the equilibrium curves of this three-player war in Figure~\ref{Three-player Example}, which shows the typical case in numerical examples: one player exits instantly, and a single player becomes active at each later activation time.

The time at which a type provides places the types of all players on one scale, the \emph{order of provision}; this scale is set by the equilibrium and need not be the order of the indifference levels $W_i(v)$. Players are ordered by timing as well. I call the order of the players' activation times the \emph{order of activation}, in which I place the instant-exit player, if there is one, before the other players active from time $0$. In Figure~\ref{Three-player Example}, the order of activation is player 3, player 2, player 1.

Instant exit and delayed activation arise only if not all players are identical. Two players with the same cost, discount rate and value distribution have the same curve, because exchanging their strategies maps the equilibrium to an equilibrium, and by Theorem~\ref{exi and uni} their strategies then agree up to null sets of types (as in the proof of Lemma~\ref{group structure}(i)). When all players are identical, they therefore share one curve and one upper bound, so Lemma~\ref{Characterization}(iii) makes all of them active from time $0$, and Lemma~\ref{Characterization}(iv), which allows at most one instant-exit player, rules out instant exit. When either feature arises, the order of provision places a positive-probability set of one player's types before every type of another player, except possibly that player's highest type.

The active players' indifference conditions explain why instant-exit types never wait and why an initially inactive player does. Formally, rearranging (\ref{main equations}) gives (\ref{inter equations}), which shows the trade-off that each active player $i$ faces at $t$:
\begin{equation}\label{inter equations}
-\sum_{k\in A(t),\,k\neq i}\frac{d}{dt}\ln F_k(\Phi_k(t))=W_i(\Phi_i(t)),\qquad i\in A(t).
\end{equation}
Since inactive players do not provide, the left side of (\ref{inter equations}) is the opponents' provision hazard $h_{-i}(t)$, so (\ref{inter equations}) is the indifference condition of player $i$'s current type $\Phi_i(t)$. Consider first an instant-exit type of player $i$. He values the good more than type $u_i=\Phi_i(0)$, who is just indifferent at the hazard at which his opponents provide right after time $0$. Since that hazard only falls afterwards, waiting never pays an instant-exit type, and he provides at once. When Section~\ref{existence and uniqueness} builds the equilibrium backward in time, the instant-exit types are those of player $i$ that remain once the curves of all his opponents have reached their upper bounds, so that no opponent has types left that could provide before them. After time $0$, by contrast, individual optimality requires that the types revealed at each moment make each other indifferent between providing and waiting. I call this \emph{mutual indifference}.

Mutual indifference gives the provision hazard explicitly, and with it the hazard rate at which every active player provides. At every $t>0$ that is not an activation time, each active player's hazard rate appears on the left side of (\ref{inter equations}) for each of the other $M(t)-1$ active players, so summing (\ref{inter equations}) over $A(t)$ gives the first equation below. Player $i\in A(t)$ then provides at hazard rate $h(t)-W_i(\Phi_i(t))$, which is (\ref{main equations}) in the shorter second form:
\begin{equation}\label{aggregate rate}
h(t)=\frac{1}{M(t)-1}\sum_{k\in A(t)}W_k(\Phi_k(t)),\qquad \Phi_i'(t)=\frac{F_i(\Phi_i(t))}{f_i(\Phi_i(t))}\big[W_i(\Phi_i(t))-h(t)\big],\quad i\in A(t).
\end{equation}

The provision hazard falls over time. As the high types of every active player provide and the curves fall, every active player's current type values the good less and is indifferent between providing and waiting at a lower hazard of his opponents' provision. So in equilibrium the opponents of every active player provide more and more slowly, and the provision hazard falls: the war slows down. Formally, $h$ is strictly decreasing on $(0,+\infty)$ in every equilibrium (Lemma~\ref{hazard lemma}(a) in Appendix~\ref{appendix equilibrium analysis}): on each stage, every term of the sum in (\ref{aggregate rate}) falls with the active curves, and at each positive activation time, $h$ falls by a jump, as I explain below.

The provision hazard also explains why an initially inactive player waits. Before his activation time, his opponents provide at a hazard above the indifference level of even his highest type, so they are likely enough to provide soon to keep all his types waiting. Formally, since player $i$ does not provide before $d_i$, his opponents' provision hazard $h_{-i}(t)$ equals $h(t)$ for every $t\in(0,d_i)$, and in every equilibrium $h(t)>W_i(\overline{v}_i)$ there (Lemma~\ref{hazard lemma}(c)); by (\ref{payoff slope}), each of his types gains by waiting. Yet player $i$ becomes active while $h$ still exceeds $W_i(\overline{v}_i)$. His opponents' provision hazard falls to $W_i(\overline{v}_i)$, as (\ref{inter equations}) requires, only by a jump at $d_i$, which the activation itself causes. In equilibrium, the players who become active at $d_i$, player $i$ among them, start to provide at a positive hazard rate just after $d_i$ (see the proof of Theorem~\ref{exi and uni}), and thereby add to the opponents' provision hazard of every \emph{incumbent}, that is, every player already active. An incumbent's current type stays indifferent only if his opponents' provision hazard does not jump at $d_i$, so every incumbent provides at a lower hazard rate, and the incumbents together slow down by more than the newly active players add to the provision hazard. Hence $h$ falls at $d_i$, and the incumbents' curves kink, as in Figure~\ref{Three-player Example}. Section~\ref{existence and uniqueness} shows how $d_i$ is determined.

The idea that asymmetry affects the equilibrium by changing the set of active players resembles one in \citet*{bergstrom1986private}, who develop a complete-information model without timing. They find that considerable redistribution changes the set of contributors to the provision of public goods, either shrinking or enlarging it. Unlike their model, mine incorporates incomplete information and timing, which together can make the set of active players change endogenously over time.
\subsection{Existence and Uniqueness}\label{existence and uniqueness}

I employ a backward construction to show both existence and essential uniqueness. The first step is to consider the last stage, on which all $N$ players are active. This stage starts at the last activation time $\overline{d}:=\max_{i\in I_N}d_i$.

I extend the definition of both functions $F_i$ and $f_i$ to the domain $(\overline{v}_i,+\infty)$ for every player $i$, with the constant density $f_i(\overline{v}_i)$, so that solutions can continue past the upper bounds, which equilibrium curves never do. With this extension, (\ref{main equations}) is well posed when $A(t)$ is replaced by any fixed set of at least two players and $M(t)$ by its size. Since $f_i$ is continuous and positive, the resulting system, written in the log-survival coordinates $\ln F_i(\Phi_i)$, has a continuously differentiable right side, so every initial value with $\Phi_i>\underline{v}_i$ for every player $i$ in the set yields a unique maximal solution (Lemma~\ref{log survival} in Appendix~\ref{appendix preliminaries}).

When every player is active, (\ref{main equations}) becomes the \emph{all-active system}
\begin{equation}\label{all-active}
\Phi_i'(t)=\frac{F_i(\Phi_i(t))}{f_i(\Phi_i(t))}\Big[W_i(\Phi_i(t))-\frac{1}{N-1}\sum_{k\in I_N}W_k(\Phi_k(t))\Big],\qquad i\in I_N .
\end{equation}
A solution $\Phi$ of (\ref{all-active}) \emph{converges to} $c:=(c_1,\dots,c_N)$ if $\Phi_i(t)>c_i$ for every $i$ and $t$, and $\Phi(t)\to c$ as $t\to+\infty$.

\begin{Lemma}\label{asymp property}
The all-active system (\ref{all-active}) has a solution $\Gamma=(\Gamma_i)_{i\in I_N}$ on $[0,+\infty)$, the \emph{saddle path}, with $\Gamma_i(0)\leq\overline{v}_i$ for every $i$, with equality for at least one $i$, such that:
\begin{enumerate}[(i)]
\item Uniqueness: $\Gamma$ converges to $c$, and every solution of (\ref{all-active}) that converges to $c$ is a time shift of $\Gamma$: it equals $\Gamma(\cdot-\tau)$, wherever both are defined, for some $\tau$.
\item Monotonicity: every $\Gamma_i$ is strictly decreasing.
\item Sensitivity: $c$ is a saddle point of (\ref{all-active}): its linearization at $c$ has $N-1$ unstable directions and one stable direction, spanned by a vector with positive components. Hence, from an initial value $m\in\times_{i\in I_N}[c_i,\overline{v}_i]$ at time $0$, the solution converges to $c$ if and only if $m=\Gamma(t)$ for some $t\geq0$.
\end{enumerate}
\end{Lemma}

\emph{Proof}. See Appendix~\ref{proof asymp property}.
\medskip

Lemma~\ref{asymp property}(iii) says that, when every player is active, the curves can converge to the costs only along the saddle path. Recall that in equilibrium every type above cost provides at a finite time, but types just above cost wait arbitrarily long (Lemma~\ref{Characterization}(i) and (ii)). Every curve $\Phi_i$ therefore converges to $c_i$ while staying above it. In other words, the share of player $i$'s remaining types that can still provide, $1-F_i(c_i)/F_i(\Phi_i(t))$, approaches zero without ever reaching it.

To see why the costs are a saddle point, consider curves near the costs but displaced from the saddle path. First, suppose all curves are slightly higher, as if the war were slightly earlier on the saddle path. Near $c$, the saddle path keeps the players' shares, to first order, in one particular proportion, so the shares, though all larger, still stand in this proportion. Since every player has provided less than on the saddle path, each player's current type values the good more and is willing to keep waiting only if his opponents provide faster. To first order, mutual indifference then requires all players to provide faster, at hazard rates that are the same multiple of their shares, so every share shrinks at the same rate relative to its size, and the shares approach zero in this proportion. The displacement therefore only shifts the path in time, which is why the direction that corresponds to this proportion is stable.

Second, suppose player $i$'s curve is slightly above the saddle path while his opponents' curves are on it. Player $i$ has provided less than on the saddle path, so his current type is willing to keep waiting only if his opponents provide faster, and they then do. But the current type of each opponent $j$ remains indifferent only if his opponents' provision hazard $h_{-j}(t)$ equals his indifference level $W_j(\Phi_j(t))$, which is the same as on the saddle path because his curve is on it. When $N=2$, player $i$ is $j$'s only opponent and so must provide at the same hazard rate as on the saddle path. When $N\geq3$, player $i$ must provide at a strictly lower hazard rate to offset the faster provision of $j$'s other opponents. Either way, player $i$ provides no faster than on the saddle path, while his opponents provide faster. Measured against the saddle path, he therefore falls further behind them, so the displacement feeds on itself, and the curves do not converge to the costs. This is why every other direction is unstable.

Formally, in the log-survival coordinates, the linearization of (\ref{all-active}) at $c$ weights each player $k$'s coordinate by $\frac{r_k}{c_k}\frac{F_k(c_k)}{f_k(c_k)}$. Since the unweighted matrix is symmetric, with one negative eigenvalue and $N-1$ positive ones, Sylvester's law of inertia gives, for positive weights, exactly one stable direction and $N-1$ unstable ones, as in \citet[Lemma~3.12]{fibich2012asymmetric} for asymmetric first-price auctions. With two players, the saddle path links the two players' curves by an explicit equation (Appendix~\ref{proof asymp property}). It is the path that \citet[Lemma~11(i)]{ponsati1995war} select, by linearizing at the costs, in their version of this model with two players and a common discount rate.

The stubborn-type assumption $\underline{v}_i<c_i$ drives uniqueness: it makes $F_i(c_i)>0$, which turns the costs into a saddle point. Without it, uniqueness can fail. For example, consider two players with $\underline{v}_1>c_1$ and $\underline{v}_2>c_2$, so that (\ref{all-active}) has no rest point $\Phi$ with $\Phi_1>\underline{v}_1$ and $\Phi_2>\underline{v}_2$. First, there exist degenerate equilibria in which all types of one player provide instantly. Second, multiplicity occurs even among the equilibria that satisfy the conditions of Lemma~\ref{Characterization}.\footnote{Here, $c_i$ should be replaced by $\underline{v}_i$ in clauses (i), (ii) and (v).} Indeed, for every initial value with $\underline{v}_k<\Phi_k(0)\leq\overline{v}_k$ for $k=1,2$, with equality for at least one $k$, the solution of (\ref{all-active}) satisfies $\Phi_i'(t)\leq-W_j(\underline{v}_j)F_i(\Phi_i(t))/f_i(\Phi_i(t))<0$ for $i=1,2$ and $j\neq i$. The profile with these curves therefore satisfies Lemma~\ref{Characterization}(ii) and (v), which no longer restrict the players' behavior. Each such profile is an equilibrium: by (\ref{payoff slope}) and (\ref{inter equations}), the expected payoff of type $v$ of player $i$ rises while $\Phi_i(t)>v$ and falls afterwards. For two players with a common discount rate and smooth densities, \citet[Proposition~1(iii)]{ponsati1995war} find this continuum and the degenerate equilibria.

Now, I summarize the backward construction with which I prove the existence and essential uniqueness of equilibrium. The proof takes three steps as follows.

First, in every equilibrium the last stage follows the saddle path. On this stage every player is active, so the curves solve the all-active system (Lemma~\ref{Characterization}(vi)), and they converge to $c$ (Lemma~\ref{Characterization}(i), (ii) and (v)). By Lemma~\ref{asymp property}(i), they are a time shift of $\Gamma$.

Second, going back in time along the saddle path, the curves rise until they \emph{just touch} their upper bounds, that is, until every curve is at or below its upper bound and at least one equals it. On the saddle path this happens at time $0$, and in every equilibrium it happens at $\overline{d}$, because no player has an atom after time $0$ and at most one player exits instantly (Lemma~\ref{Characterization}(iv)). Since $\Gamma$ is strictly decreasing (Lemma~\ref{asymp property}(ii)), $\Gamma(0)$ is the only point of the saddle path at which the curves just touch their upper bounds, so on the last stage, $\Phi(t)=\Gamma(t-\overline{d})$ for $t\geq\overline{d}$.

Third, the earlier stages follow by backward construction. The players whose curves touch their upper bounds at $\overline{d}$ become active at $\overline{d}$, and before $\overline{d}$ only the incumbents are active. If at least two incumbents remain, the values of their curves at $\overline{d}$ are initial values that determine these curves on the previous stage, by the well-posedness above. The players who become active at $\overline{d}$ start to provide at a positive hazard rate, because every curve of the saddle path falls at a positive rate (see the proof of Lemma~\ref{asymp property}). As Section~\ref{characterization} explains, the incumbents provide at lower hazard rates once these players become active, so just before $\overline{d}$ their curves fall faster than just after it. Going back in time, they keep rising until, at a finite earlier time, they in turn just touch their upper bounds (Lemma~\ref{backward}(b) and (c) in Appendix~\ref{proof asymp property}), and this time is the previous activation time.

Likewise, I iterate this backward construction to determine each earlier activation time and thereby gradually reduce the number of active players. This process stops when the number of players below their upper bounds decreases to either zero or one, which corresponds to an equilibrium without or with one instant-exit player, respectively. The time at which it stops becomes time $0$. Every step of this construction is forced by Lemma~\ref{Characterization} and the well-posedness above. In this way, I explicitly construct the unique profile that satisfies the conditions of Lemma~\ref{Characterization}. I then state my result:

\begin{Theorem}\label{exi and uni}
There exists an essentially unique equilibrium.
\end{Theorem}

\emph{Proof}. See Appendix~\ref{proof exi and uni}.
\medskip
\subsection{Asymmetry and Equilibrium Strategy}\label{comparative statics}

In this subsection, I compare two players in the same war and ask how the differences between their costs, discount rates and revelation rates determine which of the two provides earlier at each value. Player $i$'s \emph{revelation rate} $f_i(v)/F_i(v)$ is the density of his type at $v$ conditional on the revelation that his type does not exceed $v$, that is, the reversed hazard rate of his value distribution.

\begin{Proposition}\label{Asy Comp Sta}
	Consider the equilibrium. Let $\alpha\neq\beta$ be two players and $t_0\geq\max\{d_\alpha,d_\beta\}$. Each condition on revelation rates below is imposed at every value in $(\max\{c_\alpha,c_\beta\},\allowbreak\min\{\overline{v}_\alpha,\overline{v}_\beta\})$.
	\begin{enumerate}[(i)]
		\item Let $r_\alpha=r_\beta$ and $c_\alpha=c_\beta$. If $f_\alpha/F_\alpha>f_\beta/F_\beta$, then $\Phi_\alpha(t)<\Phi_\beta(t)$ for every $t\geq t_0$; if $f_\alpha/F_\alpha=f_\beta/F_\beta$, then $\Phi_\alpha(t)=\Phi_\beta(t)$ for every $t\geq t_0$.
		\item Let $r_\alpha=r_\beta$ and $f_\alpha/F_\alpha=f_\beta/F_\beta$. If $c_\alpha>c_\beta$, then $\Phi_\alpha(t)>\Phi_\beta(t)$ for every $t\geq t_0$.
		\item Let $c_\alpha=c_\beta$ and $f_\alpha/F_\alpha=f_\beta/F_\beta$. If $r_\alpha>r_\beta$, then $\Phi_\alpha(t)<\Phi_\beta(t)$ for every $t\geq t_0$.
	\end{enumerate}
\end{Proposition}

\emph{Proof}. See Appendix~\ref{proof Asy Comp Sta}.
\medskip

Since each player's types provide in decreasing order of value, the inequality $\Phi_\alpha(t)<\Phi_\beta(t)$ for every $t\geq\max\{d_\alpha,d_\beta\}$ implies that, at every value that both players can have and that exceeds both costs, player $\alpha$'s type provides earlier than player $\beta$'s, except possibly at the lower of the two upper bounds if both types with this value provide at time $0$.\footnote{Let $t_0=\max\{d_\alpha,d_\beta\}$. At $t_0$, one of the two curves still equals its upper bound, because a player who becomes active at a positive time has no atom then (Lemma~\ref{Characterization}(iv)) and at most one player exits instantly. Take a value $v$ that both players can have and that exceeds both costs. If $v\leq\Phi_\alpha(t_0)$, player $\alpha$'s type $v$ provides when his curve reaches $v$, at a time $t\geq t_0$ at which $\Phi_\beta(t)>v$, that is, before player $\beta$'s type $v$. If $v>\Phi_\alpha(t_0)$, player $\alpha$'s curve is below its upper bound at $t_0$, which leaves $\Phi_\beta(t_0)=\overline{v}_\beta$. Player $\alpha$'s type $v$ then provides by $t_0$, and player $\beta$'s type $v$ at $t_0$ or later, at the same time only if $t_0=0$ and $v=\overline{v}_\beta$.} I then say that player $\alpha$ \emph{provides earlier at every value}.

The last two clauses of Proposition~\ref{Asy Comp Sta} convey straightforward intuitions, which work through the indifference level $W_i(v)$. Recall that every active player's opponents provide more and more slowly over time. A higher cost reduces the net gain that waiting delays and raises the cost that waiting saves, and so lowers $W_i(v)$. Each type of the high-cost player then waits until his opponents' provision has slowed to this lower level, and the player provides later at every value. A higher discount rate, which corresponds to impatience, increases the opportunity cost of waiting, and so raises $W_i(v)$. Each type of the impatient player then waits only until his opponents' provision has slowed to this higher level, and the player provides earlier at every value. The proof needs one more step, because the two players face different opponents' provision hazards. At a moment when both players are active and their curves are at the same value, the player whose type has the lower indifference level faces opponents who provide more slowly. Since, for each of the two players, his own hazard rate and his opponents' provision hazard add up to $h(t)$, this player provides at the higher hazard rate, and with equal revelation rates his curve falls faster. In clause (ii), the high-cost player's curve can thus cross the other's only from above, and because it converges to the higher cost, it lies above the other's whenever both are active. In clause (iii), if the impatient player's curve were ever at or above the other's while both are active, it would stay there. From then on, his current type would have the higher indifference level, so he would provide at the lower hazard rate. This is impossible, because the share of his remaining types that must eventually provide would be at least as large as the other player's.

The interpretation of the first clause, which concerns the revelation rate, is less straightforward, because what the clause implies depends on what else is held fixed. For example, if player $\alpha$'s and $\beta$'s upper bounds are equal, Proposition~\ref{Asy Comp Sta}(i) implies that the player with a uniformly higher revelation rate provides earlier at every value and also precedes the other player in the order of activation. In contrast, if the two players are equally likely to be stubborn, this player still provides earlier at every value, but the other player precedes him in the order of activation.\footnote{Let player $\alpha$ have the uniformly higher revelation rate, let $r$ and $c$ be the two players' common discount rate and cost, and take $t_0=\max\{d_\alpha,d_\beta\}$; by clause (i), $\Phi_\alpha(t)<\Phi_\beta(t)$ for every $t\geq t_0$. A player who becomes active at a positive time has no atom then (Lemma~\ref{Characterization}(iv)), and his curve still equals his upper bound at that time. With equal upper bounds, $\Phi_\alpha(t_0)<\Phi_\beta(t_0)\leq\overline{v}_\beta=\overline{v}_\alpha$, which leaves two possibilities: $d_\alpha<t_0=d_\beta$, or $t_0=0$ and player $\alpha$ exits instantly. With equal stubborn probabilities, $F_\alpha(c)=F_\beta(c)$. Since both curves converge to $c$, integrating from $t\geq t_0$ to $+\infty$ the difference between the equations $\frac{d}{dt}\ln F_i(\Phi_i(t))=W_i(\Phi_i(t))-h(t)$ for $i=\alpha,\beta$, which hold by (\ref{aggregate rate}) at every $t>t_0$ that is not an activation time, gives $\ln F_\alpha(\Phi_\alpha(t))-\ln F_\beta(\Phi_\beta(t))=\frac{r}{c}\int_t^{+\infty}(\Phi_\beta(s)-\Phi_\alpha(s))\,ds>0$. At $t=t_0$, this says that player $\beta$ is more likely than player $\alpha$ to have provided by $t_0$. Hence $\Phi_\beta(t_0)<\overline{v}_\beta$, and either $d_\beta<t_0=d_\alpha$, or $t_0=0$ and player $\beta$ exits instantly.\label{equal bounds compstat}}

This discrepancy between the two cases arises because a higher revelation rate means different things in them. In the first case, a higher revelation rate makes the player who has it more likely than the other to have a value above $v$, for every $v$ between the common cost $c$ and the common upper bound. Since he also provides earlier at every value, he is more likely than the other player to have provided by every time after he becomes active, so the other player free rides on him. Because the upper bounds are equal, he also precedes the other player in the order of activation. In the second case, a higher revelation rate requires the player who has it to have the lower upper bound, because his distribution function, equal to the other's at $c$, grows proportionally faster above $c$. While both are active, the player with the higher revelation rate has the lower current type at each moment, by clause (i), so this type stays indifferent only if his opponents provide more slowly than the other player's opponents do. Since, for each of the two players, his own hazard rate and his opponents' provision hazard add up to $h(t)$, the player with the higher revelation rate provides at the higher hazard rate. Yet the probabilities that the two players have provided converge over time to the same limit, $1-F_\alpha(c)=1-F_\beta(c)$. The player with the lower revelation rate, who provides at the lower hazard rate once both are active, must therefore have provided with positive probability by the time the other player becomes active, with values above the other player's upper bound. So he precedes the other player in the order of activation.

\subsection{An Important Case: UTD War}\label{ltd war}

To illustrate behavior and welfare insights, I introduce a special yet significant family of wars, called the \emph{upper-truncated-distribution war} (\emph{UTD war}). This class of war is important for two reasons: (i) apart from this case, there are very few mathematically tractable examples, and (ii) it helps me relate my model to the symmetric benchmark.

In each UTD war, players have identical costs $c$ and discount rates $r$, but their value distributions are all upper-truncated distributions of the same benchmark distribution. Formally, let $F$ be a distribution on $[\underline{v},\overline{v}]$, with $\underline{v}<c$ and a density $f$ that is continuous and bounded away from $0$, and let $(\overline{v}_i)_{i\in I_N}$ be the players' upper bounds, each of which is no greater than $\overline{v}$ and greater than $c$. Then, player $i$'s value distribution is the upper-truncated distribution of $F$ given upper bound $\overline{v}_i$, that is, $F_i(v):=F(v|v\leq\overline{v}_i)$ for all $i$. Since each $F_i$ depends on $F$ only below $\overline{v}_i$, I take $\overline{v}=\max_{i\in I_N}\overline{v}_i$ in each war without loss of generality, so that $\overline{v}$ is also the highest type across players. I call such a war a UTD war generated from $F$. All players have the same indifference level $W(v):=\frac{r}{c}(v-c)$.

A UTD war is easy to analyze because the ratios $F_i(v)/f_i(v)$ on the right side of (\ref{main equations}) are identical across players on their overlapping domain, where they equal $F(v)/f(v)$. So the behavior of all players who are active at the same time is characterized by symmetric equations, and it is easy to verify that the equilibrium is \emph{locally symmetric}: active players' types with the same value provide at the same time, so that all active players share a common curve.\footnote{To see that (\ref{LTD solution}) below is an equilibrium, note that on a stage with $M$ active players, each of them provides at hazard rate $W(\Phi(t))/(M-1)$, where $\Phi$ is their common curve, so that the opponents of an active player provide at hazard rate $W(\Phi(t))$, and the opponents of a player who is not yet active provide at a higher rate. Since $W(\Phi(t))$ falls over time, (\ref{payoff slope}) implies that the expected payoff of a type $v>c$ rises until $\Phi$ reaches $v$ and falls afterwards. The types of the player with the highest upper bound that exceed $\Phi(0)$, the second-highest upper bound, therefore provide at once, while every other type $v\in(c,\Phi(0))$ does better by providing when $\Phi$ reaches $v$ than at time $0$, where he would at best share the cost with that player's instant exit. Types at or below $c$ never provide, which is optimal, since their indifference level is not positive and their expected payoff never falls. Almost every type plays a best response, and by Theorem~\ref{exi and uni} this is the equilibrium.}

First, I explicitly write down what the equilibrium of a UTD war looks like. Without loss of generality, let $\overline{v}_1\geq\overline{v}_2\geq\dots\geq\overline{v}_N$, so that player~1 has the highest upper bound. Formally, let $\Phi_{\mathrm{UTD}}(t\mid M,v)$ denote the solution of $\Phi'(t)=-\frac{1}{M-1}\frac{F(\Phi(t))}{f(\Phi(t))}W(\Phi(t))$, the symmetric $M$-player version of (\ref{main equations}), with the initial condition $\Phi_{\mathrm{UTD}}(0\mid M,v)=v$, and let $\Phi_{\mathrm{UTD}}^{-1}(w\mid M,v)$ be the time at which it reaches $w\in(c,v]$. Then, the equilibrium is given by:
\begin{equation}\label{LTD solution}
\begin{aligned}
&d_1=d_2=0,\qquad d_M=d_{M-1}+\Phi_{\mathrm{UTD}}^{-1}(\overline{v}_M\mid M-1,\overline{v}_{M-1}),\quad M=3,\dots,N,\\
&T_1(v)=0,\quad v\in(\overline{v}_2,\overline{v}_1],\\
&\Phi_1(t)=\dots=\Phi_M(t)=\Phi_{\mathrm{UTD}}(t-d_M\mid M,\overline{v}_M),\quad t\in[d_M,d_{M+1}),\ M=2,\dots,N,
\end{aligned}
\end{equation}
where $d_{N+1}:=+\infty$ and, on $[d_M,d_{M+1})$, $\Phi_i(t)=\overline{v}_i$ for every $i>M$. Here $d_M$ is player $M$'s activation time, and tied upper bounds give equal activation times and empty intervals $[d_M,d_{M+1})$.

Since $\Phi_{\mathrm{UTD}}(\cdot\mid M,v)$ solves a separable equation, the time it needs to fall from $v$ to $w\in(c,v]$ is
\begin{equation*}
(M-1)\int_w^v\frac{f(u)}{F(u)W(u)}\,du.
\end{equation*} Summing these times over the stages of (\ref{LTD solution}) gives, for every player $i$ and every type $v\in(c,\overline{v}_i]$ (Appendix~\ref{proof UTD times}),
\begin{equation}\label{UTD times}
T_i(v)=\sum_{j=2}^{N}\int_v^{\max\{v,\overline{v}_j\}}\frac{f(u)}{F(u)W(u)}\,du .
\end{equation}
Each player other than player~1 whose upper bound exceeds $v$ contributes one term. Over the values between $v$ and his upper bound, he shares the hazard that keeps every other active player's current type indifferent, so the common curve falls more slowly. The highest upper bound enters only when another player's upper bound equals it.

\begin{figure}[htbp]
	\centering
	\includegraphics[width=.6\textwidth]{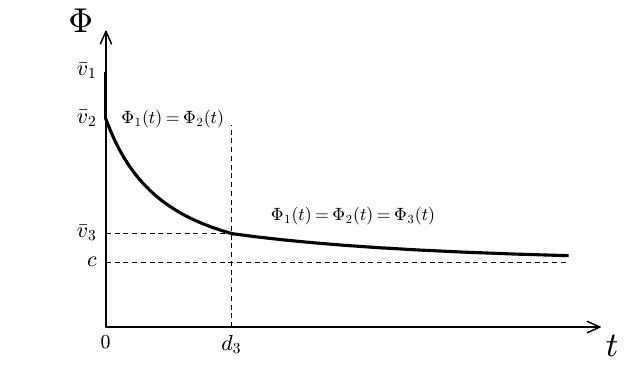}
	\caption{\footnotesize This figure shows the equilibrium of a three-player UTD war with $\overline{v}_1>\overline{v}_2>\overline{v}_3>c$. The player with the highest upper bound, player~1, provides at time $0$ when his value exceeds the second-highest upper bound, $\overline{v}_2$. Before $d_3$, two players are active, and at $d_3$, the third becomes active as well. Active players' types with the same value provide at the same time, so the active curves coincide. The common curve of players~1 and~2 kinks at $d_3$ and is flatter after it, because these players provide at lower hazard rates once player~3 becomes active. All curves converge to $c$.}
	\label{LTD Example}
\end{figure}

By local symmetry, the order of provision is the order of values, whichever player holds them, and players become active in the order of their upper bounds. In Figure~\ref{LTD Example}, I depict the equilibrium of a three-player UTD war.

Another noteworthy feature is that players' behavior is \emph{asymmetrically dependent} on each other: a change in one player's upper bound leaves unchanged the stopping times of types at or above the higher of his old and new bounds and changes those of lower types above $c$. More formally, consider a UTD war in which player $k$'s upper bound takes one of two values $a<b$ in $(c,\overline{v}]$, while some other player's upper bound is at least $b$. Then take any player and any type $v>c$ that he has in both games. If $v\geq b$, this type provides at the same time in both games. If $v<b$, it provides strictly later in the game in which $k$'s upper bound is $b$, by $\int_{\max\{v,a\}}^{b}\frac{f(u)}{F(u)W(u)}\,du$, a delay that is the same for all $v\leq a$. Indeed, since another player has the highest upper bound in both games, the two games differ only in $k$'s term of (\ref{UTD times}). The player with the highest upper bound is an exception: as long as his bound stays the highest, its variation changes no type's stopping time, only his probability of instant exit.

Asymmetric dependence also explains the leading role of the highest type across players, $\max_i\overline{v}_i$, in determining efficiency, which I measure by the \emph{expected discount factor} $E[e^{-rt_m}]$ of the provision time. Stopping times are set by the lower upper bounds, so a rise in the highest one only adds provision at time $0$, and no type waits longer in response. On the other hand, when a lower upper bound, say player $k$'s, rises without passing the highest one, the rise has a direct effect on efficiency and an offsetting effect.\footnote{When a player's upper bound rises but does not exceed the highest one, I call the offset in a given measure \emph{complete} if the measure does not change, \emph{partial} if it still rises, and \emph{excessive} if it falls. Proposition~\ref{U-LTD expect rate} shows that the offset is complete for $E[e^{-\rho rt_m}]$ in every uniform UTD war, both defined below. With other distributions it can be partial or excessive. For example, let $N=3$, $c=r=1$, $\underline{v}=0$ (so that $\rho=1$), $\overline{v}_1=3$, $\overline{v}_2=2.5$ and $F(v)=(e^{\theta v}-1)/(e^{3\theta}-1)$ on $[0,3]$. Numerically, raising $\overline{v}_3$ from $1.2$ to $2.4$ raises $E[e^{-rt_m}]$ from $0.426$ to $0.463$ when $\theta=-1$ (partial offset) and lowers it from $0.821$ to $0.805$ when $\theta=1$ (excessive offset).} His new high types make early provision more likely, and he becomes active earlier unless he is active from time $0$ in both games. But every type below his new upper bound that provides waits longer, which works against the direct effect of the rise.

Now I use the uniform distribution to show the leading role of the highest type across players in determining efficiency. I call a UTD war generated from a uniform distribution $F$ a \emph{uniform UTD war}. In it, $\Phi_{\mathrm{UTD}}$ has the closed form
\begin{equation}\label{sym U-LTD solution}
\Phi_{\mathrm{UTD}}(t\mid M,v)=\underline{v}+\frac{c-\underline{v}}{1-\lambda(v)e^{-\rho rt/(M-1)}},\qquad \lambda(v):=\frac{v-c}{v-\underline{v}},\quad \rho:=1-\frac{\underline{v}}{c}.
\end{equation}
Combining (\ref{LTD solution}) and (\ref{sym U-LTD solution}), I obtain the equilibrium.

In uniform UTD wars, I measure efficiency by the \emph{expected discount factor at rate $\rho r$}, $E[e^{-\rho rt_m}]$.\footnote{A more natural measure is the expected discount factor $E[e^{-rt_m}]$, used above for general UTD wars and in Section~\ref{large societies and asymmetry}: a type $v_i$ who never provides earns $v_i$ times the expected discount factor of the other players' first stopping time $\min_{j\neq i}T_j(v_j)$, and in large societies this factor is close to $E[e^{-rt_m}]$. The invariance in Proposition~\ref{U-LTD expect rate} carries over to $E[e^{-rt_m}]$ when $\underline{v}=0$, since then $\rho=1$, but not otherwise. For example, in a two-player uniform UTD war with $c=r=1$, $\underline{v}=1/2$ and $\overline{v}_1=2$, $E[e^{-rt_m}]$ is about $0.549$ at $\overline{v}_2=2$ and $0.591$ at $\overline{v}_2=3/2$.} One important property of uniform UTD wars is that in equilibrium $E[e^{-\rho rt_m}]$ is determined by the highest upper bound, for given $c$ and $\underline{v}$. Therefore, it is unaffected by the number of players and by the other players' upper bounds, although both necessarily determine how each type behaves. I present this property in the following proposition:
\begin{Proposition}\label{U-LTD expect rate}
	In the equilibrium of every uniform UTD war, $E[e^{-\rho rt_m}]=\lambda\big(\max_{i\in I_N}\overline{v}_i\big)$.
\end{Proposition}

\emph{Proof}. See Appendix~\ref{proof U-LTD expect rate}.
\medskip

Proposition~\ref{U-LTD expect rate} shows that, for $E[e^{-\rho rt_m}]$, the offset described above is exactly complete: a change in a lower upper bound that does not pass the highest one changes the stopping times of all types between $c$ and the higher of the old and new bounds just enough to cancel the direct effect of the change. I explain the cancellation at two levels. Consider again the rise in player $k$'s upper bound, suppose that his new upper bound does not exceed the second-highest upper bound before the rise, and look at the values between his old and his new upper bound, where the change acts.

After the rise, player $k$'s curve falls together with the common curve through these values, so one more player is active while the common curve passes them. Since each active player may hold a value in this range, one more player makes provision within these values more likely. This is the direct effect. But by mutual indifference, the opponents of each active player together provide at the hazard that keeps his current type indifferent. Since player $k$ now supplies part of this hazard, every other active player provides at a lower hazard rate. So the common curve falls more slowly through these values and reaches every lower value above $c$ later. This is the offset.

The proof shows why the cancellation is exact. Given that nobody has provided by the time the common curve reaches a value $v$, the expected discount factor at rate $\rho r$ of the remaining wait is $\lambda(v)$, whatever the number of active players. So a change that only alters the number of active players over some range of values leaves this conditional factor unchanged. Since the common curve starts at $\overline{v}_2$, the remaining wait given no provision at time $0$ has expected discount factor $\lambda(\overline{v}_2)$ at rate $\rho r$. Adding player~1's instant exit, which happens with probability $1-F_1(\overline{v}_2)$, gives $E[e^{-\rho rt_m}]=1-F_1(\overline{v}_2)+F_1(\overline{v}_2)\lambda(\overline{v}_2)=\lambda(\overline{v}_1)$.

Another insight shown by Proposition~\ref{U-LTD expect rate} is that a loss always occurs: the loss $1-E[e^{-\rho rt_m}]$ exceeds the probability $\prod_iF_i(c)$ that nobody ever provides, so part of the loss comes from delay alone. For given $c$ and $\underline{v}$, the loss vanishes only in the limit as the highest upper bound grows without bound. Section~\ref{large societies and asymmetry} revisits these insights in large societies, whose players share $c$ and $r$. As the population grows with fixed group proportions and value distributions, the \emph{cost of delay} $1-E[e^{-rt_m}]$ converges to $c/\max_i\overline{v}_i$, whatever the shapes of these distributions. In uniform UTD wars with $\underline{v}=0$, so that $\rho=1$, Proposition~\ref{U-LTD expect rate} gives this value exactly, for every $N$.
\section{Large Societies and Asymmetry}\label{large societies and asymmetry}

In this section, I investigate how ex-ante asymmetry influences welfare in large societies, and in particular how individuals with different value distributions contribute to welfare. To make the analysis tractable, I follow the seminal work of \citet{bliss1984dragon}, who study a large-population symmetric war of attrition. I derive the counterpart for asymmetric settings and show that the highest type across players plays the leading role.

I introduce some definitions and notation. A \emph{society} consists of groups that may differ from one another, and in each group, all players are ex ante homogeneous. Formally, an $N$-player society is divided into $L$ groups, indexed by $\ell\in\{1,\dots,L\}$; group $\ell$ takes up the proportion $p_\ell$ of the total population, with $\sum_{\ell=1}^{L}p_\ell=1$, and has $n_\ell:=p_\ell N\geq2$ players, whose values are drawn from $F_\ell$ on $[\underline{v}_\ell,\overline{v}_\ell]$. The notation is that of Section~\ref{model}, with the subscript $\ell$ denoting a group. Throughout this section, all players have the same provision cost $c$ and discount rate $r$, hence the same indifference level $W(v)=\frac{r}{c}(v-c)$. Since the members of a group are identical, the uniqueness in Theorem~\ref{exi and uni} implies that they all use the same strategy (Lemma~\ref{group structure}(i) in Appendix~\ref{appendix large societies}); let $\Phi_\ell$ be the common curve of group $\ell$ and $d_\ell$ its activation time. Write $\overline{v}:=\max_\ell\overline{v}_\ell$ for the highest type across players, and $\overline{v}(\tau):=\max_\ell\Phi_\ell(\tau)$ for the highest type across players who has not provided by $\tau$.

To begin with, I study welfare as the population grows:

\begin{Lemma}\label{Aligned Welfare}
As $N\to\infty$ with the groups' proportions and distributions fixed:
\begin{enumerate}[(i)]
	\item the expected discount factor $E[e^{-rt_m}]$ converges to $1-c/\overline{v}$;
	\item the expected payoff of type $v$ of a player in group $\ell$ converges to $v(1-c/\overline{v})$, uniformly in $v$ and $\ell$.
\end{enumerate}
Both errors are at most of order $1/N$.
\end{Lemma}

\emph{Proof}. See Appendix~\ref{proof Aligned Welfare}.
\medskip

The symmetric version of Lemma~\ref{Aligned Welfare} is the counterpart of the limit result of \citet[pp.~7--8]{bliss1984dragon}:

\begin{Corollary}\label{Sym Welfare}
Consider a symmetric $N$-player war with cost $c$, discount rate $r$ and value distribution $F$ on $[\underline{v},\overline{v}]$. As $N\to\infty$, the expected discount factor $E[e^{-rt_m}]$ converges to $1-c/\overline{v}$, and the expected payoff of type $v$ converges to $v(1-c/\overline{v})$.
\end{Corollary}

\emph{Proof}. This is Lemma~\ref{Aligned Welfare} with $L=1$.
\medskip

To see the correspondence, divide the payoffs of each type $v>0$ by his value: the good is then worth one to every such type, and providing costs $c/v$, which is at least $c/\overline{v}$, the lowest cost of any such type. By Lemma~\ref{Aligned Welfare}(i), the expected discount factor converges to one minus this lowest cost, as in \citet{bliss1984dragon}. Their Theorem~6, in which the first best is attained, is the special case of a zero lowest cost, which a finite $\overline{v}$ rules out.

Lemma~\ref{Aligned Welfare} gives two insightful results. First, as the population grows, the expected discount factor and every type's expected payoff converge to limits that depend on the value distributions only through the highest type across players. Second, no matter how large the population grows, inefficiency always occurs: the cost of delay converges to $c/\overline{v}>0$.

The proof of Lemma~\ref{Aligned Welfare} provides more information. At every fixed time, the provision hazard $h(t)$ converges to $W(\overline{v})$, the indifference level of the highest type across players, and $t_m$ converges in distribution to an exponential random variable with this rate, whose expected discount factor is $1-c/\overline{v}$. In the limit, provision thus arrives at a steady hazard, and the cost of delay $c/\overline{v}$ is pure delay: it comes from waiting, not from the risk that the good is never provided.

Why does inefficiency persist? In a large society every curve falls slowly, because a player in group $\ell$ provides at a hazard of at most $1/n_\ell$ times the provision hazard, which stays bounded. In terms of behavior, every type below $\overline{v}$ free rides for longer and longer, and only types closer and closer to $\overline{v}$ provide early. But there are more and more players, and together these few types still provide at a hazard that does not vanish. Each of these types provides when he is indifferent between providing at once, which gives him about $\overline{v}-c$, and waiting for the others, which gives him about $\overline{v}x/(x+r)$ if they provide at a constant hazard $x$; the two are equal exactly when $x=W(\overline{v})$. The curves therefore fall at the pace that keeps the provision hazard near $W(\overline{v})$.

A complementary lemma shows how a difference in upper bounds determines the order in which two groups become active in large societies:

\begin{Lemma}\label{decisive upper bound}
Let groups $\ell$ and $\ell'$ have upper bounds $\overline{v}_\ell>\overline{v}_{\ell'}$. If $n_{\ell'}>(\overline{v}_\ell-c)/(\overline{v}_\ell-\overline{v}_{\ell'})$, then $d_\ell<d_{\ell'}$. Moreover, with the groups' proportions and distributions fixed, $\liminf_{N\to\infty}(d_{\ell'}-d_\ell)/N>0$.
\end{Lemma}

\emph{Proof}. See Appendix~\ref{proof decisive upper bound}.
\medskip

The threshold ensures that, once group $\ell'$ is active, the provision hazard is too low to keep the types of group $\ell$ close to $\overline{v}_\ell$ waiting. If these types were still waiting when group $\ell'$ becomes active, they would face opponents who provide too slowly to keep them waiting, so group $\ell$ becomes active first. Formally, by mutual indifference and because each member of group $\ell'$ provides at most a share $1/n_{\ell'}$ of the provision hazard, the provision hazard while group $\ell'$ is active is at most $\frac{n_{\ell'}}{n_{\ell'}-1}W(\overline{v}_{\ell'})$, and the condition of Lemma~\ref{decisive upper bound} says exactly that this cap lies below $W(\overline{v}_\ell)$. Lemma~\ref{decisive upper bound} implies, in particular, that when $N$ is large, all groups are active from time $0$ only if they share the same upper bound.\footnote{Groups with equal upper bounds need not become active together. If groups $\ell$ and $\ell'$ share the upper bound $\hat{v}$ and group $\ell$ has a strictly higher revelation rate, $f_\ell/F_\ell>f_{\ell'}/F_{\ell'}$ on $(c,\hat{v})$, then $d_\ell<d_{\ell'}$ for every $N$ (by footnote~\ref{equal bounds compstat}, since no player in a society exits instantly, Lemma~\ref{group structure}(i)). Numerically, the difference does not vanish as $N$ grows: with $c=r=1$, two groups of equal size, and values uniform on $[0,2]$ and on $[-1,2]$, the group with values on $[0,2]$ becomes active first, and the difference approaches about $0.10$.\label{equal bounds footnote}}

Lemma~\ref{decisive upper bound} gives the large-society counterpart of the order of activation in UTD wars, in which players become active in the order of their upper bounds (Section~\ref{ltd war}). In large societies, moreover, a group with a lower upper bound becomes active only after a delay at least of order $N$. Until then, its members face opponents who provide at a hazard above the indifference level of the group's highest type, as every initially inactive player does (Section~\ref{characterization}). By the time the group becomes active, the good has therefore been provided with a probability that converges to one. This is another demonstration of how the highest type across players determines the expected discount factor.

Curiously, Lemmas~\ref{Aligned Welfare} and~\ref{decisive upper bound} do not depend on the shapes of the distributions, not even on the density at the highest type. The reason is mutual indifference. At every moment, the opponents of each active player provide at the indifference level of his current type. With many players, his own share of the provision hazard is small, so the provision hazard itself is close to that level, for every active player. The current types of all active groups therefore nearly coincide with $\overline{v}(t)$, and the provision hazard is pinned near $W(\overline{v}(t))$ whatever the densities (Lemma~\ref{hazard facts}(b) in Appendix~\ref{appendix large societies}). The shapes do matter for when each type above $c$ provides, for example for the order in which two groups with the same upper bound become active (footnote~\ref{equal bounds footnote}). Since $\overline{v}(t)$ is the highest type across players who has not provided by $t$, the shapes also matter for how fast $\overline{v}(t)$ falls.

The insights of Lemma~\ref{Aligned Welfare} extend to every moment after the war of attrition begins, with $\overline{v}(\tau)$ in place of $\overline{v}$:

\begin{Theorem}\label{U-LTD equivalency}
As $N\to\infty$ with the groups' proportions and distributions fixed, uniformly in $\tau\geq0$:
\begin{enumerate}[(i)]
	\item the expected discount factor of the remaining wait, $E[e^{-r(t_m-\tau)}\mid t_m\geq\tau]$, differs from $1-c/\overline{v}(\tau)$ by a term of order at most $1/N$;
	\item the expected payoff of type $v\leq\Phi_\ell(\tau)$ of a player in group $\ell$, discounted to $\tau$ and given that the good has not been provided before $\tau$, differs from $v\big(1-c/\overline{v}(\tau)\big)$ by a term of order at most $1/N$, uniformly in $v$ and $\ell$ as well.
\end{enumerate}
\end{Theorem}

\emph{Proof}. See Appendix~\ref{proof U-LTD equivalency}.
\medskip
\section{Discussion}\label{discussion}

\subsection{Related Literature}

The closest paper is \citet{kambe2019n}, who studies an $N$-player war of attrition over the provision of a public good. Like this paper, he lets the players differ ex ante and finds a unique equilibrium path in which at most one player provides at time 0 with positive probability, the rate at which the good is provided never rises, and some players can become active only later, which cannot happen with two players. The main difference is the information: he considers public values and binary private costs, whereas here each player's value is private and continuously distributed. In his model, a player is summarized by one index of value, cost and patience, together with the probability of an infinite cost; here, which of two players provides earlier at every value also depends on the shapes of their value distributions, through the revelation rate. The type with an infinite cost, who never provides, is the counterpart of my stubborn types. But in his model every type who can provide has done so by a finite time; here, with continuously distributed values, types just above cost provide arbitrarily late. This difference changes the dynamics: in his model an active player can run out of types who provide while the others fight on, whereas here no player ever runs out. It also changes the technique: he builds the equilibrium backward from this finite time, whereas here no such time exists, and the saddle point at the costs plays the role of his end time. Finally, he motivates his study of provision at time 0 by noting that it avoids delay and so improves efficiency. In uniform UTD wars, by contrast, the probability of instant exit can change while my measure of efficiency does not, and in large societies I go beyond efficiency and derive each type's limit payoff.

My model is the asymmetric extension of \citet{bliss1984dragon}, in which identical players privately informed about their costs wait for one another to provide a public good. With identical players, the equilibrium of my model is symmetric, so every player is active from time 0 and none exits instantly, which shows that the stratified pattern comes from asymmetry. Other wars of attrition with identical players and private information include \citet{riley1980strong}, \citet{alesina1991stabilizations}, \citet{weesie1994incomplete}, \citet{bulow1999generalized} and \citet{sahuguet2006volunteering}. With complete information, wars of attrition among players who differ have been studied as exit from declining industries \citep{ghemawat1985exit,ghemawat1990devolution,whinston1988exit} and as volunteering over time \citep{weesie1993asymmetry,bilodeau1996toilet}, a timing version of the volunteer's dilemma of \citet{diekmann1985volunteer}; with two players and spillovers, see \citet{beth2024}. Complete-information models without timing also show that asymmetry can decide which players are active \citep*{bergstrom1986private,hillman1989politically}. Here, as Section~\ref{characterization} notes, waiting reveals types, so the set of active players can change over time. Models in which a third party intervenes in a two-player war of attrition \citep{casella1996can,powell2017taking}, or in which two groups bargain \citep{ponsati1996multiperson}, involve more than two players, but the war is fought between two sides.

Wars of attrition can have a continuum of equilibria \citep{riley1980strong,nalebuff1985asymmetric}, also with complete information \citep*{hendricks1988war}. Uniqueness usually comes from types who never concede or from other perturbations: committed types \citep*{kornhauser1989reputation,kambe1999bargaining,abreu2000bargaining}, exit failure and other perturbations of the game \citep{myatt2005instant}, or, as here and in \citet{fudenberg1986theory}, payoffs that make never conceding optimal for some types. With two players and a common discount rate, my model is, after a change of units, the case of \citet{ponsati1995war} in which both players have tough types, here stubborn types, except that they assume smooth densities and nonnegative values. Uniqueness with $N$ asymmetric players and continuous types is known for first-price auctions \citep{lebrun1999first,maskin2003uniqueness}, in which equilibrium inverse bid functions solve a system of differential equations and a mass of types at the bottom of the value distributions helps pin down the solution. My contribution is the counterpart for wars of attrition, where the boundary condition is at infinite time and the stubborn types make the costs a saddle point. The computation showing that this saddle point has exactly one stable direction is the one \citet[Lemma~3.12]{fibich2012asymmetric} use for asymmetric first-price auctions. In these auctions, however, the equilibrium leaves the saddle point along its unstable directions, so with more than two bidders the saddle point alone does not select the equilibrium.

Two-player wars of attrition with incomplete information and players who differ are the most widely studied \citep*{riley1980strong,nalebuff1985asymmetric,fudenberg1986theory,kornhauser1989reputation,ponsati1995war,abreu2000bargaining,myatt2005instant,horner2011war}, and instant exit, a term I take from \citet{myatt2005instant}, already arises there \citep{nalebuff1985asymmetric,ponsati1995war,riley1999asymmetric,abreu2000bargaining}. For comparing players, \citet[Proposition~3]{fudenberg1986theory} show, in a symmetric duopoly, that a cost distribution with a higher ratio of density to distribution function, which puts more weight on high costs, slows selection down: firms stay in longer. Proposition~\ref{Asy Comp Sta}(i) compares two players in the same war through the same ratio, the revelation rate, and, like \citet[Proposition~2]{myatt2005instant}, orders their strategies type by type. Unlike in his setting, however, this ranking does not by itself decide which of the two players exits instantly or becomes active earlier (Section~\ref{comparative statics}).

Finally, my large-society results are the counterpart, for groups that differ, of the limit result of \citet{bliss1984dragon}, and they hold whatever the shapes of the value distributions.

\subsection{An Application}\label{insights}


Consider the Mekong River massacre, an attack on Chinese cargo ships in October 2011.\footnote{The introduction of the \href{https://en.wikipedia.org/wiki/Mekong_River_massacre}{Wikipedia article} on this event reads: ``... on the morning of 5 October 2011, ... two Chinese cargo ships were attacked on a stretch of the Mekong River in the Golden Triangle region on the borders of Myanmar (Burma) and Thailand. All 13 crew members on the two ships were killed and dumped in the river. ... In response, China ... reached an agreement with Myanmar, Thailand, and Laos to jointly patrol the river. The event was also the impetus for the Naypyidaw Declaration and other anti-drug cooperation efforts in the region.''} A surprising consequence of this tragedy was that China took the lead in building a security system that aims to repress illegal activities in the Golden Triangle, a region long known for drug trafficking. Before the event, the Golden Triangle hurt Thailand, Laos, and Myanmar, the countries that share it, while China, despite its greater enforcement capacity, was not directly concerned. Leadership in building this system can be seen as a privately provided public good, and the situations before and after the massacre as two wars of attrition that differ in the distribution of China's value of the good. The attack on Chinese ships raised how much China could value regional security, including the security of Chinese commercial activities in Southeast Asia.\footnote{About half a year after the event, a \href{https://www.chicagotribune.com/lifestyles/ct-xpm-2012-05-10-sns-rt-china-mekongmurder-tvl4e8ga36l-20120510-story.html}{report} in the Chicago Tribune noted that ``[t]he patrols, ostensibly conducted with Myanmar, Laos, and Thailand, have been seen as an expansion of Beijing's growing role in regional security, extending its law enforcement down the highly strategic waterway.''} China could now value regional security more than any other country could, so the highest value that any player can have rose as well. In the model, such a rise both makes China the only player who exits instantly in UTD wars (Section~\ref{ltd war}) and improves efficiency in uniform UTD wars (Proposition~\ref{U-LTD expect rate}). This may help explain why the overdue security cooperation followed within months.

\bibliographystyle{aer}
\bibliography{multiplayer}

\appendix
\section{Preliminaries}\label{appendix preliminaries}

This section fixes notation for the proofs and proves two facts that several of them use. The curves $\Phi_j$, the activation times $d_j$, the sets $A(t)$ of active players, the provision hazard $h$, the opponents' provision hazards $h_{-i}$ and the indifference levels $W_i$ are those of Section~\ref{model}; they are defined for every strategy profile $(T_i)_{i\in I_N}$, which need not be an equilibrium. For each player $j$ and $t\geq0$, let $Q_j(t):=F_j(\Phi_j(t))=\operatorname{Prob}(T_j(v_j)>t)$ be the probability that player $j$ has not provided by $t$. Let $T_{-i}:=\min_{j\neq i}T_j(v_j)$ be the first stopping time of player $i$'s opponents, and $Q_{-i}(t):=\operatorname{Prob}(T_{-i}>t)=\prod_{j\neq i}Q_j(t)=1-F_{-i}^{\min}(t)$. For $t\geq0$, let $Q_j(t-):=\operatorname{Prob}(T_j(v_j)\geq t)$ and $Q_{-i}(t-):=\operatorname{Prob}(T_{-i}\geq t)$; for $t>0$ these are the left limits of $Q_j$ and $Q_{-i}$, so that $\operatorname{Prob}(T_{-i}=t)=Q_{-i}(t-)-Q_{-i}(t)$, and $Q_j(0-)=Q_{-i}(0-)=1$. Let $x_j:=\ln Q_j$; then $\eta_j:=-x_j'$ is player $j$'s hazard rate, so that $h=\sum_{j\in I_N}\eta_j$ and $h_{-i}=h-\eta_i$ where these derivatives exist.

For each player $k$, $H_k:=F_k/f_k$ is the reciprocal of his revelation rate. With the extension of Section~\ref{existence and uniqueness}, $f_k(v)=f_k(\overline{v}_k)$ and $F_k(v)=1+f_k(\overline{v}_k)(v-\overline{v}_k)$ for $v>\overline{v}_k$, so $f_k$ is continuous and positive on $[\underline{v}_k,+\infty)$, $F_k$ is continuously differentiable there with derivative $f_k$, and $H_k$ is continuous on $[\underline{v}_k,+\infty)$ and positive on $(\underline{v}_k,+\infty)$. For vectors $a$ and $b$ with the same index set, $a\leq b$ means $a_j\leq b_j$ for every $j$, and $a\ll b$ means $a_j<b_j$ for every $j$. For $I\subseteq I_N$, let $c_I:=(c_j)_{j\in I}$, $\underline{v}_I:=(\underline{v}_j)_{j\in I}$ and $\overline{v}_I:=(\overline{v}_j)_{j\in I}$, and for a vector $\gamma$ indexed by $I$ and a set $I'\subseteq I$, let $\gamma_{I'}:=(\gamma_j)_{j\in I'}$. For $I\subseteq I_N$ with $|I|\geq2$ and $\gamma\gg\underline{v}_I$, let $h_I(\gamma):=\frac{1}{|I|-1}\sum_{k\in I}W_k(\gamma_k)$, and call
\begin{equation}\label{system on I}
\gamma_j'(t)=H_j(\gamma_j(t))\big[W_j(\gamma_j(t))-h_I(\gamma(t))\big],\qquad j\in I,
\end{equation}
the \emph{system on $I$}. For a vector $\gamma$ indexed by a set that contains $I$, $h_I(\gamma)$ stands for $h_I(\gamma_I)$; in particular, $h_I(\Phi(t))=h_I\big((\Phi_j(t))_{j\in I}\big)$. A \emph{solution} of (\ref{system on I}) on an interval is a function $\gamma$ with $\gamma(t)\gg\underline{v}_I$ that is differentiable on the interval, with one-sided derivatives at the endpoints that belong to it, and that satisfies (\ref{system on I}) there. By Lemma~\ref{Characterization}(vi) and (\ref{aggregate rate}), in the interior of each stage of an equilibrium the curves of the players in $A(t)$ solve the system on $A(t)$, and $h(t)=h_{A(t)}(\Phi(t))$; the system on $I_N$ is the all-active system (\ref{all-active}).

\begin{alemma}[Log-survival coordinates]\label{log survival}
	Let $I\subseteq I_N$ with $M:=|I|\geq2$. For $x\in\mathbb{R}^I$, let $\xi_k(x_k):=F_k^{-1}(e^{x_k})$ for $k\in I$, $\xi(x):=(\xi_k(x_k))_{k\in I}$, and
	\begin{equation*}
	G_j(x):=W_j(\xi_j(x_j))-h_I(\xi(x)),\qquad j\in I.
	\end{equation*}
	Then $G$ is continuously differentiable on $\mathbb{R}^I$, with
	\begin{equation*}
	\frac{\partial G_j}{\partial x_k}(x)=\Big(\mathbbm{1}\{j=k\}-\frac{1}{M-1}\Big)\frac{r_k}{c_k}H_k(\xi_k(x_k)).
	\end{equation*}
	A function $\gamma\gg\underline{v}_I$ solves (\ref{system on I}) on an interval if and only if its \emph{log-survival coordinates} $x:=(\ln F_j(\gamma_j))_{j\in I}$ solve $x'=G(x)$ there. Consequently, (\ref{system on I}) has a unique maximal solution through every $(t_0,\gamma^0)$ with $\gamma^0\gg\underline{v}_I$, and near each finite endpoint of its interval this solution lies outside any given compact subset of $\times_{j\in I}(\underline{v}_j,+\infty)$. Moreover, if $\gamma\gg\underline{v}_I$ is continuous on an interval $J$ and solves (\ref{system on I}) in the interior of $J$, then $\gamma$ is the restriction to $J$ of the maximal solution through $(t,\gamma(t))$ for any $t\in J$.
\end{alemma}

\begin{proof}
Since $f_k$ is continuous and positive on $[\underline{v}_k,+\infty)$ and $F_k$ maps $(\underline{v}_k,+\infty)$ onto $(0,+\infty)$, the inverse function theorem makes $\xi_k$ a continuously differentiable bijection from $\mathbb{R}$ onto $(\underline{v}_k,+\infty)$, with $\xi_k'(x_k)=e^{x_k}/f_k(\xi_k(x_k))=H_k(\xi_k(x_k))>0$; its inverse $v\mapsto\ln F_k(v)$ is continuously differentiable with derivative $f_k(v)/F_k(v)>0$. So $W_k\circ\xi_k$ is continuously differentiable with derivative $\frac{r_k}{c_k}H_k\circ\xi_k$, which gives the displayed partial derivatives, and they are continuous.

If $\gamma$ solves (\ref{system on I}), then $x_j:=\ln F_j(\gamma_j)$ satisfies $x_j'=\frac{f_j(\gamma_j)}{F_j(\gamma_j)}\gamma_j'=W_j(\gamma_j)-h_I(\gamma)=G_j(x)$. Conversely, if $x'=G(x)$, then $\gamma_j:=\xi_j(x_j)$ satisfies $\gamma_j'=H_j(\gamma_j)G_j(x)$, which is (\ref{system on I}). Since $G$ is continuously differentiable, $x'=G(x)$ has a unique maximal solution through every point (Picard--Lindel\"of theorem), and if its interval has a finite endpoint, then $|x(t)|\to+\infty$ there (extension theorem). The change of coordinates $x\mapsto\xi(x)$ is a homeomorphism from $\mathbb{R}^I$ onto $\times_{j\in I}(\underline{v}_j,+\infty)$, which gives the sentence of the lemma that begins with ``Consequently''.

For the last sentence, $\gamma$ coincides on the interior of $J$ with the maximal solution through any interior point, by uniqueness. If an endpoint $t_e$ of $J$ belonged to $J$ but not to the interval of this maximal solution, the latter would lie, near $t_e$, outside the compact set $\{\gamma(t):t\in J,\ |t-t_e|\leq\delta\}$, where $\delta>0$ is so small that these times form a closed interval, although it equals $\gamma$ there. So $t_e$ belongs to the interval of this maximal solution, and $\gamma(t_e)$ is the value of the maximal solution at $t_e$ by continuity.
\end{proof}

Fix a player $i$. A type $v$ of player $i$ who chooses $t$ obtains $v$ at time $\min\{T_{-i},t\}$ and, if $T_{-i}\geq t$, pays his share of the cost at $t$. So (\ref{expected gain}) can be written as
\begin{equation}\label{payoff split}
R_i(t|v)=v\phi_i(t)-c_ie^{-r_it}\mu_i(t)\quad\text{for }t\in[0,+\infty),\qquad R_i(+\infty|v)=v\phi_i(+\infty),
\end{equation}
where
\begin{equation*}
\phi_i(t):=E\big[e^{-r_i\min\{T_{-i},t\}}\big],\qquad\mu_i(t):=Q_{-i}(t)+E\bigg[\frac{\mathbbm{1}\{T_{-i}=t\}}{1+\#\{j\neq i:T_j(v_j)=t\}}\bigg].
\end{equation*}
Neither $\phi_i$ nor $\mu_i$ depends on $v$, $\phi_i$ takes values in $[0,1]$, and since at most $N-1$ opponents provide at $t$,
\begin{equation*}
Q_{-i}(t)+\operatorname{Prob}(T_{-i}=t)/N\leq\mu_i(t)\leq Q_{-i}(t)+\operatorname{Prob}(T_{-i}=t);
\end{equation*}
in particular, $\mu_i(t)=Q_{-i}(t)$ if no opponent of $i$ has an atom at $t$. By Fubini's theorem, $\phi_i(t)=1-r_i\int_0^te^{-r_is}Q_{-i}(s)\,ds$, since $e^{-r_i\min\{T_{-i},t\}}=1-r_i\int_0^te^{-r_is}\mathbbm{1}\{T_{-i}>s\}\,ds$.

\begin{alemma}[Payoff derivative]\label{payoff derivative}
	Fix a player $i$ and a value $v$. At every $t>0$ at which every $Q_j$, $j\neq i$, is positive and differentiable, $R_i(\cdot|v)$ is differentiable and satisfies (\ref{payoff slope}). If every $Q_j$, $j\neq i$, is positive and locally absolutely continuous on $(0,+\infty)$, then $R_i(\cdot|v)$ is locally absolutely continuous on $(0,+\infty)$, so (\ref{payoff slope}) holds at almost every $t>0$ and $R_i(\cdot|v)$ is the integral of its derivative.
\end{alemma}

\begin{proof}
Fix $t>0$ at which every $Q_j$, $j\neq i$, is positive and differentiable. Then $Q_{-i}$ is differentiable at $t$, with $Q_{-i}'(t)=Q_{-i}(t)\sum_{j\neq i}Q_j'(t)/Q_j(t)=-Q_{-i}(t)h_{-i}(t)$. As $Q_{-i}(s)$ and $Q_{-i}(s-)$ both equal $Q_{-i}(t)+Q_{-i}'(t)(s-t)+o(|s-t|)$, the mass $\operatorname{Prob}(T_{-i}=s)=Q_{-i}(s-)-Q_{-i}(s)$ is $o(|s-t|)$ as $s\to t$, and it is $0$ at $s=t$. By the bounds on $\mu_i$, so is $\mu_i(s)-Q_{-i}(s)$, and $\phi_i'(t)=-r_ie^{-r_it}Q_{-i}(t)$ because $Q_{-i}$ is continuous at $t$. Differentiating (\ref{payoff split}) at $t$ gives
\begin{equation*}
\frac{dR_i(t|v)}{dt}=e^{-r_it}Q_{-i}(t)\big[c_ih_{-i}(t)-r_i(v-c_i)\big],
\end{equation*}
which is (\ref{payoff slope}), since $Q_{-i}=1-F_{-i}^{\min}$.

If every $Q_j$, $j\neq i$, is positive and locally absolutely continuous on $(0,+\infty)$, so is the finite product $Q_{-i}$, and, as each such $Q_j$ is continuous there, no opponent has an atom in $(0,+\infty)$, so $\mu_i=Q_{-i}$ there. Hence $R_i(\cdot|v)=v\phi_i-c_ie^{-r_i\cdot}Q_{-i}$ is locally absolutely continuous on $(0,+\infty)$, $\phi_i$ being Lipschitz, and the first part applies wherever every $Q_j$ is differentiable, which is almost everywhere.
\end{proof}

\section{Proofs for Section~\texorpdfstring{\ref{equilibrium analysis}}{3}}\label{appendix equilibrium analysis}

The proofs below use two properties of every profile that satisfies conditions (i)--(vi) of Lemma~\ref{Characterization}.

\begin{alemma}[One-sided derivatives]\label{one-sided derivatives}
	Let a profile satisfy conditions (i)--(vi) of Lemma~\ref{Characterization}, and let $d$ be an activation time. Then, for every player $j\in A(d)$, the curve $\Phi_j$ has a right derivative at $d$, and it satisfies (\ref{main equations}) at $t=d$. If $d>0$, then, for every player $j$ with $d_j<d$, the curve $\Phi_j$ has a left derivative at $d$, and it satisfies (\ref{main equations}) at $t=d$ with $A_-:=\{l\in I_N:d_l<d\}$ in place of $A(t)$ and its number of players in place of $M(t)$. Consequently, at every $t\geq0$, with right derivatives, every $j\in A(t)$ provides at hazard rate $\eta_j(t)=h_{A(t)}(\Phi(t))-W_j(\Phi_j(t))$, every $j\notin A(t)$ at hazard rate $0$, and $h(t)=h_{A(t)}(\Phi(t))$; that is, (\ref{inter equations}) and (\ref{aggregate rate}) hold at every $t\geq0$. At every positive activation time $d$, $\lim_{t\uparrow d}h(t)=h_{A_-}(\Phi(d))$.
\end{alemma}

\begin{proof}
Let $d'$ be the next activation time after $d$, with $d':=+\infty$ if $d$ is the last one. On $(d,d')$, the set of active players is $A(d)$, and by (vi), every $\Phi_j$ with $j\in A(d)$ is differentiable, with $\Phi_j'(s)$ equal to the right side of (\ref{main equations}), which is a continuous function of $(\Phi_k(s))_{k\in A(d)}$. Each $\Phi_k$ with $k\in A(d)$ is continuous on $[d_k,+\infty)\supseteq[d,+\infty)$ by (v). So, for $t\in(d,d')$, the mean value theorem gives $(\Phi_j(t)-\Phi_j(d))/(t-d)=\Phi_j'(s_t)$ for some $s_t\in(d,t)$, and as $t\downarrow d$, $\Phi_j'(s_t)$ converges to the right side of (\ref{main equations}) at $t=d$. If $d>0$, let $d''$ be the activation time before $d$, which exists because $0$ is an activation time by (iii). On $(d'',d)$, the set of active players is $A_-$, which contains $A(0)$ and so has at least two players by (iii), and the same argument on $(d'',d)$, with $t\uparrow d$, gives the left derivatives.
Finally, by (vi) and the right derivatives just obtained, at every $t\geq0$, every $j\in A(t)$ has
\begin{equation*}
\eta_j(t)=-\frac{f_j(\Phi_j(t))}{F_j(\Phi_j(t))}\Phi_j'(t)=h_{A(t)}(\Phi(t))-W_j(\Phi_j(t)),
\end{equation*}
and every $j\notin A(t)$ has $\eta_j(t)=0$ because $\Phi_j=\overline{v}_j$ before $d_j$. Summing over $j$ gives $h(t)=h_{A(t)}(\Phi(t))$ for every $t\geq0$; here $M(t)\geq2$ by (iii). The left limit at $d>0$ follows from (\ref{aggregate rate}) on the stage that ends at $d$ and the continuity of the curves.
\end{proof}

\begin{alemma}[Provision hazard]\label{hazard lemma}
	Let a profile satisfy conditions (i)--(vi) of Lemma~\ref{Characterization}, as the canonical representative of every equilibrium does, and let $h$ be its provision hazard. Then:
	\begin{enumerate}[(a)]
		\item $h$ is strictly decreasing on $(0,+\infty)$;
		\item at every positive activation time $d$, $h$ jumps down weakly: $\lim_{t\uparrow d}h(t)\geq h(d)$;
		\item for every player $i$ and every $t\in(0,d_i)$, $h(t)>W_i(\overline{v}_i)$.
	\end{enumerate}
\end{alemma}

\begin{proof}
By Lemma~\ref{one-sided derivatives}, $h(t)=h_{A(t)}(\Phi(t))$ for every $t>0$. Let $d$ be an activation time and $d'$ the next one, with $d':=+\infty$ if $d$ is the last. Then $A(t)=A(d)$ for $t\in[d,d')$, and every $\Phi_k$ with $k\in A(d)$ is continuous and strictly decreasing on $[d_k,+\infty)\supseteq[d,+\infty)$ by (v). So $h_{A(d)}(\Phi)$ is continuous and strictly decreasing on $[d,+\infty)$, and $h=h_{A(d)}(\Phi)$ on $[d,d')\cap(0,+\infty)$. I prove (b) first, since (a) uses it.

(b) Let $d>0$ be an activation time, let $A_-:=\{j\in I_N:d_j<d\}$ be the set of active players on the stage that ends at $d$, and let $\Delta:=A(d)\setminus A_-=\{k\in I_N:d_k=d\}$. Then $\lim_{t\uparrow d}h(t)=h_{A_-}(\Phi(d))$ (Lemma~\ref{one-sided derivatives}), and $|A_-|\geq2$ by (iii). Every $k\in\Delta$ has $d_k=d>0$ and, by (iv), no atom at $d$, so $\Phi_k(d)=u_k=\overline{v}_k$. As $\Phi_k$ is decreasing on $[d,+\infty)$, its right derivative at $d$ is nonpositive, so Lemma~\ref{one-sided derivatives} gives $W_k(\overline{v}_k)\leq h(d)$. Hence
\begin{equation*}
(M(d)-1)h(d)=\sum_{j\in A_-}W_j(\Phi_j(d))+\sum_{k\in\Delta}W_k(\overline{v}_k)\leq(|A_-|-1)h_{A_-}(\Phi(d))+|\Delta|h(d),
\end{equation*}
that is, $(|A_-|-1)h(d)\leq(|A_-|-1)h_{A_-}(\Phi(d))$, so $h(d)\leq h_{A_-}(\Phi(d))=\lim_{t\uparrow d}h(t)$.

(a) By the first paragraph, $h$ is strictly decreasing on each $[d,d')\cap(0,+\infty)$, with $d$ and $d'$ as defined there, and at each positive activation time $d$, its left limit $h_{A_-}(\Phi(d))$ is below its value at every earlier time of the preceding stage and, by (b), at least $h(d)$. So $h$ is strictly decreasing on $(0,+\infty)$.

(c) If $t\in(0,d_i)$, then $d_i>0$ is a positive activation time at which player $i$ becomes active, so by (a), (b) and the proof of (b), $h(t)>\lim_{s\uparrow d_i}h(s)\geq h(d_i)\geq W_i(\overline{v}_i)$.
\end{proof}

\subsection{Proof of Lemma~\texorpdfstring{\ref{Characterization}}{1}}\label{proof Characterization}

\begin{proof}
Throughout the proof, I use the notation of Section~\ref{model} and Appendix~\ref{appendix preliminaries}, including $Q_j$, $Q_j(t-)$, $T_{-i}$, $Q_{-i}$, $x_j$, $\eta_j$, $H_j$, $\phi_i$, $\mu_i$ and (\ref{payoff split}). Player $j$ \emph{provides in} a set $B\subseteq[0,+\infty)$ if $F_j(\{v_j:T_j(v_j)\in B\})>0$. The \emph{support} $S_j$ of player $j$ is the set of times $t>0$ such that player $j$ provides in $(t-\epsilon,t+\epsilon)$ for every $\epsilon\in(0,t)$; it is closed in $(0,+\infty)$. A type $v$ of player $i$ is \emph{optimal at} $t\in[0,+\infty]$ if $R_i(t|v)=\sup_{s\in[0,+\infty]}R_i(s|v)$. One consequence of (\ref{payoff split}) is used repeatedly: if $\operatorname{Prob}(s\leq T_{-i}\leq t)=0$ for some $s<t<+\infty$, then $\phi_i(s)-\phi_i(t)=(e^{-r_is}-e^{-r_it})Q_{-i}(t)$ and $\mu_i(s)=\mu_i(t)=Q_{-i}(t)$, so that for every $v$,
\begin{equation}\label{vacancy gain}
R_i(s|v)-R_i(t|v)=(v-c_i)(e^{-r_is}-e^{-r_it})Q_{-i}(t);
\end{equation}
that is, when no opponent provides in $[s,t]$, providing at $s$ rather than at $t$ only saves discounting. Similarly, comparing the cases $T_{-i}<s$, $T_{-i}=s$, $s<T_{-i}<t$ and $T_{-i}\geq t$ shows that, for $0\leq s<t<+\infty$ and $v\geq c_i$,
\begin{equation}\label{delay bound}
R_i(t|v)-R_i(s|v)\geq c_ie^{-r_it}\operatorname{Prob}(s<T_{-i}<t)-(v-c_i)(e^{-r_is}-e^{-r_it})Q_{-i}(s).
\end{equation}
 For a profile $(T_i)_{i\in I_N}$, let $O_i$ be the set of types $v_i$ of player $i$ that are optimal at $T_i(v_i)$.

I first prove the necessity. Fix an equilibrium $(T_i)_{i\in I_N}$, so that $F_i(O_i)=1$ for every $i$.

\emph{Step 1 (clause (i), stubborn types).} For all $i\in I_N$, the weakly dominant strategy for type $v_i<c_i$ of player $i$ is $T_i(v_i)=+\infty$, since for every $t\in[0,+\infty)$, $0\leq\phi_i(t)-\phi_i(+\infty)=E[(e^{-r_it}-e^{-r_iT_{-i}})\mathbbm{1}\{T_{-i}>t\}]\leq e^{-r_it}Q_{-i}(t)\leq e^{-r_it}\mu_i(t)$, so that, by (\ref{payoff split}),
\begin{equation*}
R_i(+\infty|v_i)-R_i(t|v_i)=c_ie^{-r_it}\mu_i(t)-v_i[\phi_i(t)-\phi_i(+\infty)]\geq e^{-r_it}\min\{c_i-v_i,c_i\}\mu_i(t)\geq0.
\end{equation*}

The dominance is strict at $t$ whenever $\mu_i(t)>0$, that is, whenever $\operatorname{Prob}(T_{-i}\geq t)=\prod_{j\neq i}Q_j(t-)>0$. So a type $v_i<c_i$ that is optimal at a finite $t$ needs $Q_j(t-)=0$ for some $j\neq i$. This cannot happen, because every player is stubborn with positive probability. Let $\hat{t}_j:=\inf\{t\geq0:Q_j(t)=0\}$, and suppose that $\hat{t}:=\min_{j\in I_N}\hat{t}_j<+\infty$, attained by player $j$. Since $Q_j(\hat{t})=0$, almost every type of player $j$ provides by $\hat{t}$, including a set of optimal types $v_j<c_j$, which has probability $F_j(c_j)>0$ by the stubborn-type assumption. Each of them, optimal at some $t\leq\hat{t}$, needs a player $k\neq j$ with $Q_k(t-)=0$; then $\hat{t}_k\leq t\leq\hat{t}$, so $t=\hat{t}$ and $Q_k(\hat{t}-)=0$. Then almost every type of player $k$ provides before $\hat{t}$, including a set of optimal types $v_k<c_k$ of probability $F_k(c_k)>0$, and each of them, optimal at some $t'<\hat{t}$, needs a player $l\neq k$ with $Q_l(t'-)=0$, so that $\hat{t}_l\leq t'<\hat{t}$, which contradicts the definition of $\hat{t}$. Hence $Q_j(t)>0$ for every $j$ and every finite $t$, the dominance is strict at every finite $t$, and every stubborn type in $O_i$ other than $c_i$ chooses $+\infty$. Consequently, for all $i,j\in I_N$ and every $t\in[0,+\infty)$,
\begin{equation*}
Q_j(t)\geq F_j(c_j)>0,\qquad \Phi_j(t)\geq c_j,\qquad Q_{-i}(t)\geq\prod_{j\neq i}F_j(c_j)>0.
\end{equation*}

\emph{Step 2 (clause (i), types above cost).} In contrast, for type $v_i>c_i$ in $O_i$, $T_i(v_i)$ cannot be $+\infty$, since otherwise, this type can be better off by changing his stopping time to a finite point, say $\tau$, whereby he gains at least $e^{-r_i\tau}[(v_i-c_i)Q_{-i}(\tau)-v_i\operatorname{Prob}(\tau<T_{-i}<+\infty)]$ if no opponent of $i$ has an atom at $\tau$, which holds for all but countably many $\tau$. Indeed, $\mu_i(\tau)=Q_{-i}(\tau)$, and by the identity for $\phi_i(t)-\phi_i(+\infty)$ in Step 1, $\phi_i(\tau)-\phi_i(+\infty)\geq e^{-r_i\tau}[Q_{-i}(\tau)-\operatorname{Prob}(\tau<T_{-i}<+\infty)]$; so, by (\ref{payoff split}) and $v_i>0$, $R_i(\tau|v_i)-R_i(+\infty|v_i)=v_i[\phi_i(\tau)-\phi_i(+\infty)]-c_ie^{-r_i\tau}Q_{-i}(\tau)$ is at least the gain above. This gain is positive for large $\tau$, because $\operatorname{Prob}(\tau<T_{-i}<+\infty)\to0$ as $\tau\to+\infty$ while $Q_{-i}(\tau)\geq\prod_{j\neq i}F_j(c_j)>0$ by Step 1. With Step 1, this gives $F_i(\{v_i:T_i(v_i)=+\infty\})=F_i(c_i)$, so that $Q_i(t)\to F_i(c_i)$ and $\Phi_i(t)\to c_i$ as $t\to+\infty$.

\emph{Step 3 (monotonicity in clause (v)).} I show that $T_i(\cdot)$ is nonincreasing on $O_i$ for all $i\in I_N$. In equilibrium, types $v_i$ and $\tilde{v}_i$ of player $i$ in $O_i$ both reach optimality, which gives:
\begin{equation*}
\begin{aligned}
R_i(T_i(v_i)|v_i)\geq R_i(T_i(\tilde{v}_i)|v_i), \\
R_i(T_i(\tilde{v}_i)|\tilde{v}_i)\geq R_i(T_i(v_i)|\tilde{v}_i).
\end{aligned}
\end{equation*}
Adding these two inequalities and using (\ref{payoff split}), whose second term does not depend on the type, I have $(v_i-\tilde{v}_i)[\phi_i(T_i(v_i))-\phi_i(T_i(\tilde{v}_i))]\geq 0$ for all $i\in I_N$ and $v_i$, $\tilde{v}_i\in O_i$. This means that $\phi_i(T_i(\cdot))$ is nondecreasing on $O_i$. Since $\phi_i(t)=1-r_i\int_0^te^{-r_is}Q_{-i}(s)\,ds$ (Appendix~\ref{appendix preliminaries}), $\phi_i'(t)=-r_ie^{-r_it}Q_{-i}(t)\leq-r_ie^{-r_it}\prod_{j\neq i}F_j(c_j)<0$ at almost every $t$, by Step 1, implying that $\phi_i(\cdot)$ is strictly decreasing on $[0,+\infty]$, where $\phi_i(+\infty)=\lim_{t\to+\infty}\phi_i(t)$, and thus $T_i(\cdot)$ is nonincreasing on $O_i$.

Monotonicity locates the optimal types. Let $v\in O_i$ with $T_i(v)=s<+\infty$. Every $v'\in O_i$ with $v'\geq v$ has $T_i(v')\leq s$, and every $v'\in O_i$ with $v'\leq v$ has $T_i(v')\geq s$. Since $F_i(O_i)=1$, it follows that $Q_i(s)\leq F_i(v)\leq Q_i(s-)$, that is,
\begin{equation}\label{sandwich}
\Phi_i(s)\leq v\leq F_i^{-1}(Q_i(s-)).
\end{equation}

\emph{Step 4 (clause (iv)).}

\begin{alemma}[Atoms]\label{no atom}
	In every equilibrium:
	\begin{enumerate}[(a)]
		\item at every $t\in[0,+\infty)$, at most one player has an atom;
		\item no player has an atom at any $t\in(0,+\infty)$.
	\end{enumerate}
\end{alemma}

\begin{proof}
I prove (a) by contradiction. Suppose there are two or more players who provide at some $t\in[0,+\infty)$ with positive probabilities. By Step 1, for any one of them, say player $k$, almost every type $v_k$ that provides at $t$ is optimal and satisfies $v_k\in(c_k,\overline{v}_k]$. Such a type can be better off by slightly delaying the stopping time. Formally, define $p_k:=\operatorname{Prob}(T_{-k}=t)$, the probability that the first stopping time of $k$'s opponents is $t$, which is positive because another player has an atom at $t$ and $Q_j(t)>0$ for every $j$ (Step 1). Type $v_k$ can be better off by waiting until $t+\epsilon$ for some small $\epsilon>0$ to gain at least $e^{-r_kt}[p_kc_k/N-(1-p_k)(1-e^{-r_k\epsilon})(v_k-c_k)]$, which is positive for every small enough $\epsilon>0$, uniformly in $v_k\leq\overline{v}_k$. Indeed, the delay changes nothing if $T_{-k}<t$; if $T_{-k}=t$, type $v_k$ obtains $v_ke^{-r_kt}$ instead of at most $(v_k-c_k/N)e^{-r_kt}$, because he shares the cost with at most $N-1$ opponents; and if $T_{-k}>t$, he obtains at least $(v_k-c_k)e^{-r_k(t+\epsilon)}$ instead of $(v_k-c_k)e^{-r_kt}$. This contradicts the optimality of $v_k$. Thus, at most one player has an atom at $t$.

Suppose that player $i$ has an atom at some $t^*>0$, of probability $q:=F_i(\{v_i:T_i(v_i)=t^*\})>0$. By (a), no other player has an atom at $t^*$. I first show that no opponent of $i$ provides just before $t^*$: a type that does gains by waiting until just after $t^*$, because he then free rides on player $i$'s atom. Fix $j\neq i$, $\delta\in(0,t^*)$, a type $v\in(c_j,\overline{v}_j]$ and a time $s\in(t^*-\delta,t^*)$. By (\ref{delay bound}) with $t=t^*+\delta$, since $\operatorname{Prob}(s<T_{-j}<t^*+\delta)\geq\operatorname{Prob}(T_{-j}=t^*)\geq q\prod_{l\neq i,j}F_l(c_l)$ by Step 1, $s>t^*-\delta$, $v\leq\overline{v}_j$ and $Q_{-j}(s)\leq1$, providing at $t^*+\delta$ rather than at $s$ gains at least
\begin{equation*}
c_je^{-r_j(t^*+\delta)}q\prod_{l\neq i,j}F_l(c_l)-(\overline{v}_j-c_j)\big(e^{-r_j(t^*-\delta)}-e^{-r_j(t^*+\delta)}\big),
\end{equation*}
which is positive for $\delta$ small enough, uniformly in $v$ and $s$. Fix $\delta\in(0,t^*)$ small enough for every $j\neq i$, which is possible because there are finitely many players. Then no type in $O_j\cap(c_j,\overline{v}_j]$ chooses a stopping time in $(t^*-\delta,t^*)$, and by Step 1 no type in $O_j\cap[\underline{v}_j,c_j)$ does, so player $j$ does not provide in $(t^*-\delta,t^*)$. Hence $\operatorname{Prob}(t^*-\delta<T_{-i}\leq t^*)=0$, and (\ref{vacancy gain}) with $s=t^*-\delta/2$ and $t=t^*$ shows that every type $v_i>c_i$ strictly prefers $t^*-\delta/2$ to $t^*$: by providing earlier, he loses no chance that an opponent provides first and saves discounting. By Step 1, almost every type that provides at $t^*$ is such a type, so the atom, of probability $q>0$, consists of types outside $O_i$, a set of probability zero, a contradiction.
\end{proof}

By Lemma~\ref{no atom}, every $Q_j$, and hence every $\Phi_j$, is continuous on $(0,+\infty)$, and $\mu_k(t)=Q_{-k}(t)$ for every $t>0$. So, by (\ref{payoff split}), $R_k(t|v)=v\phi_k(t)-c_ke^{-r_kt}Q_{-k}(t)$ is jointly continuous in $(t,v)$ on $(0,+\infty)\times\mathbb{R}$, and by (\ref{sandwich}), every $v\in O_k$ with $T_k(v)=s\in(0,+\infty)$ satisfies $v=\Phi_k(s)$. I also use two facts about supports. For an open interval $J\subseteq(0,+\infty)$, player $k$ provides in $J$ if and only if $J\cap S_k\neq\emptyset$: if $J\cap S_k=\emptyset$, every point of $J$ has a neighborhood in which player $k$ does not provide, and countably many of these neighborhoods cover $J$. If player $k$ does not provide in an interval $(a,b)\subseteq(0,+\infty)$, then $Q_k$ is constant on $[a,b]$, because he has no atom at a positive time, and so $\eta_k=0$ on $(a,b)$.

\emph{Step 5 (two tools).}

\begin{alemma}[No single provider]\label{no single provider}
	In every equilibrium, there is no interval $(a,b)$ with $0\leq a<b$ in which exactly one player provides. Consequently, every $t\in S_j$ lies in $S_k$ for some $k\neq j$.
\end{alemma}

\begin{proof}
Suppose that player $j$ provides in $(a,b)$ and no other player does. By Step 1, almost every type $v$ of player $j$ with $T_j(v)\in(a,b)$ is in $O_j$ and higher than $c_j$; take such a type $v$, with $T_j(v)=t\in(a,b)$, and let $s\in(a,t)$. Since no opponent of $j$ provides in $(a,b)$, $\operatorname{Prob}(s\leq T_{-j}\leq t)=0$, and (\ref{vacancy gain}) gives $R_j(s|v)-R_j(t|v)=(v-c_j)(e^{-r_js}-e^{-r_jt})Q_{-j}(t)>0$, which contradicts $v\in O_j$. For the second claim, let $t\in S_j$ and suppose that $t\notin S_k$ for every $k\neq j$. Since the supports are closed in $(0,+\infty)$, some interval $(t-\epsilon,t+\epsilon)\subseteq(0,+\infty)$ meets no $S_k$ with $k\neq j$, so player $j$ is the only player who provides in it (Step 4), which contradicts the first claim.
\end{proof}

\begin{alemma}[Current types]\label{OSP}
	In every equilibrium, for every player $k$ and every $t\in S_k$, the type $\Phi_k(t)$ is optimal at $t$.
\end{alemma}

\begin{proof}
Let $V_k(v):=\sup_{s\in[0,+\infty]}R_k(s|v)$. By (\ref{payoff split}), $V_k$ is a supremum of affine functions of $v$ with slopes $\phi_k(s)\in[0,1]$, so it is Lipschitz continuous with constant $1$. Let $t\in S_k$. For every $\epsilon\in(0,t)$, player $k$ provides in $(t-\epsilon,t+\epsilon)$, so some $v_\epsilon\in O_k$ has $s_\epsilon:=T_k(v_\epsilon)\in(t-\epsilon,t+\epsilon)$, and $v_\epsilon=\Phi_k(s_\epsilon)$ by Step 4. As $\epsilon\to0$, $v_\epsilon\to\Phi_k(t)$, because $\Phi_k$ is continuous on $(0,+\infty)$. Since $R_k(s_\epsilon|v_\epsilon)=V_k(v_\epsilon)$, the joint continuity of $R_k$ on $(0,+\infty)\times\mathbb{R}$ and the continuity of $V_k$ give $R_k(t|\Phi_k(t))=V_k(\Phi_k(t))$.
\end{proof}

\emph{Step 6 (Lipschitz continuity).} This part follows the spirit of the proof of \citet[Lemma~1(iv)]{fudenberg1986theory}. For every player $k$ and every $\overline{t}>\underline{t}$ with $\underline{t}\in S_k$, the type $v_k:=\Phi_k(\underline{t})$, which is optimal at $\underline{t}$ by Lemma~\ref{OSP}, prefers providing at $\underline{t}$ to waiting until $\overline{t}$ to provide. Since $c_k\leq v_k\leq\overline{v}_k$ (Step 1) and, by Lemma~\ref{no atom}, no opponent has an atom at $\overline{t}>0$, (\ref{delay bound}) with $s=\underline{t}$ and $t=\overline{t}$ gives
\begin{equation*}
\begin{aligned}
0&\geq R_k(\overline{t}|v_k)-R_k(\underline{t}|v_k)\geq c_ke^{-r_k\overline{t}}\big[Q_{-k}(\underline{t})-Q_{-k}(\overline{t})\big]-(v_k-c_k)\big(e^{-r_k\underline{t}}-e^{-r_k\overline{t}}\big)Q_{-k}(\underline{t}) \\
&\Rightarrow Q_{-k}(\underline{t})-Q_{-k}(\overline{t})\leq\frac{v_k-c_k}{c_k}\big(e^{r_k(\overline{t}-\underline{t})}-1\big)\leq\frac{r_k(\overline{v}_k-c_k)}{c_k}(\overline{t}-\underline{t})e^{r_k(\overline{t}-\underline{t})}.
\end{aligned}
\end{equation*}
The inequality above shows that the decrease of $Q_{-k}(\cdot)$ over an interval of length at most one that starts in $S_k$ is at most $r_k(\overline{v}_k-c_k)e^{r_k}/c_k$ times the length of the interval. So, for every player $j$, every $\underline{t}\in S_j$ and every $\overline{t}\in(\underline{t},\underline{t}+1]$, taking $k\neq j$ with $\underline{t}\in S_k$ (Lemma~\ref{no single provider}) and using that $Q_{-k}=Q_j\prod_{l\neq j,k}Q_l$ (Appendix~\ref{appendix preliminaries}), where the product is nonincreasing and at least $\prod_{l\neq j,k}F_l(c_l)$ (Step 1), gives
\begin{equation*}
Q_j(\underline{t})-Q_j(\overline{t})\leq\frac{Q_{-k}(\underline{t})-Q_{-k}(\overline{t})}{\prod_{l\neq j,k}F_l(c_l)}\leq L_Q(\overline{t}-\underline{t}),\qquad L_Q:=\frac{\max_{k\in I_N}r_k(\overline{v}_k-c_k)e^{r_k}/c_k}{\min_{j,k\in I_N,\,j\neq k}\prod_{l\neq j,k}F_l(c_l)}.
\end{equation*}
For arbitrary $0<s<t\leq s+1$, if $S_j\cap[s,t]=\emptyset$, player $j$ does not provide in $(s,t)$ and $Q_j(s)=Q_j(t)$ (Step 4); otherwise, with $\underline{t}:=\min(S_j\cap[s,t])$, $Q_j(s)=Q_j(\underline{t})$ and $Q_j(s)-Q_j(t)\leq L_Q(t-\underline{t})\leq L_Q(t-s)$. Therefore, every $Q_j$ is Lipschitz continuous on $(0,+\infty)$ with constant $L_Q$. Next, because $f_i(\cdot)$ is bounded away from 0, $F_i^{-1}(\cdot)$ is Lipschitz continuous, as $|F_i^{-1}(y)-F_i^{-1}(y')|\leq|y-y'|\max_{v\in[\underline{v}_i,\overline{v}_i]}(1/f_i(v))$. Thus, $\Phi_i(\cdot)=F_i^{-1}(F_i(\Phi_i(\cdot)))$ is Lipschitz continuous on $(0,+\infty)$, which indicates that $\Phi_i(\cdot)$ is differentiable almost everywhere. Likewise, $x_i=\ln F_i(\Phi_i)$ is Lipschitz continuous on $(0,+\infty)$, because $F_i(\Phi_i)\geq F_i(c_i)>0$, so $\eta_i=-x_i'\geq0$ exists at almost every $t>0$; and by Lemma~\ref{payoff derivative}, every $R_k(\cdot|v)$ is locally absolutely continuous on $(0,+\infty)$, with derivative (\ref{payoff slope}) at almost every $t>0$.

\emph{Step 7 (indifference condition).} Let $k\in I_N$, and let $t\in S_k$ be a time at which every $Q_j$ is differentiable, which holds at almost every $t\in S_k$ by Step 6. By Lemma~\ref{OSP}, the type $\Phi_k(t)$ is optimal at $t$, and $t>0$ is an interior point of $[0,+\infty]$. By Lemma~\ref{payoff derivative}, $R_k(\cdot|\Phi_k(t))$ is differentiable at $t$ and satisfies (\ref{payoff slope}), so its derivative vanishes there, and since $c_ke^{-r_kt}Q_{-k}(t)>0$ by Step 1,
\begin{equation*}
h_{-k}(t)=W_k(\Phi_k(t))\quad\text{at almost every }t\in S_k.
\end{equation*}
This is the indifference condition of type $\Phi_k(t)$ (Section~\ref{model}). Let $\mathcal{S}(t):=\{k\in I_N:t\in S_k\}$. If $l\notin\mathcal{S}(t)$, player $l$ does not provide in some neighborhood of $t$, so $\eta_l(t)=0$ (Step 4). Therefore, at almost every $t>0$, either $\mathcal{S}(t)=\emptyset$ and $h(t)=0$, or $|\mathcal{S}(t)|\geq2$ (Lemma~\ref{no single provider}) and summing $h(t)-\eta_k(t)=W_k(\Phi_k(t))$ over $k\in\mathcal{S}(t)$ gives
\begin{equation}\label{rate identity}
h(t)=h_{\mathcal{S}(t)}(\Phi(t)),
\end{equation}
where $h_I$ is the function of Appendix~\ref{appendix preliminaries}. Since $W_k(\Phi_k)\geq0$ by Step 1, it follows that $0\leq h(t)\leq\sum_{k\in I_N}W_k(\Phi_k(t))\leq\overline{h}:=\sum_{k\in I_N}W_k(\overline{v}_k)$ at almost every $t>0$.

\emph{Step 8 (provision never stops).} Let $\overline{s}:=\sup\bigcup_{k\in I_N}S_k$, with $\sup\emptyset:=0$. I show that $\overline{s}=+\infty$, that is, some player provides after every time. Suppose that $\overline{s}<+\infty$. Then no player provides in $(\overline{s},+\infty)$, so by Step 2, $Q_k(\overline{s})=F_k(c_k)$ and $\Phi_k(\overline{s})=c_k$ for every $k$. If $\overline{s}=0$, every player would then have an atom at $0$, which contradicts Lemma~\ref{no atom}(a); so $\overline{s}>0$. Let $D(t):=\sum_{k\in I_N}(\Phi_k(t)-c_k)\geq0$. It is Lipschitz continuous on $(0,+\infty)$ by Step 6, $D(\overline{s})=0$, and $D(s)>0$ for $s\in(0,\overline{s})$, since otherwise $Q_k(s)=F_k(c_k)$ for every $k$ and no player would provide after $s$. At almost every $t>0$, $\Phi_k'=-H_k(\Phi_k)\eta_k$ by the chain rule, and $h\leq\sum_kW_k(\Phi_k)\leq\max_k(r_k/c_k)D$ by Step 7. So, with $\overline{H}:=\max_{k\in I_N}\max_{v\in[\underline{v}_k,\overline{v}_k]}H_k(v)<+\infty$ and $L_D:=\overline{H}\max_{k\in I_N}r_k/c_k$,
\begin{equation*}
D'(t)\geq-\overline{H}h(t)\geq-L_DD(t).
\end{equation*}
Hence $e^{L_Dt}D(t)$, which is absolutely continuous on compact subintervals of $(0,+\infty)$, is nondecreasing there (Gronwall), and $D(\overline{s})\geq e^{-L_D(\overline{s}-s)}D(s)>0$ for $s\in(0,\overline{s})$, a contradiction.

\emph{Step 9 (clauses (ii) and (iii)).}

\begin{alemma}[No common vacancy]\label{no common vacancy}
	In every equilibrium, if for some $0\leq a<b$ no player provides in $(a,b)$, then no player provides in $(a,+\infty)$.
\end{alemma}

\begin{proof}
Suppose that no player provides in $(a,b)$, and let $b':=\sup\{t\geq b:\text{no player provides in }(a,t)\}$. Suppose that $b'<+\infty$. Then no player provides in $(a,b')$, but for every $\epsilon>0$ some player provides in $(b',b'+\epsilon)$, since no player has an atom at $b'>0$ (Lemma~\ref{no atom}). As there are finitely many players, some player $k$ provides in $(b',b'+\epsilon)$ for every $\epsilon>0$, so $b'\in S_k$, and $Q_k(b')>Q_k(b'+1)\geq F_k(c_k)$ by Step 1, that is, $v:=\Phi_k(b')>c_k$. By Lemma~\ref{OSP}, type $v$ is optimal at $b'$. Since no opponent of $k$ provides in $(a,b')$ or has an atom at $b'$, (\ref{vacancy gain}) with $s:=(a+b')/2$ gives
\begin{equation*}
R_k(s|v)-R_k(b'|v)=(v-c_k)(e^{-r_ks}-e^{-r_kb'})Q_{-k}(b')>0,
\end{equation*}
a contradiction. Hence $b'=+\infty$.
\end{proof}

Three facts follow. First, every $t>0$ lies in $S_k$ for at least two players $k$. Otherwise, by Lemma~\ref{no single provider}, $t$ lies in no support; since the supports are closed in $(0,+\infty)$, some interval $(a,b)\ni t$ with $a>0$ meets no support, so no player provides in it (Step 4), and by Lemma~\ref{no common vacancy} no player provides in $(a,+\infty)$, which contradicts Step 8. Second, at least two players have $d_k=0$, which is clause (iii). For every integer $n\geq1$, $1/n\in S_j\cap S_k$ for some $j\neq k$, so some pair $\{j,k\}$ does so for infinitely many $n$; then $\inf S_j=\inf S_k=0$, and $d_l\leq\inf S_l$ for every player $l$ by the definition of $d_l$. Third, every $S_k$ is unbounded. Suppose that $\overline{s}_k:=\sup S_k<+\infty$ for some $k$, with $\sup\emptyset:=0$. Then player $k$ does not provide in $(\overline{s}_k,+\infty)$, so $\Phi_k(\overline{s}_k)=c_k$ by Step 2, and the type $c_k$ is optimal at $\overline{s}_k$. If $\overline{s}_k>0$, this is Lemma~\ref{OSP}. If $\overline{s}_k=0$, player $k$ has an atom at $0$ that contains almost every type $v>c_k$, these types are optimal at $0$, and letting $v$ fall to $c_k$ in $R_k(0|v)\geq R_k(s|v)$, $s\in[0,+\infty]$, gives the claim, since both sides are affine in $v$. No opponent of $k$ has an atom at $\overline{s}_k$ (Lemma~\ref{no atom}), so $R_k(\cdot|c_k)$ is continuous at $\overline{s}_k$ by (\ref{payoff split}). By Step 6 and $W_k(c_k)=0$, for every $t>\overline{s}_k$,
\begin{equation*}
R_k(t|c_k)-R_k(\overline{s}_k|c_k)=\int_{\overline{s}_k}^tc_ke^{-r_ku}Q_{-k}(u)h_{-k}(u)du\geq c_ke^{-r_kt}Q_{-k}(t)\ln\frac{Q_{-k}(\overline{s}_k)}{Q_{-k}(t)}>0,
\end{equation*}
because the opponents of $k$ provide in $(\overline{s}_k,t)$ by the first fact. This contradicts the optimality of $c_k$ at $\overline{s}_k$. Since every $S_k$ is unbounded, every player provides after every time, so $Q_k(t)>F_k(c_k)$ and $\Phi_k(t)>c_k$ for every $k$ and every $t\in[0,+\infty)$.

\emph{Step 10 (clause (v): no gaps).}

\begin{alemma}[No gap]\label{no gap}
	Fix an equilibrium. A \emph{gap} of player $k$ is an interval $(a,b)$ with $0<a<b<+\infty$, $a,b\in S_k$ and $(a,b)\cap S_k=\emptyset$. For $n\geq2$, let $C(n)$ be the following statement: for every open interval $J\subseteq(0,+\infty)$ and every set $\mathcal{K}\subseteq I_N$ of $n$ players such that $S_l\cap J=\emptyset$ for every $l\notin\mathcal{K}$ and every $t\in J$ lies in $S_k$ for at least two players $k\in\mathcal{K}$,
	\begin{enumerate}[(a)]
		\item[(G)] no player in $\mathcal{K}$ has a gap with both endpoints in $J$, and
		\item[(H)] $h$ is almost everywhere equal on $J$ to a strictly decreasing function.
	\end{enumerate}
	Then $C(n)$ holds for every $n\in\{2,\dots,N\}$. Moreover, if player $k$ has an atom at $t=0$, then $\inf S_k=0$.
	Consequently, for every player $k$, $S_k=[d_k,+\infty)\cap(0,+\infty)$, and $\Phi_k$ is continuous and strictly decreasing on $[d_k,+\infty)$ and maps it onto $(c_k,u_k]$.
\end{alemma}

\begin{proof}
The idea is that no player's support has a gap. At both ends of a gap of player $k$, his current type is the same and is optimal, so the provision hazard cannot exceed this type's indifference level just after the gap opens, nor fall below it just before the gap closes. A strictly falling hazard cannot satisfy both conditions, and I show that the hazard falls strictly by induction on the number of players who provide. I first collect three tools.

\emph{Gap constraints.} Let $(a,b)$ be a gap of player $k$. Player $k$ does not provide in $(a,b)$, so $\Phi_k(a)=\Phi_k(b)=:v^*$ and $\eta_k=0$ on $(a,b)$, hence $h_{-k}=h$ almost everywhere on $(a,b)$ (Step 4). By Lemma~\ref{OSP}, type $v^*$ is optimal at $a$ and at $b$. By Step 6, $R_k(s|v^*)-R_k(a|v^*)=\int_a^sc_ke^{-r_ku}Q_{-k}(u)[h(u)-W_k(v^*)]du$ for $s\in[a,b]$, so for every $s\in(a,b)$,
\begin{equation}\label{gap constraints}
\int_a^se^{-r_ku}Q_{-k}(u)[h(u)-W_k(v^*)]du\leq0\leq\int_s^be^{-r_ku}Q_{-k}(u)[h(u)-W_k(v^*)]du.
\end{equation}
The left inequality uses only the optimality of $v^*$ at $a$ and $\eta_k=0$ on $(a,s)$, and the right one only the optimality of $v^*$ at $b$ and $\eta_k=0$ on $(s,b)$.

\emph{Pieces.} Let $J$ and $\mathcal{K}$ satisfy the hypotheses of $C(n)$. An open interval $U\subseteq J$ is a \emph{piece} if, for every $k\in\mathcal{K}$, either $U\subseteq S_k$ or $U\cap S_k=\emptyset$. Let $\mathcal{S}(U):=\{k\in\mathcal{K}:U\subseteq S_k\}$; then, by the hypotheses of $C(n)$, $\mathcal{S}(t)=\mathcal{S}(U)$ for every $t\in U$, and $|\mathcal{S}(U)|\geq2$. On a piece $U$, $\eta_l=0$ for every $l\notin\mathcal{S}(U)$ (Step 4), and the indifference condition of Step 7 holds almost everywhere for every $k\in\mathcal{S}(U)$, so (\ref{rate identity}) gives $h=h_{\mathcal{S}(U)}(\Phi)$ and $\eta_k=h_{\mathcal{S}(U)}(\Phi)-W_k(\Phi_k)$ for $k\in\mathcal{S}(U)$, almost everywhere on $U$. The function $h_{\mathcal{S}(U)}(\Phi)$ is continuous and nonincreasing on $(0,+\infty)$, as every $\Phi_k$ is. It is strictly decreasing on the closure of $U$ in $(0,+\infty)$. Otherwise it would be constant on an interval $(t_1,t_2)\subseteq U$, so would each $W_k(\Phi_k)$ and each $Q_k$ with $k\in\mathcal{S}(U)$, and then $h=0$ almost everywhere on $(t_1,t_2)$, which contradicts $h=h_{\mathcal{S}(U)}(\Phi)>0$ there; the last inequality holds because $\Phi_k>c_k$ by Step 9.

\emph{Switching.} Suppose that, for every $k\in\mathcal{K}$, $S_k\cap J$ is a subinterval of $J$ (possibly empty), closed relative to $J$. Then the endpoints of these intervals that lie in $J$, called \emph{switching times}, are finitely many, and they divide $J$ into finitely many pieces. Let $\tau$ be a switching time, let $U_-$ and $U_+$ be the pieces to its left and right, and write $\mathcal{S}_\pm:=\mathcal{S}(U_\pm)$, $h_\pm:=h_{\mathcal{S}_\pm}(\Phi(\tau))$ and $w_k:=W_k(\Phi_k(\tau))$ for every $k$. I claim that $h_+\leq h_-$. If $j\in\mathcal{S}_+\setminus\mathcal{S}_-$ (player $j$ starts providing at $\tau$), then $0\leq\eta_j=h_{\mathcal{S}_+}(\Phi)-W_j(\Phi_j)$ almost everywhere on $U_+$, so $w_j\leq h_+$ by continuity. If $k\in\mathcal{S}_-\setminus\mathcal{S}_+$ (player $k$ stops providing at $\tau$), then $\tau\in S_k$ because $S_k$ is closed, and $\eta_k=0$ on $U_+$; so the argument behind the left inequality in (\ref{gap constraints}) applies at $\tau$, with $v^*=\Phi_k(\tau)$ and $h=h_{\mathcal{S}_+}(\Phi)$ on $U_+$, and gives $w_k\geq h_+$, since otherwise the integrand would be positive near $\tau$. With $\mathcal{C}:=\mathcal{S}_-\cap\mathcal{S}_+$, the players who start providing give $(|\mathcal{S}_+|-1)h_+=\sum_{k\in\mathcal{C}}w_k+\sum_{j\in\mathcal{S}_+\setminus\mathcal{S}_-}w_j\leq\sum_{k\in\mathcal{C}}w_k+|\mathcal{S}_+\setminus\mathcal{S}_-|h_+$, that is, $\sum_{k\in\mathcal{C}}w_k\geq(|\mathcal{C}|-1)h_+$, and then the players who stop providing give
\begin{equation*}
(|\mathcal{S}_-|-1)h_-=\sum_{k\in\mathcal{C}}w_k+\sum_{k\in\mathcal{S}_-\setminus\mathcal{S}_+}w_k\geq(|\mathcal{C}|-1)h_++|\mathcal{S}_-\setminus\mathcal{S}_+|h_+=(|\mathcal{S}_-|-1)h_+.
\end{equation*}
Since $|\mathcal{S}_-|\geq2$, $h_-\geq h_+$. Hence the function $\tilde{h}$ that equals $h_{\mathcal{S}(U)}(\Phi)$ on each piece $U$ and $h_+$ at each switching time is strictly decreasing on $J$, and it equals $h$ almost everywhere on $J$.

\emph{Proof of $C(n)$.} I argue by induction on $n$. Let $J$ and $\mathcal{K}$ satisfy the hypotheses of $C(n)$.

\emph{Proof of (G).} If $n=2$, every $t\in J$ lies in both supports of the players in $\mathcal{K}$, so no player in $\mathcal{K}$ has a gap with both endpoints in $J$. Let $n\geq3$, assume $C(n-1)$, and suppose that player $i\in\mathcal{K}$ has a gap $(a,b)$ with $a,b\in J$. Then $(a,b)\subseteq J$, $S_i\cap(a,b)=\emptyset$, $S_l\cap(a,b)=\emptyset$ for every $l\notin\mathcal{K}$, and every $t\in(a,b)$ lies in $S_k$ for at least two players $k\in\mathcal{K}':=\mathcal{K}\setminus\{i\}$. So $C(n-1)$(H), applied to $(a,b)$ and $\mathcal{K}'$, gives a strictly decreasing function $\tilde{h}$ on $(a,b)$ with $\tilde{h}=h$ almost everywhere on $(a,b)$. As $\tilde{h}$ is decreasing and $h\leq\overline{h}$ almost everywhere (Step 7), $\tilde{h}\leq\overline{h}$ on $(a,b)$, so $\tilde{h}(a+):=\lim_{t\downarrow a}\tilde{h}(t)$ is finite. Let $v^*:=\Phi_i(a)$. If $\tilde{h}(a+)>W_i(v^*)$, then $\tilde{h}>W_i(v^*)$ on some $(a,t_0)\subseteq(a,b)$, and the left inequality in (\ref{gap constraints}) fails at $s=t_0$. Otherwise $\tilde{h}(t)<\tilde{h}(a+)\leq W_i(v^*)$ for every $t\in(a,b)$, and the right inequality fails at every $s\in(a,b)$. Either way, this is a contradiction.

\emph{Proof of (H).} By (G), every $S_k\cap J$ with $k\in\mathcal{K}$ is a subinterval of $J$. Otherwise there are $t_1<t_2<t_3$ in $J$ with $t_1,t_3\in S_k$ and $t_2\notin S_k$, and since $S_k$ is closed, $\big(\max(S_k\cap[t_1,t_2]),\min(S_k\cap[t_2,t_3])\big)$ is a gap of player $k$ with both endpoints in $J$. The interval $S_k\cap J$ is closed relative to $J$, because $S_k$ is closed in $(0,+\infty)$. So the switching tool applies and gives (H).

\emph{Atom at $0$.} Let player $k$ have an atom at $0$, of probability $q>0$; by Lemma~\ref{no atom}(a), no other player has an atom at $0$. Let $v^0:=\Phi_k(0)=F_k^{-1}(1-q)<\overline{v}_k$, and suppose that $\underline{s}_k:=\inf S_k>0$; then $\underline{s}_k<+\infty$ and $\underline{s}_k\in S_k$ by Step 9. Player $k$ does not provide in $(0,\underline{s}_k)$, so $\Phi_k=v^0$ on $[0,\underline{s}_k]$ (Step 4), $v^0=\Phi_k(\underline{s}_k)>c_k$ by Step 9, and $\eta_k=0$ on $(0,\underline{s}_k)$. Three facts lead to a contradiction. First, $v^0$ is optimal at $0$. Indeed, every $v\in O_k$ with $v>v^0$ provides at $0$, since types in $O_k$ that provide at a time $s>0$ equal $\Phi_k(s)\leq v^0$ (Step 4) and types in $O_k$ that never provide are at most $c_k$ (Step 2); so $R_k(0|v)\geq R_k(s|v)$ for every $s\in[0,+\infty]$ and almost every $v>v^0$, and letting $v$ fall to $v^0$ gives the claim, since both sides are affine in $v$. Second, $v^0=\Phi_k(\underline{s}_k)$ is optimal at $\underline{s}_k$ by Lemma~\ref{OSP}. Third, no opponent of $k$ has an atom at $0$, so $R_k(\cdot|v^0)$ is continuous at $0$ by (\ref{payoff split}), and by Step 6, $R_k(s|v^0)-R_k(0|v^0)=\int_0^sc_ke^{-r_ku}Q_{-k}(u)[h(u)-W_k(v^0)]du$ for $s\in(0,\underline{s}_k]$, where the integrand is bounded because $h\leq\overline{h}$. Now every $t\in(0,\underline{s}_k)$ lies in $S_l$ for at least two players $l\neq k$ (Step 9). If $N=2$, this is impossible. If $N\geq3$, $C(N-1)$(H), applied to $(0,\underline{s}_k)$ and $I_N\setminus\{k\}$, gives a strictly decreasing function $\tilde{h}$ on $(0,\underline{s}_k)$ equal to $h$ almost everywhere. As in the proof of (G), if $\tilde{h}(0+)>W_k(v^0)$, then $R_k(t_0|v^0)>R_k(0|v^0)$ for some $t_0\in(0,\underline{s}_k)$, which contradicts the first fact; otherwise $\tilde{h}<W_k(v^0)$ on $(0,\underline{s}_k)$, so $R_k(\underline{s}_k|v^0)<R_k(0|v^0)$, which contradicts the second. Hence $\inf S_k=0$.

\emph{Conclusion.} The hypotheses of $C(N)$ with $J=(0,+\infty)$ and $\mathcal{K}=I_N$ hold by Step 9. By $C(N)$(G) and the argument in the proof of (H), every $S_k$ is a subinterval of $(0,+\infty)$, closed relative to $(0,+\infty)$, and it is unbounded (Step 9); so $S_k=[\underline{s}_k,+\infty)\cap(0,+\infty)$ with $\underline{s}_k:=\inf S_k$. If player $k$ has no atom at $0$, then $d_k=\underline{s}_k$, because, by Lemma~\ref{no atom}, he provides in $[0,t]$ if and only if he provides in $(0,t)$, that is, if and only if $\underline{s}_k<t$. If he has an atom at $0$, then $d_k=0=\underline{s}_k$ by the atom-at-$0$ claim. So $S_k=[d_k,+\infty)\cap(0,+\infty)$. Hence player $k$ provides in every nondegenerate interval in $(d_k,+\infty)$, and $Q_k$ is strictly decreasing on $[d_k,+\infty)$; it is also continuous there (Lemma~\ref{no atom}, and $Q_k$ is right-continuous at $0$). So $\Phi_k$ is continuous and strictly decreasing on $[d_k,+\infty)$, it maps $[d_k,+\infty)$ onto $(c_k,u_k]$ by Steps 2 and 9, and $\Phi_k(t)=\overline{v}_k$ for $t<d_k$.
\end{proof}

\emph{Step 11 (clauses (i)--(v)).} For every player $k$, define
\begin{equation*}
T_k^*(v):=\begin{cases}
+\infty & \text{for }v\in[\underline{v}_k,c_k],\\
\Phi_k^{-1}(v) & \text{for }v\in(c_k,u_k],\\
d_k & \text{for }v\in(u_k,\overline{v}_k],
\end{cases}
\end{equation*}
where $\Phi_k^{-1}$ is the inverse of $\Phi_k$ on $[d_k,+\infty)$. The function $T_k^*$ is continuous and strictly decreasing on $(c_k,u_k]$, and it agrees with $T_k$ almost everywhere. Indeed, let $v\in O_k$. If $T_k(v)=+\infty$, then $v\leq c_k$ by Step 2, and $T_k^*(v)=+\infty$. If $T_k(v)=s\in(0,+\infty)$, then $v=\Phi_k(s)$ by Step 4, and either $s\geq d_k$, so that $T_k^*(v)=s$, or $s<d_k$ and $v=\overline{v}_k$. If $T_k(v)=0$, then either $d_k>0$, and such types form a null set because $\Phi_k(0)=\overline{v}_k$, or $d_k=0$, $v\geq\Phi_k(0)$ by (\ref{sandwich}), and $T_k^*(v)=d_k=0$. Let $T^*:=(T_k^*)_{k\in I_N}$. Since $T_k^*=T_k$ almost everywhere for every $k$, $T^*$ has the same distributions of stopping times as $(T_k)_{k\in I_N}$, so it is an equilibrium (Section~\ref{model}) with the same $Q_k$, $\Phi_k$, $d_k$, $u_k$ and $S_k$. I now show that $T^*$ satisfies (i)--(vi): clauses (i)--(v) in this step and clause (vi) in Step 12. Clause (i) holds by construction. Clause (ii) holds because $\Phi_k(t)>c_k$ for every $t$ and $\Phi_k(t)\to c_k$ as $t\to+\infty$ (Steps 2 and 9). Clause (iii) is shown in Step 9. Clause (iv) is Lemma~\ref{no atom}, because an instant-exit player is a player with an atom at $0$. Clause (v) holds by construction and Lemma~\ref{no gap}: $T_k^*$ is nonincreasing, equal to $d_k$ on $[u_k,\overline{v}_k]$, and continuous and strictly decreasing on $(c_k,u_k]$, where its inverse is $\Phi_k$ on $[d_k,+\infty)$. Moreover, a profile that satisfies (i) and (v) is determined by its curves: they fix $d_k$ and $u_k$ (Section~\ref{model}), and its strategy for player $k$ equals $+\infty$ on $[\underline{v}_k,c_k]$ and $d_k$ on $[u_k,\overline{v}_k]$, and maps each $v\in(c_k,u_k]$ to the unique $t\geq d_k$ with $\Phi_k(t)=v$. Since profiles that agree almost everywhere have the same curves, no profile other than $T^*$ that satisfies (i) and (v) agrees with the equilibrium almost everywhere. Finally, for every $t>0$, player $k$ is active at $t$ if and only if $t\geq d_k$, that is, if and only if $t\in S_k$; so $A(t)=\mathcal{S}(t)$ for every $t>0$.

\emph{Step 12 (clause (vi)).} Let $(d,d')$ be the interior of a stage. There $A(t)=\mathcal{S}(t)$ (Step 11), so, at almost every $t\in(d,d')$, Step~7 and (\ref{rate identity}) give $\eta_k(t)=h_{A(t)}(\Phi(t))-W_k(\Phi_k(t))$ for every $k\in A(t)$, which is (\ref{main equations}) because $\eta_k=-\frac{d}{dt}\ln F_k(\Phi_k)$.

The equation holds at every $t\in(d,d')$. Let $x:=(x_j)_{j\in A(d)}$, and let $G$ be the vector field of Lemma~\ref{log survival} for the set $A(d)$, which equals $A(t)$ for every $t\in(d,d')$. So $x'=-(\eta_j)_{j\in A(d)}=G(x)$ almost everywhere on $(d,d')$. Each $\Phi_k$ with $k\in A(d)$ is continuous on $[d_k,+\infty)\supseteq(d,d')$ (Lemma~\ref{no gap}), so $t\mapsto G(x(t))$ is continuous on $(d,d')$. Since each $x_j$ is Lipschitz continuous on $(0,+\infty)$ (Step 6), $x(t)=x(s)+\int_s^tG(x(u))du$ for all $s,t\in(d,d')$. So $x$ is continuously differentiable on $(d,d')$ and solves $x'=G(x)$ there, and by Lemma~\ref{log survival}, the curves $(\Phi_j)_{j\in A(d)}$ solve (\ref{main equations}) at every $t\in(d,d')$. This completes the proof of the necessity: $T^*$ satisfies (i)--(vi), the equilibrium agrees with it except on sets of types of probability zero, and by Step 11 no other profile that satisfies (i)--(vi) does, so $T^*$ is the canonical representative of the equilibrium.

Now, I prove the sufficiency. By the identification of profiles in Section~\ref{model}, it suffices to show that, in a profile satisfying (i)--(vi), $T_i(v_i)$ maximizes $R_i(\cdot|v_i)$ over $[0,+\infty]$ for almost every type $v_i$ of every player $i$. By Lemma~\ref{payoff derivative} and (\ref{payoff slope}), at every $t>0$ other than an activation time, $dR_i(t|v_i)/dt$ has the sign of $h_{-i}(t)-W_i(v_i)$. This difference equals $\frac{r_i}{c_i}(\Phi_i(t)-v_i)$ for $t>d_i$, by (\ref{inter equations}). For $t<d_i$, it equals $h(t)-W_i(v_i)$, which exceeds $W_i(\overline{v}_i)-W_i(v_i)\geq0$ by Lemma~\ref{hazard lemma}(c). Since $\Phi_i$ is strictly decreasing on $[d_i,+\infty)$ and $\Phi_i(T_i(v_i))=v_i$ for every $v_i\in(c_i,\overline{v}_i]$ with $T_i(v_i)>0$, $dR_i(t|v_i)/dt\gtrless0$ when $t\lessgtr T_i(v_i)$.

It remains to compare with $t=0$ and $t=+\infty$, type by type. By (iv), $\mu_i(t)=Q_{-i}(t)$ for $t>0$, so by (\ref{payoff split}), $R_i(+\infty|v_i)=\lim_{t\to+\infty}R_i(t|v_i)$ and $R_i(0|v_i)=v_i-c_i\mu_i(0)\leq v_i-c_iQ_{-i}(0)=\lim_{t\downarrow0}R_i(t|v_i)$, with equality if no opponent of $i$ has an atom at $0$. By (i), (iv) and (v), each $Q_j=F_j(\Phi_j)$ is positive and continuous on $(0,+\infty)$, and it is locally absolutely continuous there, because $\Phi_j$ is continuous and, by (vi), continuously differentiable on the closure of each stage, with one-sided derivatives at activation times (Lemma~\ref{one-sided derivatives}). So $R_i(\cdot|v_i)$ is locally absolutely continuous on $(0,+\infty)$ by Lemma~\ref{payoff derivative}, and the signs above make it strictly monotone between the relevant times. If $v_i\leq c_i$, $h_{-i}(t)-W_i(v_i)$ is positive at every $t>0$ other than an activation time: for $t>d_i$ because $\Phi_i(t)>c_i\geq v_i$, and for $t<d_i$ by Lemma~\ref{hazard lemma}(c); so $R_i(\cdot|v_i)$ increases strictly on $(0,+\infty)$, and $T_i(v_i)=+\infty$ is optimal. If $v_i>c_i$ and $T_i(v_i)>0$, which covers every $v_i\in(c_i,u_i)$ and, when $d_i>0$, also $u_i=\overline{v}_i$, then $R_i(\cdot|v_i)$ increases strictly on $(0,T_i(v_i)]$ and decreases strictly on $[T_i(v_i),+\infty)$, so $T_i(v_i)$ is optimal. If $T_i(v_i)=0$, then $d_i=0$, $v_i\geq u_i=\Phi_i(0)$, and $R_i(\cdot|v_i)$ decreases strictly on $(0,+\infty)$, so $t=0$ is optimal if no opponent of $i$ has an atom at $0$. Otherwise, by (iv), an opponent of $i$ is the instant-exit player, and player $i$ has no atom at $0$, so $u_i=\Phi_i(0)=\overline{v}_i$ and the only type of player $i$ that provides at $0$ is $\overline{v}_i$. Under cost sharing, $R_i(0|\overline{v}_i)$ falls short of $\lim_{t\downarrow0}R_i(t|\overline{v}_i)$, which no time attains, but this is a single type. Thus, $T_i(\cdot)$ gives the globally optimal choice for almost every type when the rival strategy profile is fixed.
\end{proof}
\subsection{Proof of Lemma~\texorpdfstring{\ref{asymp property}}{2}}\label{proof asymp property}

The proof uses the following lemma, which holds for every set of at least two players and is used again in the proof of Theorem~\ref{exi and uni}.

\begin{alemma}[Backward monotonicity]\label{backward}
	Let $I\subseteq I_N$ with $M:=|I|\geq2$.
	\begin{enumerate}[(a)]
		\item $c_I$ is the only rest point $\gamma\gg\underline{v}_I$ of (\ref{system on I}).
		\item Backward monotonicity: if a solution $\gamma$ of (\ref{system on I}) on an interval that contains a time $b$ has $\gamma'(b)\ll0$, then $\gamma'(t)\ll0$ at every $t\leq b$ at which $\gamma$ is defined.
		\item Just touch: if, in addition, $c_I\ll\gamma(b)\ll\overline{v}_I$, then there is exactly one time $a<b$ in the interval of the maximal solution through $(b,\gamma(b))$, which I also denote by $\gamma$, such that $\gamma(t)\ll\overline{v}_I$ for every $t\in(a,b]$ and the curves \emph{just touch} their upper bounds at $a$: $\gamma(a)\leq\overline{v}_I$, with equality in at least one component. I call $a$ the \emph{just-touch time} of $\gamma$ from $b$.
	\end{enumerate}
\end{alemma}

\begin{proof}
By Lemma~\ref{log survival}, every solution is the restriction of a unique maximal solution, so in (b) it suffices to consider the maximal solution through $(b,\gamma(b))$. In (b) and (c), $\gamma$ denotes this maximal solution, and $(t^-,t^+)$ is its interval.

(a) Since $H_j>0$ on $(\underline{v}_j,+\infty)$, a point $\gamma\gg\underline{v}_I$ is a rest point if and only if $W_j(\gamma_j)=h_I(\gamma)$ for every $j\in I$, that is, $\mathcal{A}_Mw=0$ for $w:=(W_k(\gamma_k))_{k\in I}$, where $\mathcal{A}_M:=(M-1)\,\mathrm{Id}-\mathbf{1}\mathbf{1}^{\top}$, with $\mathbf{1}$ the vector of ones, has diagonal entries $M-2$ and off-diagonal entries $-1$. The matrix $\mathcal{A}_M$ has the eigenvalue $-1$ with the eigenvector $\mathbf{1}$ and the eigenvalue $M-1$ on the orthogonal complement of $\mathbf{1}$, so $\det(\mathcal{A}_M)=-(M-1)^{M-1}\neq0$, and the only solution is $w=0$, that is, $\gamma=c_I$. Conversely, $c_I$ is a rest point.

(b) Let $x:=(\ln F_j(\gamma_j))_{j\in I}$, which solves $x'=G(x)$ by Lemma~\ref{log survival}, and let $y:=x'$, so that $y_j=W_j(\gamma_j)-h_I(\gamma)$ and $\gamma_j'=H_j(\gamma_j)y_j$ has the sign of $y_j$. Since $G$ is continuously differentiable, $y=G(x)$ is continuously differentiable and solves the linear equation $y'(t)=DG(x(t))y(t)$, whose coefficients $\frac{\partial G_j}{\partial x_k}(x(t))$ are continuous in $t$ and negative for $k\neq j$. Fix $t_0\in(t^-,b)$ and let $q(s):=-y(b-s)$ for $s\in[0,b-t_0]$. Then $q'(s)=\Theta(s)q(s)$ with $\Theta(s):=-DG(x(b-s))$, whose off-diagonal entries are positive, and $q(0)=-y(b)\gg0$ because $\gamma'(b)\ll0$. I claim that $q\gg0$ on $[0,b-t_0]$. Otherwise, let $s_0$ be the first time at which a component of $q$ vanishes, and $j$ such a component; then $s_0>0$, $q\geq0$ on $[0,s_0]$, and since the off-diagonal entries of $\Theta$ are positive, $q_j'\geq\theta_{jj}q_j$ there, where $\theta_{jj}$ is the $j$th diagonal entry of $\Theta$, so $s\mapsto q_j(s)e^{-\int_0^s\theta_{jj}}$ is nondecreasing and $0=q_j(s_0)e^{-\int_0^{s_0}\theta_{jj}}\geq q_j(0)>0$, a contradiction. So $y\ll0$ and $\gamma'\ll0$ on $[t_0,b]$. As $t_0$ is arbitrary, (b) follows.

(c) Let $\mathcal{T}:=\{t\in(t^-,b]:\gamma_j(t)\geq\overline{v}_j\text{ for some }j\in I\}$. By (b), every $\gamma_j$ is strictly decreasing on $(t^-,b]$, so $\gamma_j(t)\geq\gamma_j(b)>c_j$ there. Suppose that $\mathcal{T}=\emptyset$. Then $\gamma(t)$ lies in the compact set $\times_{j\in I}[\gamma_j(b),\overline{v}_j]\subseteq\times_{j\in I}(\underline{v}_j,+\infty)$ for every $t\in(t^-,b]$, so $t^-=-\infty$ by Lemma~\ref{log survival}, and each bounded monotone $\gamma_j$ converges to some $\gamma_j^*\in[\gamma_j(b),\overline{v}_j]$ as $t\to-\infty$. The right side of (\ref{system on I}) is continuous in $\gamma$, so $\gamma(t)-\gamma(t-1)$, its integral over $[t-1,t]$, tends both to $0$ and to its value at $\gamma^*$ as $t\to-\infty$. Hence $\gamma^*\gg\underline{v}_I$ is a rest point, and $\gamma^*=c_I$ by (a), which contradicts $\gamma_j^*\geq\gamma_j(b)>c_j$. So $\mathcal{T}\neq\emptyset$. Let $a:=\sup\mathcal{T}$. As $\mathcal{T}$ is closed in $(t^-,b]$, $a\in\mathcal{T}$, and $a<b$ because $\gamma(b)\ll\overline{v}_I$. For $t\in(a,b]$, $t\notin\mathcal{T}$, so $\gamma(t)\ll\overline{v}_I$; by continuity $\gamma(a)\leq\overline{v}_I$, with equality in some component because $a\in\mathcal{T}$. If two times $a'<a''$ both had these properties, then $a''\in(a',b]$ would give $\gamma(a'')\ll\overline{v}_I$, which contradicts the equality at $a''$.
\end{proof}

\begin{proof}[Proof of Lemma~\ref{asymp property}]
Throughout, (\ref{all-active}) is the system on $I_N$ with the extended $F_i$ and $f_i$. I read Lemma~\ref{asymp property} for solutions with values in $\times_{i\in I_N}(\underline{v}_i,\overline{v}_i]$, where $F_i$ and $f_i$ are originally defined. Let $G$ be the vector field of Lemma~\ref{log survival} for $I=I_N$, and let $x^*:=(\ln F_i(c_i))_{i\in I_N}$ be the log-survival coordinates of $c$, which are finite because $F_i(c_i)>0$ by the stubborn-type assumption. By Lemma~\ref{backward}(a), $x^*$ is the only rest point of $x'=G(x)$. I show that $c$ is a saddle point (Step~1), that at most one orbit converges to $c$ (Step~2), and that this orbit falls in every coordinate and just touches the upper bounds at a finite time (Step~3); Step~4 derives (i)--(iii).

\emph{Step 1: $c$ is a saddle point.} For each $k$, let $\varpi_k:=\frac{r_k}{c_k}H_k(c_k)$, which is positive because $F_k(c_k)>0$, and let $\mathcal{R}:=\operatorname{diag}(r_1/c_1,\dots,r_N/c_N)$ and $\mathcal{H}:=\operatorname{diag}(H_1(c_1),\dots,H_N(c_N))$, so that $\mathcal{R}\mathcal{H}=\operatorname{diag}(\varpi_1,\dots,\varpi_N)$. By Lemma~\ref{log survival}, $DG(x^*)=\frac{1}{N-1}\mathcal{A}_N\mathcal{R}\mathcal{H}$, with $\mathcal{A}_N$ as in the proof of Lemma~\ref{backward}(a). This matrix is similar to the symmetric matrix $\frac{1}{N-1}(\mathcal{R}\mathcal{H})^{1/2}\mathcal{A}_N(\mathcal{R}\mathcal{H})^{1/2}$, which is congruent to $\frac{1}{N-1}\mathcal{A}_N$. Being similar to a symmetric matrix, $DG(x^*)$ is diagonalizable with real eigenvalues, and by Sylvester's law of inertia exactly one of them is negative, $\nu<0$, while the other $N-1$ are positive, as for $\mathcal{A}_N$. So $x^*$ is a hyperbolic rest point with one stable direction and $N-1$ unstable ones. If $\zeta$ is an eigenvector for $\nu$, then $(\varpi_j-\nu)\zeta_j=\frac{1}{N-1}\sum_{k\in I_N}\varpi_k\zeta_k$ for every $j$; the right side does not depend on $j$, and it is nonzero, since $\varpi_j-\nu>0$ and otherwise $\zeta=0$. So I can normalize $\zeta_j:=1/(\varpi_j-\nu)>0$. In the coordinates $\Phi$, the right side of (\ref{all-active}) is $g(\gamma):=\big(H_j(\gamma_j)[W_j(\gamma_j)-h_{I_N}(\gamma)]\big)_{j\in I_N}$. Each bracket is affine in $\gamma$ and vanishes at $c$, and each $H_j$ is continuous, so $g(c+\epsilon)=\frac{1}{N-1}\mathcal{H}\mathcal{A}_N\mathcal{R}\,\epsilon+o(|\epsilon|)$. Hence $g$ is differentiable at $c$, and $Dg(c)=\mathcal{H}\,DG(x^*)\,\mathcal{H}^{-1}$ has the eigenvalues of $DG(x^*)$, with the stable eigenvector $\mathcal{H}\zeta\gg0$. This proves the first sentence of (iii).

\emph{Step 2: the orbits that tend to $c$.} Let $E^u$ be the span of the eigenvectors of $DG(x^*)$ for its positive eigenvalues, so that $\mathbb{R}^N=\operatorname{span}\{\zeta\}\oplus E^u$, and for $p\in\mathbb{R}^N$ let $\sigma(p)$ be the coefficient of $\zeta$ when $p-x^*$ is decomposed along this sum; $\sigma$ is continuous. Since $G$ is continuously differentiable and $x^*$ is hyperbolic, the stable manifold theorem, in the form of \citet[Theorems 9.4 and 9.5]{teschl2012ordinary}, gives a neighborhood $U$ of $x^*$, numbers $s_-<0<s_+$ and a continuously differentiable function $\vartheta:(s_-,s_+)\to E^u$ with $\vartheta(0)=0$ and $\vartheta'(0)=0$, such that the set $\mathcal{M}$ of points $p\in U$ from which the solution of $x'=G(x)$ is defined and stays in $U$ for all $t\geq0$ is $\{x^*+s\zeta+\vartheta(s):s_-<s<s_+\}$, and the solution from every $p\in\mathcal{M}$ tends to $x^*$. By the definition of $\mathcal{M}$, the solution from a point of $\mathcal{M}$ stays in $\mathcal{M}$, and a solution that tends to $x^*$ lies in $U$, hence in $\mathcal{M}$, from some time on. Every $p\in\mathcal{M}$ equals $x^*+\sigma(p)\zeta+\vartheta(\sigma(p))$, so $\sigma$ is one to one on $\mathcal{M}$ and vanishes there only at $x^*$. Since $\zeta\gg0$ and $\vartheta(s)/s\to0$ as $s\to0$, there is $s_1\in(0,\min\{-s_-,s_+\}]$ such that every $p\in\mathcal{M}$ satisfies $p\gg x^*$ if $0<\sigma(p)<s_1$, and $p\ll x^*$ if $-s_1<\sigma(p)<0$.

Let $p\in\mathcal{M}$ with $\sigma(p)>0$, and let $x$ be the solution with $x(0)=p$. For $t\geq0$, $x(t)\in\mathcal{M}$, and $x(t)\neq x^*$ because, by uniqueness, only the constant solution reaches the rest point. So $t\mapsto\sigma(x(t))$ is continuous and positive on $[0,+\infty)$ and tends to $0$, and by the intermediate value theorem the solution passes through every $q\in\mathcal{M}$ with $0<\sigma(q)\leq\sigma(p)$. Hence $\mathcal{M}^+:=\{p\in\mathcal{M}:\sigma(p)>0\}$ lies on one orbit. Let $\Gamma$ be the maximal solution of (\ref{all-active}) whose log-survival coordinates pass through $\mathcal{M}^+$; its time origin is fixed in Step~3. Now let $\Phi$ be a solution of (\ref{all-active}) that converges to $c$, with log-survival coordinates $x$. There is $t_1$ such that $x(t)\in\mathcal{M}$ and $|\sigma(x(t))|<s_1$ for all $t\geq t_1$. Since $\Phi(t_1)\gg c$, $x(t_1)\gg x^*$, so $\sigma(x(t_1))>0$ by the choice of $s_1$, and $\Phi$ is the restriction of a time shift of $\Gamma$.

\emph{Step 3: monotonicity and just touch.} Let $x$ be the log-survival coordinates of $\Gamma$ and $s(t):=\sigma(x(t))$, which is positive for large $t$ and tends to $0$. Since $G$ is differentiable at $x^*$, $G(x^*)=0$, $DG(x^*)\zeta=\nu\zeta$ and $\vartheta(s)=o(s)$,
\begin{equation*}
x'(t)=G(x(t))=DG(x^*)\big(s(t)\zeta+\vartheta(s(t))\big)+o(s(t))=s(t)\big(\nu\zeta+o(1)\big)\qquad\text{as }t\to+\infty.
\end{equation*}
As $\nu\zeta\ll0$, $x'(t)\ll0$, and hence $\Gamma'(t)\ll0$, for all large $t$: near $c$, the curves of all players fall together along the stable direction. Lemma~\ref{backward}(b), applied at such a time, gives
\begin{equation}\label{Gamma decreasing}
\Gamma_j'(t)<0\qquad\text{for every }j\in I_N\text{ and every }t\text{ at which }\Gamma\text{ is defined}.
\end{equation}
So every $\Gamma_j$ is strictly decreasing, and since $\Gamma(t)\to c$, $\Gamma(t)\gg c$ at every $t$.

Since $\Gamma(t)\to c\ll\overline{v}_{I_N}$, there is $b$ with $c\ll\Gamma(b)\ll\overline{v}_{I_N}$, and $\Gamma'(b)\ll0$ by (\ref{Gamma decreasing}). Lemma~\ref{backward}(c) gives the just-touch time $a<b$ of $\Gamma$ from $b$, and $\Gamma(t)\leq\Gamma(b)\ll\overline{v}_{I_N}$ for $t\geq b$. I set the time origin of $\Gamma$ at $a$. Then $\Gamma$ is defined on an open interval $(t^-,+\infty)$ with $t^-<0$, $\Gamma(t)\ll\overline{v}_{I_N}$ for every $t>0$, and $\Gamma(0)\leq\overline{v}_{I_N}$ with $\Gamma_i(0)=\overline{v}_i$ for at least one $i$; for every $t\in(t^-,0)$, this $i$ has $\Gamma_i(t)>\overline{v}_i$. The saddle path of the lemma is the restriction of $\Gamma$ to $[0,+\infty)$; below, $\Gamma$ denotes the maximal solution.

\emph{Step 4: (i)--(iii).} Part (ii) is (\ref{Gamma decreasing}). For (i), $\Gamma$ converges to $c$ by Step~3. Let $\Phi$ be a solution of (\ref{all-active}) on an interval $J$ that converges to $c$. By Step~2, there is $\tau$ with $\Phi(t)=\Gamma(t-\tau)$ for every $t\in J$. As $\Phi(t)\leq\overline{v}_{I_N}$, while $\Gamma(t')$ has a component above its upper bound for every $t'<0$, $t-\tau\geq0$ for every $t\in J$, so $\Phi$ is a time shift of the saddle path. For (iii), the first sentence was proved in Step~1. Let $m\in\times_{i\in I_N}[c_i,\overline{v}_i]$, and let $\Phi$ be the maximal solution of (\ref{all-active}) with $\Phi(0)=m$. If $m=\Gamma(t_0)$ for some $t_0\geq0$, then $\Phi(t)=\Gamma(t_0+t)$ for $t\geq0$, which converges to $c$. Conversely, if $\Phi$ converges to $c$, then by Step~2 there is $\tau$ with $\Phi(t)=\Gamma(t-\tau)$ for $t\geq0$, and $m=\Gamma(-\tau)\leq\overline{v}_{I_N}$ forces $-\tau\geq0$. This proves (iii).
\end{proof}

\paragraph{Two players.} With $N=2$, the orbit of the saddle path can be written explicitly. For $i\in\{1,2\}$ and $v\in[c_i,\overline{v}_i]$, let
\begin{equation*}
\Psi_i(v):=\int_{c_i}^{v}W_i(y)\frac{f_i(y)}{F_i(y)}\,dy,
\end{equation*}
which is continuous and strictly increasing, with $\Psi_i(c_i)=0$. With two players, (\ref{all-active}) reads
\begin{equation*}
\Phi_1'=-H_1(\Phi_1)W_2(\Phi_2),\qquad\Phi_2'=-H_2(\Phi_2)W_1(\Phi_1),
\end{equation*}
so along every solution with values in $\times_{i}[c_i,\overline{v}_i]$,
\begin{equation*}
\frac{d}{dt}\Psi_1(\Phi_1(t))=-W_1(\Phi_1(t))W_2(\Phi_2(t))=\frac{d}{dt}\Psi_2(\Phi_2(t)).
\end{equation*}
Along the saddle path, both $\Psi_i(\Gamma_i(t))$ tend to $0$, so $\Psi_1(\Gamma_1(t))=\Psi_2(\Gamma_2(t))$ for every $t\geq0$. The saddle path thus moves toward $c$ along the locus of points $\gamma$ with $\Psi_1(\gamma_1)=\Psi_2(\gamma_2)$, so that $\Gamma_1'=-H_1(\Gamma_1)W_2\big(\Psi_2^{-1}(\Psi_1(\Gamma_1))\big)$, and it starts where this locus meets the upper bounds:
\begin{equation*}
\Gamma(0)=\begin{cases}\big(\overline{v}_1,\Psi_2^{-1}(\Psi_1(\overline{v}_1))\big)&\text{if }\Psi_1(\overline{v}_1)\leq\Psi_2(\overline{v}_2),\\
\big(\Psi_1^{-1}(\Psi_2(\overline{v}_2)),\overline{v}_2\big)&\text{otherwise.}\end{cases}
\end{equation*}
By Lemma~\ref{Characterization}(iii), both players are active from time $0$, so $\overline{d}=0$, and by (\ref{last stage}) in the proof of Theorem~\ref{exi and uni}, the equilibrium curves are $\Gamma$. As noted in Section~\ref{existence and uniqueness}, this is the path that \citet[Lemma~11(i)]{ponsati1995war} select; they do not write the equation $\Psi_1(\Gamma_1)=\Psi_2(\Gamma_2)$.

\subsection{Proof of Theorem~\texorpdfstring{\ref{exi and uni}}{1}}\label{proof exi and uni}

\begin{proof}

\emph{Reduction.} By Lemma~\ref{Characterization} and Steps~11 and~12 of its proof, every equilibrium agrees almost everywhere with exactly one profile that satisfies (i)--(vi), its canonical representative, and every such profile is an equilibrium; by Step~11, a profile that satisfies (i) and (v) is determined by its curves. It therefore suffices to construct a profile that satisfies (i)--(vi), which gives existence, and to show that every profile that satisfies (i)--(vi) has the curves of the constructed one, which gives essential uniqueness.

\emph{Curves of a profile that satisfies (i)--(vi).} Fix such a profile, with curves $\Phi$. By (i), (ii) and (v), each $\Phi_i$ equals $\overline{v}_i$ before $d_i$ and is continuous and strictly decreasing on $[d_i,+\infty)$, with $\Phi_i>c_i$ and $\Phi_i(t)\to c_i$ as $t\to+\infty$. By (iv) and (v), $\Phi_i(d_i)=u_i=\overline{v}_i$ if $d_i>0$ (no atom at $d_i$), and $\Phi_i(0)<\overline{v}_i$ for at most one player. Let $0=D_0<D_1<\dots<D_K=\overline{d}$ be the activation times, indexed from the left ($0$ is one of them by (iii)), and let $D_{K+1}:=+\infty$; on $[D_k,D_{k+1})$, the set of active players is $A(D_k)$. Two facts hold for every $k$. (F1) The set $A(D_k)\supseteq A(0)$ has at least two players by (iii), and its curves are continuous on $[D_k,D_{k+1}]\cap\mathbb{R}$ and, by (vi), solve the system on $A(D_k)$ on $(D_k,D_{k+1})$, so by Lemma~\ref{log survival} they coincide on $[D_k,D_{k+1}]\cap\mathbb{R}$ with the maximal solution through any of their values. (F2) At $D_k$, these curves just touch their upper bounds: $\Phi_j(t)<\overline{v}_j$ for every $j\in A(D_k)$ and $t>D_k$, and $\Phi_j(D_k)\leq\overline{v}_j$ for every $j\in A(D_k)$, with equality exactly for the players with $d_j=D_k$ when $k\geq1$, and for all players of $A(0)$ other than the instant-exit player (if any) when $k=0$. Equality holds for at least one player: for $k\geq1$ by the definition of $D_k$, and for $k=0$ because $A(0)$ has at least two players by (iii), of whom at most one exits instantly by (iv).

\emph{Step 1: the last stage follows the saddle path.} On $[\overline{d},+\infty)$ every player is active, so by (F1) the curves solve (\ref{all-active}) there, and they lie above $c$ and converge to $c$. Since $\Phi(\overline{d}+\cdot)$ solves (\ref{all-active}) on $[0,+\infty)$ and converges to $c$, Lemma~\ref{asymp property}(iii), applied to $m:=\Phi(\overline{d})\in\times_{i\in I_N}(c_i,\overline{v}_i]$, gives $s_0\geq0$ with $m=\Gamma(s_0)$; as the system does not depend on time, uniqueness of solutions (Lemma~\ref{log survival}) gives $\Phi(t)=\Gamma(t-\tau)$ and $t-\tau\geq0$ for every $t\geq\overline{d}$, with $\tau:=\overline{d}-s_0$. By (F2) with $k=K$, $\Phi(\overline{d})=\Gamma(\overline{d}-\tau)$ has a component equal to its upper bound, whereas $\Gamma(s)\ll\overline{v}_{I_N}$ for every $s>0$, because $\Gamma$ is strictly decreasing (Lemma~\ref{asymp property}(ii)) and $\Gamma(0)\leq\overline{v}_{I_N}$. Hence $\tau=\overline{d}$:
\begin{equation}\label{last stage}
\Phi(t)=\Gamma(t-\overline{d})\qquad\text{for every }t\geq\overline{d}.
\end{equation}

\emph{Step 2: the backward construction.} Let $a_1:=0$, $A_1:=I_N$ and $\gamma^{(1)}:=\Gamma$, the maximal solution of Step~3 of the proof of Lemma~\ref{asymp property}. Suppose that, for some $n\geq1$, a time $a_n$, a set $A_n\subseteq I_N$ with $|A_n|\geq2$ and a solution $\gamma^{(n)}$ of the system on $A_n$ are given such that $\gamma^{(n)}$ is defined on an open interval that contains $a_n$, $(\gamma^{(n)})'(a_n)\ll0$, and $c_{A_n}\ll\gamma^{(n)}(a_n)\leq\overline{v}_{A_n}$ with equality in at least one component. By that Step~3 and (\ref{Gamma decreasing}), this holds for $n=1$. Let $\Delta_n:=\{j\in A_n:\gamma^{(n)}_j(a_n)=\overline{v}_j\}$, which is nonempty, and $A_{n+1}:=A_n\setminus\Delta_n$. If $|A_{n+1}|\leq1$, the construction stops, and $n^*:=n$. Otherwise, let $\gamma^{(n+1)}$ be the maximal solution of the system on $A_{n+1}$ with $\gamma^{(n+1)}(a_n)=\gamma^{(n)}_{A_{n+1}}(a_n)$ (Lemma~\ref{log survival}). Write $M:=|A_n|$, $M':=|A_{n+1}|$ and $w_k:=W_k(\gamma^{(n)}_k(a_n))$ for $k\in A_n$. Since $A_n$ is the disjoint union of $A_{n+1}$ and $\Delta_n$ and $\gamma^{(n+1)}(a_n)=\gamma^{(n)}_{A_{n+1}}(a_n)$, $(M-1)h_{A_n}(\gamma^{(n)}(a_n))=(M'-1)h_{A_{n+1}}(\gamma^{(n+1)}(a_n))+\sum_{k\in\Delta_n}w_k$, and (\ref{system on I}) for the two systems gives, for every $i\in A_{n+1}$,
\begin{equation}\label{activation jump}
\frac{(\gamma^{(n+1)}_i)'(a_n)-(\gamma^{(n)}_i)'(a_n)}{H_i(\gamma^{(n)}_i(a_n))}=h_{A_n}(\gamma^{(n)}(a_n))-h_{A_{n+1}}(\gamma^{(n+1)}(a_n))=\frac{1}{M'-1}\sum_{k\in\Delta_n}\frac{(\gamma^{(n)}_k)'(a_n)}{H_k(\overline{v}_k)}<0,
\end{equation}
where the last equality uses this identity, $|\Delta_n|=M-M'$, and $w_k-h_{A_n}(\gamma^{(n)}(a_n))=(\gamma^{(n)}_k)'(a_n)/H_k(\overline{v}_k)$ for $k\in\Delta_n$, which is (\ref{system on I}) for $\gamma^{(n)}$ with $\gamma^{(n)}_k(a_n)=\overline{v}_k$, and the sign comes from $(\gamma^{(n)})'(a_n)\ll0$. Hence $(\gamma^{(n+1)}_i)'(a_n)<(\gamma^{(n)}_i)'(a_n)<0$ for every $i\in A_{n+1}$.

By Lemma~\ref{backward}(b), $(\gamma^{(n+1)})'(t)\ll0$ at every $t\leq a_n$ at which $\gamma^{(n+1)}$ is defined. Since $c_j<\gamma^{(n+1)}_j(a_n)<\overline{v}_j$ for every $j\in A_{n+1}$, Lemma~\ref{backward}(c) gives the just-touch time $a_{n+1}<a_n$ of $\gamma^{(n+1)}$ from $a_n$. This time lies in the open interval of $\gamma^{(n+1)}$, $(\gamma^{(n+1)})'(a_{n+1})\ll0$, and $c_{A_{n+1}}\ll\gamma^{(n+1)}(a_n)\leq\gamma^{(n+1)}(a_{n+1})\leq\overline{v}_{A_{n+1}}$ with equality in at least one component, so the hypothesis holds for $n+1$. As $|A_{n+1}|<|A_n|$ and $|A_1|=N$, the construction stops at some $n^*\leq N-1$. The construction depends only on the primitives of the game.

\emph{Step 3: uniqueness.} Let a profile satisfy (i)--(vi), with curves $\Phi$ and activation times $D_0<\dots<D_K$ as above. I show by induction on $n=1,\dots,K+1$ that the construction does not stop before step $n$ and that
\begin{equation}\label{identification}
\begin{gathered}
D_{K+1-n}=a_n+\overline{d},\qquad A(D_{K+1-n})=A_n,\qquad\text{and}\\
\Phi_j(t)=\gamma^{(n)}_j(t-\overline{d})\quad\text{for every }j\in A_n\text{ and }t\in[D_{K+1-n},D_{K+2-n}]\cap\mathbb{R}.
\end{gathered}
\end{equation}
For $n=1$, (\ref{identification}) is (\ref{last stage}), since $D_K=\overline{d}$, $a_1=0$ and $A(\overline{d})=I_N$. Suppose that it holds for some $n\leq K$, and write $D:=D_{K+1-n}>0$ and $D':=D_{K-n}$. By (\ref{identification}), $\gamma^{(n)}(a_n)=\Phi_{A(D)}(D)$, so by (F2) $\Delta_n$ is the set of players with $d_j=D$, that is, $A(D)\setminus A(D')$. Hence $A_{n+1}=A(D')$, which has at least two players by (iii), and the construction does not stop at step $n$. By (F1), the curves of $A(D')$ coincide on $[D',D]$ with the maximal solution of the system on $A(D')$ through their value at $D$, which is $\gamma^{(n+1)}(a_n)$; as the system does not depend on time, $\Phi_j(t)=\gamma^{(n+1)}_j(t-\overline{d})$ for $j\in A_{n+1}$ and $t\in[D',D]$. By (F2), $D'-\overline{d}$ then has the properties of the just-touch time of $\gamma^{(n+1)}$ from $a_n$, which is unique by Lemma~\ref{backward}(c), so $D'=a_{n+1}+\overline{d}$. This completes the induction.

With $n=K+1$, (\ref{identification}) gives $0=D_0=a_{K+1}+\overline{d}$ and $A_{K+1}=A(0)$. So $A_{K+2}$ consists of the players of $A(0)$ with $\Phi_j(0)<\overline{v}_j$, of whom there is at most one by (iv), and the construction stops at $n^*=K+1$. Hence $K=n^*-1$, $\overline{d}=-a_{n^*}$, $D_k=a_{n^*-k}-a_{n^*}$ and $A(D_k)=A_{n^*-k}$ for every $k$, and $\Phi_j(t)=\gamma^{(n)}_j(t+a_{n^*})$ for $j\in A_n$ and $t+a_{n^*}\in[a_n,a_{n-1}]\cap\mathbb{R}$, where $a_0:=+\infty$. Every profile that satisfies (i)--(vi) therefore has the curves that the construction determines, and by the reduction any two equilibria differ only on sets of types of probability zero.

\emph{Step 4: existence.} Let $n^*$ and $a_1>\dots>a_{n^*}$ be given by the construction, with $a_0=+\infty$ as above. For $n\leq n^*$ and $j\in\Delta_n$, let $d_j:=a_n-a_{n^*}$, and for the player in $A_{n^*+1}$, if there is one, let $d_j:=0$. For $n\leq n^*$, $j\in A_n$ and $t\geq0$ with $t+a_{n^*}\in[a_n,a_{n-1}]$, let $\hat{\Phi}_j(t):=\gamma^{(n)}_j(t+a_{n^*})$, and let $\hat{\Phi}_j(t):=\overline{v}_j$ for $t<d_j$. Each player of $\Delta_n$ belongs to $A_1,\dots,A_n$, and the player in $A_{n^*+1}$ to every $A_n$, so every $\hat{\Phi}_j$ is defined on $[0,+\infty)$, and the pieces agree at each $a_n-a_{n^*}$ because $\gamma^{(n+1)}(a_n)=\gamma^{(n)}_{A_{n+1}}(a_n)$. By (\ref{Gamma decreasing}), Step~2 and Lemma~\ref{asymp property}(i), each $\hat{\Phi}_j$ is continuous on $[0,+\infty)$ and strictly decreasing on $[d_j,+\infty)$, with $\hat{\Phi}_j(d_j)=\overline{v}_j$ if $d_j>0$, $\hat{\Phi}_j>c_j$ and $\hat{\Phi}_j(t)\to c_j$. Let $u_j:=\hat{\Phi}_j(d_j)$. Let $T_j(v):=+\infty$ for $v\in[\underline{v}_j,c_j]$, let $T_j(v)$ be the time $t\geq d_j$ with $\hat{\Phi}_j(t)=v$ for $v\in(c_j,u_j]$, and let $T_j(v):=d_j$ for $v\in(u_j,\overline{v}_j]$; $T_j$ is nonincreasing, hence Borel measurable. Then $\{v:T_j(v)>t\}$ is $[\underline{v}_j,\hat{\Phi}_j(t))$ for $t\geq d_j$ and $[\underline{v}_j,\overline{v}_j]$ for $t<d_j$, so the curve of $T_j$ is $\hat{\Phi}_j$, and its activation time and $u_j$ are the $d_j$ and $u_j$ above. Conditions (i), (ii) and (v) follow from the properties of $\hat{\Phi}_j$. The players with $d_j=0$ form $A_{n^*}$, which has at least two members, so (iii) holds. At a time $t>d_j$, player $j$ provides only with the type $\hat{\Phi}_j(t)$, and at $d_j$ with the types in $[u_j,\overline{v}_j]$; every $j\in\Delta_n$ has $u_j=\overline{v}_j$, so only the player in $A_{n^*+1}$, if there is one, has an atom, at time $0$, and (iv) holds. The activation times are the $a_n-a_{n^*}$, $n\leq n^*$, and in the interior of the stage from $a_n-a_{n^*}$ to $a_{n-1}-a_{n^*}$ the set of active players is $\Delta_n\cup\dots\cup\Delta_{n^*}\cup A_{n^*+1}=A_n$, whose curves solve the system on $A_n$ there; this is (vi). By Lemma~\ref{Characterization}, the profile is an equilibrium.

\emph{Consequences.} In this equilibrium, the positive activation times are the $a_n-a_{n^*}$ with $n<n^*$. Fix one, $d:=a_n-a_{n^*}$. On the stage that starts at $d$, the players of $A_n$ are active and their curves are $\gamma^{(n)}(\cdot+a_{n^*})$; on the stage that ends at $d$, only the players of $A_{n+1}$ are active, and their curves are $\gamma^{(n+1)}(\cdot+a_{n^*})$. So, by Lemma~\ref{one-sided derivatives}, $h(d)=h_{A_n}(\gamma^{(n)}(a_n))$ and $\lim_{t\uparrow d}h(t)=h_{A_{n+1}}(\gamma^{(n+1)}(a_n))$, and the middle expression of (\ref{activation jump}) is $h(d)-\lim_{t\uparrow d}h(t)$. Hence, at $d$, the provision hazard jumps down strictly, and by Step~2, the incumbents, the players of $A_{n+1}$, slow down, while the players of $\Delta_n$, who become active, provide at a positive hazard rate at once, because $(\gamma^{(n)})'(a_n)\ll0$. These are the facts that Section~\ref{characterization} states. Moreover, every initially inactive player $i$, that is, every $i\in\Delta_n$ with $n<n^*$, has $d_i=a_n-a_{n^*}$, and (\ref{system on I}) for $\gamma^{(n)}$ at $a_n$ gives
\begin{equation*}
h(d_i)=h_{A_n}(\gamma^{(n)}(a_n))=W_i(\overline{v}_i)-\frac{(\gamma^{(n)}_i)'(a_n)}{H_i(\overline{v}_i)}>W_i(\overline{v}_i).
\end{equation*}
All equilibria have the same curves, so these facts hold in every equilibrium.
\end{proof}
\subsection{Proof of Proposition~\texorpdfstring{\ref{Asy Comp Sta}}{1}}\label{proof Asy Comp Sta}

\begin{proof}
Write $\check{v}:=\min\{\overline{v}_\alpha,\overline{v}_\beta\}$, and recall that $\alpha$ and $\beta$ are active at every $t\geq t_0$. By (\ref{aggregate rate}), with right derivatives at activation times (Lemma~\ref{one-sided derivatives}), the equations for $\alpha$ and $\beta$ on $[t_0,+\infty)$, which share the provision hazard $h(t)$, are:
\begin{equation}\label{asymmetric comp equations}
\frac{d}{dt}\ln F_\alpha(\Phi_\alpha(t))=W_\alpha(\Phi_\alpha(t))-h(t),\qquad
\frac{d}{dt}\ln F_\beta(\Phi_\beta(t))=W_\beta(\Phi_\beta(t))-h(t).
\end{equation}
Each curve $\Phi_j$, $j\in\{\alpha,\beta\}$, is continuous, lies in $(c_j,\overline{v}_j]$, is strictly decreasing on $[t_0,+\infty)$, and converges to $c_j$ (Lemma~\ref{Characterization}(i), (ii) and (v)).

\emph{Comparison.} Parts (i) and (iii) rest on the following comparison. In it and in these parts, $c_\alpha=c_\beta=:c$, a scalar (unlike the vector $c$ of Lemma~\ref{asymp property}). Suppose that $r_\alpha\geq r_\beta$, that $f_\alpha/F_\alpha\geq f_\beta/F_\beta$ on $(c,\check{v})$, and that $\Phi_\alpha(\tau)\geq\Phi_\beta(\tau)$ for some $\tau\geq t_0$. Then
\begin{equation}\label{comparison result}
\ln F_\alpha(\Phi_\beta(\tau))-\ln F_\beta(\Phi_\beta(\tau))\mathrel{\leq}\ln F_\alpha(c)-\ln F_\beta(c),\qquad\text{where }c<\Phi_\beta(\tau)\leq\check{v}.
\end{equation}
To prove (\ref{comparison result}), I first show that $\alpha$'s curve stays weakly above $\beta$'s from $\tau$ on, and then integrate.

First, $\Phi_\alpha(t)\geq\Phi_\beta(t)$ for every $t\geq\tau$. Indeed, for $t\geq\tau$, $c<\Phi_\beta(t)\leq\Phi_\beta(\tau)\leq\min\{\Phi_\alpha(\tau),\overline{v}_\beta\}\leq\check{v}$, and $\Phi_\beta(t)<\check{v}$ for $t>\tau$, because $\Phi_\beta$ is strictly decreasing. So $x(t):=\ln F_\alpha(\Phi_\alpha(t))$ and $z(t):=\ln F_\alpha(\Phi_\beta(t))$ are defined and lie in $[\ln F_\alpha(c),0]$. Both are continuous, and continuously differentiable on the closure of each stage, with one-sided derivatives at activation times (Lemma~\ref{one-sided derivatives}), so they are Lipschitz continuous on bounded intervals. With $\xi_\alpha(y):=F_\alpha^{-1}(e^{y})$ as in Lemma~\ref{log survival}, the function $W_\alpha\circ\xi_\alpha$ has derivative $\frac{r_\alpha}{c}H_\alpha\circ\xi_\alpha$, where $H_\alpha=F_\alpha/f_\alpha$ is positive and continuous on $[c,\overline{v}_\alpha]$. So $W_\alpha\circ\xi_\alpha$ is nondecreasing and Lipschitz continuous on $[\ln F_\alpha(c),0]$. By (\ref{asymmetric comp equations}), $x'=W_\alpha(\xi_\alpha(x))-h$. By (\ref{asymmetric comp equations}) for $\beta$ and the chain rule, $z'=q\,[W_\beta(\Phi_\beta)-h]$, where, for $t>\tau$, $q(t)\geq1$ is the ratio of $\alpha$'s revelation rate to $\beta$'s, both evaluated at $\Phi_\beta(t)$. Hence, for $t>\tau$,
\begin{equation*}
z'\leq W_\beta(\Phi_\beta)-h\leq W_\alpha(\Phi_\beta)-h=W_\alpha(\xi_\alpha(z))-h.
\end{equation*}
The first inequality holds because $q\geq1$ and, since $\Phi_\beta$ is nonincreasing, $W_\beta(\Phi_\beta)-h=\frac{d}{dt}\ln F_\beta(\Phi_\beta)\leq0$. The second holds because $r_\alpha\geq r_\beta$ and $\Phi_\beta>c$. So, at every $t>\tau$ that is not an activation time, $\zeta:=z-x$ satisfies $\zeta'\leq W_\alpha(\xi_\alpha(z))-W_\alpha(\xi_\alpha(x))$, which is at most a constant multiple of $\max\{\zeta,0\}$, because $W_\alpha\circ\xi_\alpha$ is nondecreasing and Lipschitz continuous. The function $\max\{\zeta,0\}$ is Lipschitz continuous on bounded intervals, its derivative is at most the same multiple of $\max\{\zeta,0\}$ almost everywhere, and it vanishes at $\tau$ because $\Phi_\alpha(\tau)\geq\Phi_\beta(\tau)$. By Gronwall's inequality, it vanishes on $[\tau,+\infty)$, that is, $\Phi_\beta\leq\Phi_\alpha$ there.

Second, for the same reason, $\ln F_\alpha(\Phi_\alpha)-\ln F_\beta(\Phi_\beta)$ is Lipschitz continuous on bounded intervals in $[\tau,+\infty)$, and subtracting the two equations in (\ref{asymmetric comp equations}) eliminates $h$: its derivative is $W_\alpha(\Phi_\alpha)-W_\beta(\Phi_\beta)$, and by the first step $W_\alpha(\Phi_\alpha(t))\geq W_\alpha(\Phi_\beta(t))\geq W_\beta(\Phi_\beta(t))$ for every $t\geq\tau$. Integrate over $[\tau,\tau']$ and let $\tau'\to+\infty$. Since $\Phi_j(\tau')\to c$ for $j\in\{\alpha,\beta\}$, monotone convergence shows that the integral below is finite and nonnegative, and since $\Phi_\beta(\tau)\leq\Phi_\alpha(\tau)$,
\begin{equation}\label{integrated comparison}
\begin{aligned}
\ln F_\alpha(\Phi_\beta(\tau))-\ln F_\beta(\Phi_\beta(\tau))&\mathrel{\leq}\ln F_\alpha(\Phi_\alpha(\tau))-\ln F_\beta(\Phi_\beta(\tau))\\
&=\ln F_\alpha(c)-\ln F_\beta(c)-\int_\tau^{+\infty}\big[W_\alpha(\Phi_\alpha(t))-W_\beta(\Phi_\beta(t))\big]\,dt\\
&\mathrel{\leq}\ln F_\alpha(c)-\ln F_\beta(c),
\end{aligned}
\end{equation}
and its first and last lines give (\ref{comparison result}).

(i) Here $r_\alpha=r_\beta$ and $c_\alpha=c_\beta=c$. First, let $f_\alpha/F_\alpha=f_\beta/F_\beta$ on $(c,\check{v})$. Then $W_\alpha=W_\beta$, and $\ln F_\alpha-\ln F_\beta$ is constant on $[c,\check{v}]$. The first two hypotheses of the comparison are symmetric in $\alpha$ and $\beta$. Let $\tau\geq t_0$, and suppose that $\Phi_\alpha(\tau)\geq\Phi_\beta(\tau)$; otherwise exchange the roles of $\alpha$ and $\beta$. Since $c<\Phi_\beta(\tau)\leq\check{v}$, the first and last lines of (\ref{integrated comparison}) are equal, so every inequality in (\ref{integrated comparison}) holds with equality. The first gives $F_\alpha(\Phi_\beta(\tau))=F_\alpha(\Phi_\alpha(\tau))$, so $\Phi_\alpha(\tau)=\Phi_\beta(\tau)$ because $F_\alpha$ is strictly increasing. Hence $\Phi_\alpha(t)=\Phi_\beta(t)$ for all $t\in[t_0,+\infty)$.

Next, consider the case where $f_\alpha(v)/F_\alpha(v)>f_\beta(v)/F_\beta(v)$ for all $v\in(c,\check{v})$. Suppose that $\Phi_\alpha(\tau)\geq\Phi_\beta(\tau)$ for some $\tau\geq t_0$. Then the comparison applies and gives (\ref{comparison result}). But the derivative $f_\alpha/F_\alpha-f_\beta/F_\beta$ of $\ln F_\alpha-\ln F_\beta$ is positive on $(c,\check{v})$, and $\Phi_\beta(\tau)>c$, which together contradict (\ref{comparison result}). Hence $\Phi_\alpha(t)<\Phi_\beta(t)$ for all $t\geq t_0$.

(ii) Let $r:=r_\alpha=r_\beta$, with $c_\alpha>c_\beta$. Since $\Phi_\alpha(t)\to c_\alpha$, $\Phi_\beta(t)\to c_\beta$ and $c_\alpha>c_\beta$, there is $t^*\geq t_0$ such that $\Phi_\alpha(t)>\Phi_\beta(t)$ for all $t>t^*$. If $\Phi_\alpha$ and $\Phi_\beta$ intersect on $[t_0,t^*]$, let $s$ denote the largest such point. Then $\Phi_\alpha>\Phi_\beta$ on $(s,+\infty)$, so the right derivatives at $s$ satisfy $\Phi_\alpha'(s)\geq\Phi_\beta'(s)$. At $s$, however, both curves equal some $v\in(c_\alpha,\check{v}]$, at which the two revelation rates coincide (at $\check{v}$ by continuity), so $h(s)$ cancels and (\ref{asymmetric comp equations}) gives
\begin{equation*}
\Phi_\alpha'(s)-\Phi_\beta'(s)=H_\alpha(v)\big[W_\alpha(v)-W_\beta(v)\big]=H_\alpha(v)\,rv\Big(\frac{1}{c_\alpha}-\frac{1}{c_\beta}\Big)<0,
\end{equation*}
a contradiction. Consequently, there is no intersection point on $[t_0,+\infty)$, and therefore $\Phi_\alpha(t)>\Phi_\beta(t)$ for all $t\geq t_0$.

(iii) Here $c_\alpha=c_\beta=c$ and $r_\alpha>r_\beta$. Suppose that $\Phi_\alpha(\tau)\geq\Phi_\beta(\tau)$ for some $\tau\geq t_0$. Then the comparison applies, and the second inequality in (\ref{integrated comparison}) is strict, because $W_\alpha(\Phi_\alpha(t))\geq W_\alpha(\Phi_\beta(t))>W_\beta(\Phi_\beta(t))$ for every $t\geq\tau$. So (\ref{comparison result}) holds with strict inequality. This strict inequality contradicts the condition that the derivative $f_\alpha/F_\alpha-f_\beta/F_\beta$ of $\ln F_\alpha-\ln F_\beta$ is zero on $(c,\Phi_\beta(\tau))$. Therefore, $\Phi_\alpha(t)<\Phi_\beta(t)$ for all $t\geq t_0$.
\end{proof}

\subsection{Proofs of \texorpdfstring{(\ref{UTD times})}{(8)} and Proposition~\texorpdfstring{\ref{U-LTD expect rate}}{2}}\label{proof U-LTD expect rate}\label{proof UTD times}

The proof uses the stopping times (\ref{UTD times}), whose derivation I complete first. As in (\ref{LTD solution}), the players are labeled so that $\overline{v}_1\geq\overline{v}_2\geq\dots\geq\overline{v}_N$, $d_M$ is player $M$'s activation time, $d_{N+1}=+\infty$, and $W$ is the common indifference level. Let $\overline{v}_{N+1}:=c$.

\begin{proof}[Proof of (\ref{UTD times})]
By (\ref{LTD solution}), players $1,\dots,M$ share one curve on $[d_M,d_{M+1})$, which equals $\overline{v}_M$ at $d_M$. As noted in Section~\ref{ltd war}, this curve needs time $(M-1)\int_w^v\frac{f(u)}{F(u)W(u)}\,du$ to fall from $v$ to $w\in(c,v]$. Suppose that, for some $M\in\{2,\dots,N\}$,
\begin{equation*}
d_M=\sum_{j=2}^{M}\int_{\overline{v}_M}^{\overline{v}_j}\frac{f(u)}{F(u)W(u)}\,du,
\end{equation*}
as holds for $M=2$ because $d_2=0$. Then every player who has a type $v>c$ with $\overline{v}_{M+1}\leq v\leq\overline{v}_M$ provides at the time at which the curve reaches $v$ (Lemma~\ref{Characterization}(v)), which is
\begin{equation*}
d_M+(M-1)\int_v^{\overline{v}_M}\frac{f(u)}{F(u)W(u)}\,du=\sum_{j=2}^{M}\int_v^{\overline{v}_j}\frac{f(u)}{F(u)W(u)}\,du.
\end{equation*}
This is (\ref{UTD times}), because the terms with $j>M$ vanish ($\overline{v}_j\leq\overline{v}_{M+1}\leq v$). For $M<N$, the value $v=\overline{v}_{M+1}$ gives the formula for $d_{M+1}$, which by (\ref{LTD solution}) is the time at which the curve reaches $\overline{v}_{M+1}$; this completes the induction. Player~1's types above $\overline{v}_2$ provide at time $0$, and for them the right side of (\ref{UTD times}) is also $0$. Finally, the right side of (\ref{UTD times}) does not depend on which player with the highest upper bound is labeled player~1.
\end{proof}

\begin{proof}[Proof of Proposition~\ref{U-LTD expect rate}]
For $v\in(c,\overline{v}_1]$, let $T(v):=$ the right side of (\ref{UTD times}), the stopping time of type $v$ of every player who has this type, and let $T(c):=+\infty$. By (\ref{UTD times}), $T$ vanishes on $[\overline{v}_2,\overline{v}_1]$ and is continuous and strictly decreasing on $(c,\overline{v}_2]$. In a uniform UTD war, each $F_i$ is uniform on $[\underline{v},\overline{v}_i]$, and
\begin{equation*}
\frac{f(u)}{F(u)W(u)}=\frac{c}{r(u-\underline{v})(u-c)}=\frac{1}{\rho r}\,\frac{d}{du}\ln\lambda(u),\qquad u\in(c,\overline{v}].
\end{equation*}
For $v\in(c,\overline{v}_2]$, let $E_v[\,\cdot\,]:=E[\,\cdot\mid t_m>T(v)]$ be the expectation given that nobody has provided by the time at which type $v$ provides. The event $t_m>T(v)$ has probability at least $\prod_{i\in I_N}F_i(c)>0$, since stubborn types never provide. Let $\psi(v):=E_v\big[e^{-\rho r(t_m-T(v))}\big]$ be the expected discount factor at rate $\rho r$ of the remaining wait. I show that $\psi(v)=\lambda(v)$ for every $v\in(c,\overline{v}_2]$ (Steps 1 and 2), and then add the instant exit (Step 3).

\emph{Step 1 (one stage).} Fix $M\in\{2,\dots,N\}$ and values $y<v$ with $\overline{v}_{M+1}\leq y$ and $v\leq\overline{v}_M$. Every value in $(y,v)$ lies below the upper bounds of players $1,\dots,M$ and above those of players $M+1,\dots,N$. So, for $w\in[y,v]$, (\ref{UTD times}) and the display above give
\begin{align*}
T(w)-T(v)&=(M-1)\int_w^v\frac{f(u)}{F(u)W(u)}\,du=\frac{M-1}{\rho r}\ln\frac{\lambda(v)}{\lambda(w)},\\
e^{-\rho r(T(w)-T(v))}&=\Big(\frac{\lambda(w)}{\lambda(v)}\Big)^{M-1},\qquad w\in[y,v].
\end{align*}
Under $E_v$, the values of players $1,\dots,M$ are independent and uniform on $[\underline{v},v]$, because each of them has provided by $T(v)$ exactly when his value is at least $v$, while players $M+1,\dots,N$ do not provide before $T(y)>T(v)$. Let $v^{\max}$ be the highest value among players $1,\dots,M$. If $v^{\max}\geq y$, then $t_m=T(v^{\max})$. If $v^{\max}<y$, then, almost surely, $t_m>T(y)$ when $y>c$, and $t_m=+\infty$ when $y=c$ (which requires $M=N$). So, for $y>c$, the events $\{t_m>T(v),\ v^{\max}<y\}$ and $\{t_m>T(y)\}$ coincide up to a null set, and, with $t_m-T(v)$ written as $(T(y)-T(v))+(t_m-T(y))$, the conditional expected value of $e^{-\rho r(t_m-T(y))}$ given this event is $\psi(y)$. Since $v^{\max}$ has density $M(w-\underline{v})^{M-1}/(v-\underline{v})^M$ at $w\in[\underline{v},v]$ under $E_v$,
\begin{align*}
E_v\big[e^{-\rho r(t_m-T(v))}\mathbbm{1}\{v^{\max}\geq y\}\big]&=\int_y^v\Big(\frac{\lambda(w)}{\lambda(v)}\Big)^{M-1}\frac{M(w-\underline{v})^{M-1}}{(v-\underline{v})^M}\,dw\\
&=\frac{(v-c)^M-(y-c)^M}{(v-\underline{v})(v-c)^{M-1}},\\
E_v\big[e^{-\rho r(t_m-T(v))}\mathbbm{1}\{v^{\max}<y\}\big]&=\Big(\frac{y-\underline{v}}{v-\underline{v}}\Big)^M\Big(\frac{\lambda(y)}{\lambda(v)}\Big)^{M-1}\psi(y)\\
&=\frac{(y-\underline{v})(y-c)^{M-1}}{(v-\underline{v})(v-c)^{M-1}}\,\psi(y),
\end{align*}
where the second expectation is shown for $y>c$; for $y=c$ it is zero. If $y=c$, or if $\psi(y)=\lambda(y)=(y-c)/(y-\underline{v})$, the two expectations add up to $(v-c)/(v-\underline{v})=\lambda(v)$, so $\psi(v)=\lambda(v)$.

\emph{Step 2 (all stages).} Step 1 with $M=N$ and $y=c$ gives $\psi=\lambda$ on $(c,\overline{v}_N]$. If $\psi=\lambda$ on $(c,\overline{v}_{M+1}]$ for some $M\in\{2,\dots,N-1\}$, then Step 1 with $y=\overline{v}_{M+1}$ gives $\psi=\lambda$ on $(c,\overline{v}_M]$. By induction, $\psi(v)=\lambda(v)$ for every $v\in(c,\overline{v}_2]$; tied upper bounds give empty ranges of values, which are skipped.

\emph{Step 3 (instant exit).} With probability $1-F_1(\overline{v}_2)$, player~1's value exceeds $\overline{v}_2$, and he provides at $t=0$. Otherwise, almost surely $t_m>0=T(\overline{v}_2)$, since every type below $\overline{v}_2$ provides after time $0$, if at all, and the expected discount factor at rate $\rho r$ is then $\psi(\overline{v}_2)=\lambda(\overline{v}_2)$, by Step 2. Hence
\begin{equation*}
E[e^{-\rho rt_m}]=1-F_1(\overline{v}_2)+F_1(\overline{v}_2)\lambda(\overline{v}_2)=1-\frac{\overline{v}_2-\underline{v}}{\overline{v}_1-\underline{v}}\cdot\frac{c-\underline{v}}{\overline{v}_2-\underline{v}}=\lambda(\overline{v}_1),
\end{equation*}
which is $\lambda(\max_{i\in I_N}\overline{v}_i)$ by the labeling.
\end{proof}
\section{Proofs for Section~\texorpdfstring{\ref{large societies and asymmetry}}{4}}\label{appendix large societies}

Throughout this appendix, the society is that of Section~\ref{large societies and asymmetry}: every player has cost $c$ and discount rate $r$, group $\ell$ has $n_\ell\geq2$ members, whose values are drawn from $F_\ell$ on $[\underline{v}_\ell,\overline{v}_\ell]$ with density $f_\ell$, and $W(v)=\frac{r}{c}(v-c)$ is the common indifference level. Group $\ell$ is \emph{active} at $t$ if $d_\ell\leq t$. I write $h_{-\ell}$ for the opponents' provision hazard of a member of group $\ell$, and derivatives in $t$ are right derivatives, as in Section~\ref{model}. Every bound below holds for every $N$ and uses only $n_\ell\geq2$. Lemma~\ref{group structure} collects what the proofs take from Section~\ref{equilibrium analysis}. The proofs then rest on two facts: the provision hazard stays close to $W(\overline{v}(\tau))$, the indifference level of the highest type who has not provided by $\tau$, and drifts down slowly (Lemma~\ref{hazard facts}); and Lemma~\ref{hazard to discount} turns such a hazard into bounds on the expected discount factor.

\begin{alemma}[Group structure]\label{group structure}
In the equilibrium of every society:
\begin{enumerate}[(i)]
\item All members of group $\ell$ have the same curve $\Phi_\ell$ and the same activation time $d_\ell<+\infty$, and no player has an atom at any time. The curve $\Phi_\ell$ is continuous, equals $\overline{v}_\ell$ on $[0,d_\ell]$, is strictly decreasing on $[d_\ell,+\infty)$, stays above $c$ and converges to $c$. In particular, $\operatorname{Prob}(t_m=0)=0$ and $\overline{v}(0)=\overline{v}$.
\item Some group is active from time $0$.
\item By (i), the number of active players is $M(t)=\sum_{\ell:\,d_\ell\leq t}n_\ell$. For every $t\geq0$, $\operatorname{Prob}(t_m>t)=\prod_{\ell}F_\ell(\Phi_\ell(t))^{n_\ell}$ and
\begin{equation}\label{group hazard}
h(t)=-\frac{d}{dt}\ln\operatorname{Prob}(t_m>t)=\frac{1}{M(t)-1}\sum_{\ell:\,d_\ell\leq t}n_\ell\,W(\Phi_\ell(t)).
\end{equation}
If group $\ell$ is active at $t$, then each of its members provides at the hazard rate $\eta_\ell(t):=-\frac{d}{dt}\ln F_\ell(\Phi_\ell(t))=h(t)-W(\Phi_\ell(t))$, and $h_{-\ell}(t)=W(\Phi_\ell(t))$. If group $\ell$ is inactive at $t$, then $h_{-\ell}(t)=h(t)\geq W(\overline{v}_\ell)$.
\end{enumerate}
\end{alemma}

\begin{proof}
(i) Two members of a group are identical players, so exchanging their strategies maps every equilibrium to an equilibrium. By Theorem~\ref{exi and uni}, the two strategies then differ only on a set of types of probability zero, and the two members have the same curve and the same activation time. By Lemma~\ref{Characterization}(iv), at most one player has an atom at time $0$, and no player has an atom at a positive time; since each group has at least two members with the same curve, no player has an atom. Then $\operatorname{Prob}(T_i(v_i)>t)$ is continuous in $t$, and so is $\Phi_\ell$, which equals $\overline{v}_\ell$ on $[0,d_\ell]$. By Lemma~\ref{Characterization}(i)--(ii), every type above $c$ provides at a finite time, and $T_i(v)\to+\infty$ as $v\downarrow c$; it follows that $d_\ell<+\infty$, $\Phi_\ell>c$ and $\Phi_\ell(t)\to c$. By Lemma~\ref{Characterization}(v), $\Phi_\ell$ is strictly decreasing on $[d_\ell,+\infty)$. Finally, $\Phi_\ell(0)=\overline{v}_\ell$ for every $\ell$, and therefore $t_m>0$ almost surely and $\overline{v}(0)=\overline{v}$.

(ii) By Lemma~\ref{Characterization}(iii), at least two players have $d_i=0$, and by (i) so do all members of their groups.

(iii) The product follows from independence, and the first equality in (\ref{group hazard}) is the definition of $h$. By Lemma~\ref{Characterization}(vi) and Lemma~\ref{one-sided derivatives}, (\ref{main equations}) holds at every $t\geq0$ for every active player, and so does (\ref{aggregate rate}). In group form, (\ref{aggregate rate}) is the second equality in (\ref{group hazard}) together with $\eta_\ell=h-W(\Phi_\ell)$ for every active group. Each $\ln F_\ell(\Phi_\ell)$ is continuous, and continuously differentiable on each stage with these right derivatives, so it is the integral of its right derivative; this justifies integrating hazards below. A member of group $\ell$ has $h_{-\ell}=h-\eta_\ell=W(\Phi_\ell)$ if $\ell$ is active, and $h_{-\ell}=h$ if $\ell$ is inactive, since he then does not provide. Finally, $h(t)>W(\overline{v}_\ell)$ for $0<t<d_\ell$ by Lemma~\ref{hazard lemma}(c), and since $h$ is right-continuous by (\ref{group hazard}), $h(0)\geq W(\overline{v}_\ell)$ when $d_\ell>0$.
\end{proof}

\paragraph{Notation.} Let $\underline{n}:=\min_\ell n_\ell$ and $\delta_N:=1/(\underline{n}-1)\leq1$. Let $\underline{f}_\ell:=\min_{v\in[c,\overline{v}_\ell]}f_\ell(v)>0$ and $\kappa_N:=2\frac{r}{c}N/\min_\ell(n_\ell\underline{f}_\ell)$; when $n_\ell=p_\ell N$, $\kappa_N=2\frac{r}{c}/\min_\ell(p_\ell\underline{f}_\ell)$ does not depend on $N$, and $\delta_N$ is of order $1/N$. Let $\omega(x):=x/(x+r)$ for $x\geq0$, the expected discount factor of an exponential random time with rate $x$; then $\omega(W(v))=1-c/v$ for $v>c$.

\begin{alemma}[Group hazards]\label{hazard facts}
In the equilibrium of every society, for every $\tau\geq0$:
\begin{enumerate}[(a)]
\item $\sum_{\ell:\,d_\ell\leq\tau}n_\ell\eta_\ell(\tau)=h(\tau)$, and $0\leq\eta_\ell(\tau)\leq h(\tau)/n_\ell$ for every group $\ell$ active at $\tau$. Hence $(1-1/n_\ell)h(\tau)\leq h_{-\ell}(\tau)\leq h(\tau)$ for every group $\ell$.
\item $W(\overline{v}(\tau))\leq h(\tau)\leq(1+\delta_N)W(\overline{v}(\tau))$, and $0\leq\overline{v}(\tau)-\Phi_\ell(\tau)\leq\frac{c}{r}h(\tau)/n_\ell$ for every group $\ell$ active at $\tau$.
\item $\overline{v}(\cdot)$ is nonincreasing, and if $d_\ell\leq s\leq t$, then $\ln F_\ell(\Phi_\ell(s))-\ln F_\ell(\Phi_\ell(t))\leq(1+\delta_N)W(\overline{v}(s))(t-s)/n_\ell$.
\item $W(\overline{v}(\tau))\big(1-\kappa_N(t-\tau)/N\big)\leq h(t)\leq(1+\delta_N)W(\overline{v}(\tau))$ for every $t\geq\tau$.
\end{enumerate}
\end{alemma}

\begin{proof}
(a) Each $\eta_\ell\geq0$ because $\Phi_\ell$ is nonincreasing. By (\ref{group hazard}) and $\eta_\ell=h-W(\Phi_\ell)$, $\sum_{\ell:\,d_\ell\leq\tau}n_\ell\eta_\ell(\tau)=M(\tau)h(\tau)-(M(\tau)-1)h(\tau)=h(\tau)$, and since every term is nonnegative, $n_\ell\eta_\ell(\tau)\leq h(\tau)$. By Lemma~\ref{group structure}(iii), $h_{-\ell}=h-\eta_\ell$ for an active group and $h_{-\ell}=h$ for an inactive one.

(b) Recall that $h_{-\ell}(\tau)=W(\Phi_\ell(\tau))$ for every group $\ell$ active at $\tau$ (Lemma~\ref{group structure}(iii)). Let group $\ell$ attain the maximum in $\overline{v}(\tau)=\max_k\Phi_k(\tau)$. If $\ell$ is inactive at $\tau$, then $h(\tau)\geq W(\overline{v}_\ell)=W(\overline{v}(\tau))$ by Lemma~\ref{group structure}(iii); if it is active, then $h(\tau)\geq W(\Phi_\ell(\tau))=W(\overline{v}(\tau))$ by (a). For the upper bound, take a group $\ell$ active at $\tau$, which exists by Lemma~\ref{group structure}(ii); by (a), $h(\tau)\leq W(\Phi_\ell(\tau))/(1-1/n_\ell)\leq(1+\delta_N)W(\overline{v}(\tau))$. For the last claim, the lower bound and (a) give $W(\overline{v}(\tau))-W(\Phi_\ell(\tau))\leq h(\tau)-(1-1/n_\ell)h(\tau)=h(\tau)/n_\ell$.

(c) Every curve is nonincreasing, and so is $\overline{v}(\cdot)$. For $u\in[s,t]$, (a) and (b) give $\eta_\ell(u)\leq h(u)/n_\ell\leq(1+\delta_N)W(\overline{v}(u))/n_\ell\leq(1+\delta_N)W(\overline{v}(s))/n_\ell$. Integrating $\eta_\ell$ from $s$ to $t$ (Lemma~\ref{group structure}(iii)) gives (c).

(d) Pick a group $\ell^*$ with $\Phi_{\ell^*}(\tau)=\overline{v}(\tau)$, and let $s:=\max\{\tau,d_{\ell^*}\}$, so that $\Phi_{\ell^*}(s)=\overline{v}(\tau)$. For $t\geq s$, (c), $\delta_N\leq1$, $\overline{v}(s)\leq\overline{v}(\tau)$ and $s\geq\tau$ give $\ln F_{\ell^*}(\Phi_{\ell^*}(s))-\ln F_{\ell^*}(\Phi_{\ell^*}(t))\leq x:=2W(\overline{v}(\tau))(t-\tau)/n_{\ell^*}$, so that $F_{\ell^*}(\Phi_{\ell^*}(s))-F_{\ell^*}(\Phi_{\ell^*}(t))\leq1-e^{-x}\leq x$. By the mean value theorem on $[c,\overline{v}_{\ell^*}]$,
\begin{equation*}
\overline{v}(t)\geq\Phi_{\ell^*}(t)\geq\overline{v}(\tau)-\frac{2W(\overline{v}(\tau))(t-\tau)}{n_{\ell^*}\underline{f}_{\ell^*}},
\end{equation*}
and multiplying by $r/c$ gives $W(\overline{v}(t))\geq W(\overline{v}(\tau))\big(1-\kappa_N(t-\tau)/N\big)$, by the definition of $\kappa_N$. For $\tau\leq t<s$, $\overline{v}(t)\geq\Phi_{\ell^*}(t)=\overline{v}(\tau)$. These bounds, (b) at $t$ and $\overline{v}(t)\leq\overline{v}(\tau)$ give (d).
\end{proof}

\begin{alemma}[From hazards to discount factors]\label{hazard to discount}
Let $X\in[0,+\infty]$ be a random time with $\operatorname{Prob}(X>u)=\exp\big(-\int_0^uh_X(s)\,ds\big)$ for $u\geq0$, where $h_X\geq0$ is locally integrable. Let $a>0$, $\varepsilon\geq0$ and $b>0$.
\begin{enumerate}[(i)]
\item If $h_X\leq(1+\varepsilon)a$, then $E[e^{-rX}]\leq\omega((1+\varepsilon)a)\leq\omega(a)+\varepsilon/4$.
\item If $h_X(u)\geq a(1-bu)$ for every $u$, then $E[e^{-rX}]\geq\omega(a)-(e-1)ab/r^2-\exp\big(-r(2/(ab))^{1/2}\big)$.
\end{enumerate}
\end{alemma}

\begin{proof}
Integrating by parts,
\begin{equation*}
E[e^{-rX}]=1-r\int_0^\infty e^{-ru}\operatorname{Prob}(X>u)\,du=\omega(a)-r\int_0^\infty e^{-ru}\big[\operatorname{Prob}(X>u)-e^{-au}\big]du.
\end{equation*}

(i) Here $\operatorname{Prob}(X>u)\geq e^{-(1+\varepsilon)au}$, so $E[e^{-rX}]\leq\omega((1+\varepsilon)a)$. Since $\omega$ is concave, $\omega((1+\varepsilon)a)-\omega(a)\leq\varepsilon a\omega'(a)=\varepsilon ar/(a+r)^2\leq\varepsilon/4$.

(ii) Here $\operatorname{Prob}(X>u)\leq\min\{1,e^{-au+abu^2/2}\}$. Let $u_0:=(2/(ab))^{1/2}$. For $u\leq u_0$, $y:=abu^2/2\leq1$, so by convexity $\operatorname{Prob}(X>u)-e^{-au}\leq e^{-au}(e^y-1)\leq(e-1)y$, and $r\int_0^\infty e^{-ru}(e-1)\frac{ab}{2}u^2\,du=(e-1)ab/r^2$. For $u>u_0$, the difference is at most $1$, and $r\int_{u_0}^\infty e^{-ru}du=e^{-ru_0}$. Together, these bounds give (ii).
\end{proof}

\subsection{Proofs of Lemma~\texorpdfstring{\ref{Aligned Welfare}}{3} and Theorem~\texorpdfstring{\ref{U-LTD equivalency}}{2}}\label{proof Aligned Welfare}\label{proof U-LTD equivalency}

I prove Theorem~\ref{U-LTD equivalency} first and deduce Lemma~\ref{Aligned Welfare} from it.

\paragraph{Continuation payoffs.} Fix $\tau\geq0$, a member $i$ of group $\ell$, and a type $v\leq\Phi_\ell(\tau)$ of his, who has not provided before $\tau$. Let $T_{-i}$, the time at which an opponent of $i$ first provides, and $Q_{-i}(t)=\operatorname{Prob}(T_{-i}>t)$ be as in Appendix~\ref{appendix preliminaries}; by Lemma~\ref{group structure}(iii), $Q_{-i}(t)=\exp\big(-\int_0^{t}h_{-\ell}(s)\,ds\big)>0$. Since no player has an atom (Lemma~\ref{group structure}(i)), ties have probability zero, and stopping at $t\in[\tau,+\infty]$ gives type $v$, given $T_{-i}>\tau$ and discounted to $\tau$, the expected payoff
\begin{equation}\label{continuation payoff}
R_{\ell,\tau}(t|v):=\frac{1}{Q_{-i}(\tau)}\Big[v\int_\tau^{t}e^{-r(s-\tau)}h_{-\ell}(s)Q_{-i}(s)\,ds+(v-c)e^{-r(t-\tau)}Q_{-i}(t)\Big].
\end{equation}
For $\tau=0$, (\ref{continuation payoff}) is (\ref{expected gain}). At every $t\geq\tau$, as in (\ref{payoff slope}), $\frac{d}{dt}R_{\ell,\tau}(t|v)=c\,e^{-r(t-\tau)}\frac{Q_{-i}(t)}{Q_{-i}(\tau)}\big[h_{-\ell}(t)-W(v)\big]$. Let $V_{\ell,\tau}(v):=\sup_{t\in[\tau,+\infty]}R_{\ell,\tau}(t|v)$. By Lemma~\ref{group structure}(i) and (iii), $h_{-\ell}\geq W(\overline{v}_\ell)\geq W(v)$ before $d_\ell$, and $h_{-\ell}=W(\Phi_\ell)$ is strictly decreasing after $d_\ell$. So $R_{\ell,\tau}(\cdot|v)$ is nondecreasing until $\Phi_\ell$ falls below $v$ and strictly decreasing afterwards, and the equilibrium stopping time of type $v$ attains $V_{\ell,\tau}(v)$. Since he has not provided before $\tau$, the good has not been provided before $\tau$ exactly when $T_{-i}\geq\tau$, which differs from $T_{-i}>\tau$ by a null event; hence $V_{\ell,\tau}(v)$ is the expected payoff at $\tau$ in Theorem~\ref{U-LTD equivalency}(ii). Finally, let $\psi_\tau:=E[e^{-r(t_m-\tau)}\mid t_m>\tau]$; since no player has an atom, conditioning on $t_m\geq\tau$ gives the same value.

\begin{proof}[Proof of Theorem~\ref{U-LTD equivalency}]
I show that for every $N$, every $\tau\geq0$, every group $\ell$ and every $v\in[\underline{v}_\ell,\Phi_\ell(\tau)]$,
\begin{align}
\big|\psi_\tau-\big(1-c/\overline{v}(\tau)\big)\big|&\leq\varepsilon_N,\label{T2 discount}\\
\big|V_{\ell,\tau}(v)-v\big(1-c/\overline{v}(\tau)\big)\big|&\leq|v|\Big(\varepsilon_N+\frac{1}{e(n_\ell-1)}\Big)+(\overline{v}-c)\Big(\frac{1}{n_\ell}+\frac{\kappa_N}{rN}\Big),\label{T2 payoff}
\end{align}
where
\begin{equation*}
\varepsilon_N:=\max\Big\{\frac{\delta_N}{4},\ \frac{(e-1)\kappa_N W(\overline{v})}{r^2N}+\exp\Big(-r\Big(\frac{2N}{\kappa_N W(\overline{v})}\Big)^{1/2}\Big)\Big\}.
\end{equation*}
When $n_\ell=p_\ell N$, the terms $\varepsilon_N$, $1/(n_\ell-1)$, $1/n_\ell$ and $\kappa_N/N$ are at most of order $1/N$, and $|v|\leq\max_\ell\max\{|\underline{v}_\ell|,\overline{v}_\ell\}$. So (\ref{T2 discount}) and (\ref{T2 payoff}) prove the theorem, with errors at most of order $1/N$, uniformly in $\tau$, $v$ and $\ell$.

For (\ref{T2 discount}), let $a:=W(\overline{v}(\tau))$, so that $0<a\leq W(\overline{v})$ and $\omega(a)=1-c/\overline{v}(\tau)$. Given $t_m>\tau$, the remaining wait $X:=t_m-\tau$ satisfies $\operatorname{Prob}(X>u)=\exp\big(-\int_0^{u}h(\tau+s)\,ds\big)$ by (\ref{group hazard}), and by Lemma~\ref{hazard facts}(d), $a(1-\kappa_N u/N)\leq h(\tau+u)\leq(1+\delta_N)a$. Lemma~\ref{hazard to discount}(i) with $\varepsilon=\delta_N$ gives $\psi_\tau\leq\omega(a)+\delta_N/4$. Lemma~\ref{hazard to discount}(ii) with $b=\kappa_N/N$ gives $\psi_\tau\geq\omega(a)-(e-1)a\kappa_N/(r^2N)-\exp\big(-r(2N/(a\kappa_N))^{1/2}\big)$, and this error increases in $a$, so it is at most the second term in $\varepsilon_N$.

For (\ref{T2 payoff}), I first compare $\psi_\tau$ with $\widetilde{\psi}_\tau:=E[e^{-r(T_{-i}-\tau)}\mid T_{-i}>\tau]$, which leaves out player $i$'s own provision. Let $\Lambda(s):=\int_\tau^{s}h(u)\,du$, so that $\operatorname{Prob}(t_m>s\mid t_m>\tau)=e^{-\Lambda(s)}$ and, by the first display in the proof of Lemma~\ref{hazard to discount},
\begin{equation*}
\psi_\tau-\widetilde{\psi}_\tau=r\int_\tau^\infty e^{-r(s-\tau)}\Big[\frac{Q_{-i}(s)}{Q_{-i}(\tau)}-e^{-\Lambda(s)}\Big]ds.
\end{equation*}
By Lemma~\ref{hazard facts}(a), $e^{-\Lambda(s)}\leq Q_{-i}(s)/Q_{-i}(\tau)\leq e^{-(1-1/n_\ell)\Lambda(s)}$ for $s\geq\tau$. Since $1-e^{-x}\leq x$ and $xe^{-x}\leq1/e$ for $x\geq0$, it follows that
\begin{equation*}
0\leq\frac{Q_{-i}(s)}{Q_{-i}(\tau)}-e^{-\Lambda(s)}\leq e^{-(1-\frac{1}{n_\ell})\Lambda(s)}\big(1-e^{-\Lambda(s)/n_\ell}\big)\leq\frac{\Lambda(s)}{n_\ell}e^{-(1-\frac{1}{n_\ell})\Lambda(s)}\leq\frac{1}{e(n_\ell-1)},
\end{equation*}
and hence $0\leq \psi_\tau-\widetilde{\psi}_\tau\leq1/(e(n_\ell-1))$.

Next, never providing gives $V_{\ell,\tau}(v)\geq R_{\ell,\tau}(+\infty|v)=v\widetilde{\psi}_\tau$. For $t\geq\tau$, integrating the derivative of $R_{\ell,\tau}(\cdot|v)$ from $t$ to $+\infty$ and using $Q_{-i}\leq Q_{-i}(\tau)$ on $[\tau,+\infty)$ gives
\begin{equation*}
R_{\ell,\tau}(t|v)\leq v\widetilde{\psi}_\tau+c\int_\tau^\infty e^{-r(s-\tau)}\big[W(v)-h_{-\ell}(s)\big]^+ds.
\end{equation*}
Since $v\leq\overline{v}(\tau)$, $W(v)\leq a$, and by Lemma~\ref{hazard facts}(a) and (d), $h_{-\ell}(s)\geq(1-1/n_\ell)a\big(1-\kappa_N(s-\tau)/N\big)$; so the bracket is at most $a/n_\ell+a\kappa_N(s-\tau)/N$. Integrating, and using $\frac{c}{r}a=\overline{v}(\tau)-c\leq\overline{v}-c$, gives
\begin{equation*}
0\leq V_{\ell,\tau}(v)-v\widetilde{\psi}_\tau\leq(\overline{v}-c)\Big(\frac{1}{n_\ell}+\frac{\kappa_N}{rN}\Big).
\end{equation*}
The triangle inequality, the two bounds just shown and (\ref{T2 discount}) give (\ref{T2 payoff}).
\end{proof}

\begin{proof}[Proof of Lemma~\ref{Aligned Welfare}]
At $\tau=0$, $\operatorname{Prob}(t_m>0)=1$, $\Phi_\ell(0)=\overline{v}_\ell$ and $\overline{v}(0)=\overline{v}$ (Lemma~\ref{group structure}(i)), so $\psi_0=E[e^{-rt_m}]$, and $V_{\ell,0}(v)$ is the expected payoff of every type $v\in[\underline{v}_\ell,\overline{v}_\ell]$. Lemma~\ref{Aligned Welfare} is therefore the case $\tau=0$ of (\ref{T2 discount}) and (\ref{T2 payoff}).

The proof also gives the facts stated after Lemma~\ref{Aligned Welfare}. By Lemma~\ref{hazard facts}(d) at $\tau=0$, $W(\overline{v})(1-\kappa_N t/N)\leq h(t)\leq(1+\delta_N)W(\overline{v})$ for every $t\geq0$, so the provision hazard converges to $W(\overline{v})$ at every fixed time, uniformly on bounded time intervals. Hence, by (\ref{group hazard}), $\operatorname{Prob}(t_m>t)=\exp\big(-\int_0^th(s)\,ds\big)\to e^{-W(\overline{v})t}$ for every $t\geq0$, so $t_m$ converges in distribution to an exponential random variable with rate $W(\overline{v})$, and $\operatorname{Prob}(t_m=+\infty)=\prod_\ell F_\ell(c)^{n_\ell}\leq(\max_\ell F_\ell(c))^N$ converges to zero. Finally, every curve moves slowly. For $t\geq d_\ell$, Lemma~\ref{hazard facts}(c) with $s=d_\ell$ gives $-\ln F_\ell(\Phi_\ell(t))\leq2W(\overline{v})t/n_\ell$, and as in the proof of Lemma~\ref{hazard facts}(d), $\overline{v}_\ell-\Phi_\ell(t)\leq2W(\overline{v})t/(n_\ell\underline{f}_\ell)$; this is of order $1/N$ at every fixed time.
\end{proof}

\subsection{Proof of Lemma~\texorpdfstring{\ref{decisive upper bound}}{4}}\label{proof decisive upper bound}

\begin{proof}[Proof of Lemma~\ref{decisive upper bound}]
Let $\overline{v}_\ell>\overline{v}_{\ell'}$ and $n_{\ell'}>(\overline{v}_\ell-c)/(\overline{v}_\ell-\overline{v}_{\ell'})$, and put $v^*:=c+\frac{n_{\ell'}}{n_{\ell'}-1}(\overline{v}_{\ell'}-c)$, so that $W(v^*)=\frac{n_{\ell'}}{n_{\ell'}-1}W(\overline{v}_{\ell'})$. Since $(n_{\ell'}-1)(\overline{v}_\ell-v^*)=n_{\ell'}(\overline{v}_\ell-\overline{v}_{\ell'})-(\overline{v}_\ell-c)$, the condition on $n_{\ell'}$ is equivalent to $v^*<\overline{v}_\ell$.

At $t=d_{\ell'}$, group $\ell'$ is active, so Lemmas~\ref{group structure}(iii) and~\ref{hazard facts}(a) give $(1-1/n_{\ell'})h(t)\leq h_{-\ell'}(t)=W(\Phi_{\ell'}(t))\leq W(\overline{v}_{\ell'})$, that is, $h(t)\leq W(v^*)$. By Lemma~\ref{hazard facts}(b), also $h(t)\geq W(\overline{v}(t))\geq W(\Phi_\ell(t))$. Hence
\begin{equation}\label{cap at activation}
\Phi_\ell(d_{\ell'})\leq v^*<\overline{v}_\ell.
\end{equation}
If $d_{\ell'}\leq d_\ell$, then $\Phi_\ell(d_{\ell'})=\overline{v}_\ell$ by Lemma~\ref{group structure}(i), which contradicts (\ref{cap at activation}). So $d_\ell<d_{\ell'}$.

For the size of the difference, apply Lemma~\ref{hazard facts}(c) with $s=d_\ell$ and $t=d_{\ell'}$. Since $F_\ell(\Phi_\ell(d_\ell))=1$ and $\overline{v}(d_\ell)\leq\overline{v}$, (\ref{cap at activation}) gives
\begin{equation*}
\ln\frac{1}{F_\ell(v^*)}\leq-\ln F_\ell(\Phi_\ell(d_{\ell'}))\leq(1+\delta_N)W(\overline{v})\,\frac{d_{\ell'}-d_\ell}{n_\ell}.
\end{equation*}
As $N\to\infty$ with fixed proportions, $n_{\ell'}=p_{\ell'}N$ eventually exceeds the threshold, $n_\ell=p_\ell N$, $\delta_N\to0$ and $v^*\downarrow\overline{v}_{\ell'}$, so
\begin{equation*}
\liminf_{N\to\infty}\frac{d_{\ell'}-d_\ell}{N}\geq\frac{p_\ell\ln\big(1/F_\ell(\overline{v}_{\ell'})\big)}{W(\overline{v})}>0,
\end{equation*}
because $\overline{v}_{\ell'}<\overline{v}_\ell$ gives $F_\ell(\overline{v}_{\ell'})<1$.
\end{proof}

I now prove the claim after Lemma~\ref{decisive upper bound}: if $\overline{v}_\ell>\overline{v}_{\ell'}$, then the probability that the good has been provided by the time group $\ell'$ becomes active converges to one as $N\to\infty$. By Lemma~\ref{group structure}(iii), $h\geq W(\overline{v}_{\ell'})$ on $[0,d_{\ell'})$, so $\operatorname{Prob}(t_m\geq d_{\ell'})=\operatorname{Prob}(t_m>d_{\ell'})\leq\exp\big(-W(\overline{v}_{\ell'})d_{\ell'}\big)$, which converges to zero because $d_{\ell'}\geq d_{\ell'}-d_\ell$ and $\liminf_{N\to\infty}(d_{\ell'}-d_\ell)/N>0$.

\end{document}